\documentclass[12pt]{article}  
\usepackage[a4paper,textwidth=490.0pt,textheight=703.1pt]{geometry}
\usepackage{slashed}
\usepackage{graphicx}
\usepackage{epsfig}
\usepackage{amsmath}
\usepackage{amssymb}
\usepackage{ulem}
\usepackage{mdwlist}
\usepackage{color}                                     
\usepackage{float}
\usepackage[rflt]{floatflt}
\usepackage{slashed}
\usepackage{cite}

\usepackage{array}

\usepackage[english]{babel}
\usepackage[latin1]{inputenc}
\usepackage[T1]{fontenc}
\usepackage{ae}

\usepackage{url}

\usepackage{amsmath, amsthm, amssymb}

\newtheorem{thm}{Theorem}[section]

\usepackage{array}

\usepackage[english]{babel}
\usepackage[latin1]{inputenc}
\usepackage[T1]{fontenc}
\usepackage{ae}

\newtheorem{definition}[thm]{Definition}

\newcommand{\gsim}{\raisebox{-0.07cm   }
{$\, \stackrel{>}{{\scriptstyle\sim}}\, $}}
\newcommand{\GeV}{\rm GeV}

\newcommand{\Li}{{\rm Li}}

\usepackage{rotating}

\usepackage{graphicx}

\newcounter{mmacnt}
\def\restartmma{\setcounter{mmacnt}{0}}
\restartmma \catcode`|=\active
\def|#1|{\mathrm{#1}}
\catcode`|=12
\newenvironment{mma}{
 \par\smallskip
 \catcode`|=\active
 \parskip=0pt\parindent=0pt 
 \small
 \def\In##1\\{%
   \def\linebreak{\hfill\break\null\qquad}%
   \refstepcounter{mmacnt}
   \hangindent=2.5em\hangafter=0
   \leavevmode
   \llap{\tiny\sffamily In[\arabic{mmacnt}]:=\kern.5em}%
   \mathversion{bold}\footnotesize$\displaystyle##1$\normalsize
   \mathversion{normal}\par
 }%
 \def\Print##1\\{%
   \def\linebreak{\hfill\break}%
   \hangindent=2.5em\hangafter=0
   \leavevmode ##1\par}%
 \def\Out##1\\{%
   \def\linebreak{$\hfill\break\null\hfill$}%
   \kern\abovedisplayskip\par
   \hangindent=2.5em\hangafter=0
   \leavevmode
   \llap{\tiny\sffamily Out[\arabic{mmacnt}]=\kern.5em}
   \footnotesize$\displaystyle##1$\normalsize\hfill\null\par
   \kern\belowdisplayskip
 }%
 \def\Warning##1##2\\{%
   \def\linebreak{\hfill\break}%
   \hangindent=2.5em\hangafter=0
   \leavevmode
   {\scriptsize##1 : ##2}\par}%
}{%
 \par\smallskip
}

\usepackage{color}

\newenvironment{fshaded}{%
\MakeFramed {\FrameRestore}
}%
{\endMakeFramed}

\usepackage{tikz}
\usetikzlibrary{matrix}

\allowdisplaybreaks[4]

\begin{document}
\setlength{\baselineskip}{0.515cm}
\sloppy
\thispagestyle{empty}
\begin{flushleft}
DESY 15--171
\hfill 
\\
DO--TH 15/14  \\
May 2016\\
\end{flushleft}

\setcounter{table}{0}

\mbox{}
\vspace*{\fill}
\begin{center}

{\large\bf The Complete \boldmath $O(\alpha_s^2)$ Non-Singlet Heavy Flavor Corrections}

\vspace*{2mm}
{\large \bf \boldmath to the Structure Functions $g_{1,2}^{ep}(x,Q^2)$,
  $F_{1,2,L}^{ep}(x,Q^2)$, $F_{1,2,3}^{\nu(\bar{\nu})}(x,Q^2)$} 

\vspace*{2mm}
{\large \bf \boldmath and the Associated Sum Rules}

\vspace{4cm}
\large
Johannes~Bl\"umlein,
Giulio~Falcioni\footnote{HiggsTools Fellow}, and
Abilio~De Freitas

\vspace{1.5cm}
\normalsize   
{\it  Deutsches Elektronen--Synchrotron, DESY,}\\
{\it  Platanenallee 6, D-15738 Zeuthen, Germany}
\\

\end{center}
\normalsize
\vspace{\fill}
\begin{abstract}
\noindent 
We calculate analytically the flavor non-singlet $O(\alpha_s^2)$ massive Wilson coefficients for the inclusive 
neutral current non-singlet structure functions $F_{1,2,L}^{ep}(x,Q^2)$ and $g_{1,2}^{ep}(x,Q^2)$  and charged 
current non-singlet structure functions $F_{1,2,3}^{\nu(\bar{\nu})p}(x,Q^2)$, at general virtualities $Q^2$ in 
the deep-inelastic region. Numerical results are presented. We illustrate the transition from low 
to large virtualities for these observables, which may be contrasted to basic assumptions made in the so-called 
variable flavor number scheme. We also derive the corresponding results for the Adler sum rule, the unpolarized 
and polarized Bjorken sum rules and the Gross-Llewellyn Smith sum rule. There are no logarithmic corrections 
at large scales $Q^2$ and the effects of the power corrections due to the heavy quark mass are of the size of the 
known $O(\alpha_s^4)$ corrections in the case of the sum rules. The complete charm and bottom corrections 
are compared to the approach using asymptotic representations in the region $Q^2 \gg m_{c,b}^2$. We also 
study the target mass corrections to the above sum rules. 
\end{abstract}

\vspace*{\fill}
\noindent
\numberwithin{equation}{section}

\newpage
\section{Introduction}
\label{sec:1}

\vspace*{1mm}
\noindent
Deep-inelastic scattering provides one of the most direct methods to measure the strong coupling
constant from precision data on the scaling violations of the nucleon structure functions 
\cite{Blumlein:2012bf,Bethke:2011tr}. The present accuracy of these data also allows to measure the 
mass of the charm, cf. \cite{Alekhin:2012vu}, and bottom quarks due to the heavy flavor contributions. 
The Wilson coefficients are known to 2-loop order in semi-analytic form 
\cite{Laenen:1992zk,Riemersma:1994hv,Bierenbaum:2009zt} in the tagged-flavor case\footnote{For a precise
implementation in Mellin space, see~\cite{Alekhin:2003ev}.}, i.e. for the subset in which the hadronic 
final state contains at least one heavy quark, having been produced in the hard scattering process. 
The corresponding reduced cross section does not correspond to the notion of structure functions, since 
those are purely inclusive quantities and terms containing  massless final states contribute as well. 
The heavy flavor contribution to inclusive deep-inelastic structure functions are described by five Wilson 
coefficients in the case of pure photon exchange \cite{Buza:1995ie,Buza:1996wv,Bierenbaum:2009mv}. In the 
asymptotic case $Q^2 \gg m^2$, where $Q^2 = -q^2$ denotes the virtuality of the exchanged gauge boson 
and $m$ the 
mass of the heavy quark, analytic expressions for the Wilson coefficients have been calculated. A series of 
Mellin moments has been computed to 3-loop order in \cite{Bierenbaum:2009mv}. All logarithmic 
3-loop corrections
\cite{Behring:2014eya} as well as 
all $N_F$ terms are known \cite{Ablinger:2010ty,Blumlein:2012vq}. Four out of five Wilson coefficients contributing 
to the unpolarized deep inelastic structure functions have been calculated to 3-loop order for general values of 
Mellin $N$ \cite{Ablinger:2010ty,Ablinger:2014vwa,Ablinger:2014nga} in the asymptotic region $Q^2 \gg m^2$. 
In the flavor non-singlet 
case also the asymptotic 3-loop contributions to the combinations of the polarized structure functions 
$g_{1(2)}^{\rm NS}$ \cite{Behring:2015zaa} and the unpolarized charged current structure 
function 
$xF_3^{\bar{\nu}p} + xF_3^{{\nu}p}$ have been computed \cite{Behring:2015roa}.

In the present paper, we calculate the complete 2-loop non-singlet heavy flavor corrections to the deep 
inelastic charged current structure functions $F_{1,2,3}^{\nu p}$ and the neutral current structure functions
$F_{1,2}^{ep}$ and $g_1^{ep}$ and a series of sum rules in the deep inelastic region, $Q^2 \gsim m_c^2$.
In the asymptotic case $Q^2 \gg m^2$ 
the corresponding Wilson coefficients have been calculated in \cite{Behring:2014eya,Blumlein:2014fqa,Behring:2015zaa,Behring:2015roa} 
to $O(\alpha_s^2)$ and in \cite{Ablinger:2014vwa,Behring:2015zaa,Behring:2015roa} to $O(\alpha_s^3)$. 
Here the massless Wilson coefficients \cite{Vermaseren:2005qc,Moch:2008fj}
to $O(\alpha_s^3)$ enter. 
In the tagged flavor case the 
corresponding corrections to $O(\alpha_s^2)$ have been calculated in \cite{Buza:1995ie,Buza:1996xr} and in the asymptotic charged 
current case in \cite{Buza:1997mg}\footnote{This result has been corrected in Ref.~\cite{Blumlein:2014fqa}.}. 

The associated sum 
rules are the Adler sum rule \cite{Adler:1965ty}, the unpolarized Bjorken sum rule \cite{Bjorken:1967px}, the polarized Bjorken sum 
rule \cite{Bjorken:1969mm}, and the Gross--Llewellyn Smith sum rule \cite{Gross:1969jf}. A central observation in the inclusive case 
is that there are no logarithmic corrections for the associated sum rules at large $Q^2$, which are present in the 
tagged flavor case \cite{Blumlein:1998sh,vanNeerven:1999ec}, however. The complete massive $O(\alpha_s^2)$ corrections to 
the structure functions
improves the accuracy towards lower values of $Q^2$. In the case of the sum rules, the 
corresponding 
contributions are found to be of the order of the known massless 4-loop corrections.
We will also consider
the target mass corrections to the sum rules, since they are relevant in the region of low $Q^2$.

The paper is organized as follows. In Section~\ref{sec:2} we present a general outline on the massive Wilson coefficients for the 
structure functions which will be considered. The $O(\alpha_s^2)$ corrections to the polarized non-singlet neutral current 
structure functions $g_1^{ep, \rm NS}$ and $g_2^{ep, \rm NS}$ are derived in detail in Section~\ref{sec:3} 
as an example.
In Section~\ref{sec:4} we discuss the corrections to the neutral current structure functions 
$F_{1(2)}^{ep, \rm NS}$, and in Section~\ref{sec:5} those to the non-singlet charged current structure 
functions $F_{1,2,3}^{\nu(\bar{\nu})p, \rm NS}$. Detailed numerical results are presented for all the 
seven 
non-singlet structure functions for experimental use. The heavy 
flavor $O(\alpha_s^2)$ corrections and target mass corrections to the 
associated sum rules are computed in Section~\ref{sec:6}, comparing  massless and massive effects 
and numerical results are presented for the target mass corrections. 
Section~\ref{sec:7} contains the 
conclusions. The Appendices contain technical parts of the calculation.
\section{The Wilson Coefficients}
\label{sec:2}

\vspace*{1mm}
\noindent
We consider the heavy flavor corrections to deep-inelastic structure functions, which are inclusive
observables, i.e. the hadronic final state in the corresponding differential scattering cross sections
is summed over completely. Under this condition the Kinoshita-Lee-Nauenberg theorem 
\cite{Kinoshita:1962ur,Lee:1964is} is valid and no infrared singularities, which have to be eventually 
cured by arbitrary cuts, are present \cite{Bierenbaum:2009mv}. As we consider deep-inelastic 
scattering, both the scales $Q^2$ and $W^2 = Q^2 (1-x)/x + M^2$ have to be large enough, to probe 
the interior of the nucleon. Here $M$ denotes the nucleon mass, $x = Q^2/(Sy)$ is the Bjorken variable,
with $S = (p+l)^2$, and $y= p.q/p.l$ the inelasticity, and $p$ and $l$ the incoming nucleon and 
lepton 4-momenta. One usually demands $W^2, Q^2 
\gsim 4~\GeV^2$. To fully avoid the region of higher twist terms, a cut $W^2 \gsim 12.5~\GeV^2$ 
\cite{Blumlein:2006be} is necessary. 

The structure functions are then given by
\begin{eqnarray}
F_i(x,Q^2) &=& F_i^{\rm massless}(x,Q^2) + F_i^{\rm massive}(x,Q^2), 
\label{eq:Fi}
\end{eqnarray}
where $F_i^{\rm massless}(x,Q^2)$ is the fully massless part of the structure function and
$F_i^{\rm massive}(x,Q^2)$ contains contributions due to a heavy quark mass $m_c$ or $m_b$. 
Both quantities are inclusive.  $F_i^{\rm massive}(x,Q^2)$ does not correspond to the so-called 
tagged flavor case, demanding a heavy
quark in the hadronic final state.\footnote{The 
request to tag heavy quarks in the final state usually leads to jet-cone definitions and thus to 
additional unphysical logarithmic contributions. In the past the idea to rather 
compute tagged heavy flavor structure functions up to next-to-leading order (NLO) was motivated by 
experimental measurements \cite{Laenen:1992zk,Buza:1995ie,Buza:1996xr,Buza:1997mg}.} In the 
asymptotic case $Q^2 \gg m^2$, the Wilson coefficients contributing to (\ref{eq:Fi}) were calculated
for the non-singlet neutral current structure functions $g_{1,2}^{\rm NS}$ and $F_{1,2}^{\rm NS}$ and 
the non-singlet charged current structure function $F_{3}^{\rm NS}$ 
\cite{Blumlein:2014fqa,Behring:2015zaa,Behring:2015roa} to $O(\alpha_s^3)$ (NNLO).

The unpolarized and polarized neutral current non-singlet structure functions in the case of pure photon 
exchange are given by 
\begin{eqnarray}
F_{i}^{\rm NS}(x,Q^2) &=& r_i
\sum_{k=1}^{N_F} e_k^2 \left[C_{F_i,q}^{\rm NS}\left(x,N_F,\frac{Q^2}{\mu^2}\right) 
+ L_{F_i,q}^{\rm NS}\left(x, N_F+1,\frac{Q^2}{m^2}, \frac{m^2}{\mu^2}\right)\right]
\nonumber\\ && \hspace*{5cm}
\otimes \left[f_k(x,\mu^2,N_F) + f_{\bar{k}}(x,\mu^2,N_F) \right]
\\
g_{i}^{\rm NS}(x,Q^2) &=& \frac{1}{2}
\sum_{k=1}^{N_F} e_k^2 \left[\Delta C_{g_i,q}^{\rm NS}\left(x,N_F,\frac{Q^2}{\mu^2}\right) 
+ \Delta L_{g_i,q}^{\rm NS}\left(x, N_F+1,\frac{Q^2}{m^2}, \frac{m^2}{\mu^2}\right)\right]
\nonumber\\ && \hspace*{5cm}
\otimes 
\left[\Delta f_k(x,\mu^2,N_F) + \Delta f_{\bar{k}}(x,\mu^2,N_F) \right],
\end{eqnarray}
with $i = 1,2$. 
Here $N_F$ is the number of active flavors, $e_k$ the electric 
charge of the massless quarks, and $r_1 = \tfrac{1}{2}, r_2 = x$; $C_{q}^{\rm NS}$ and
$\Delta C_{q}^{\rm NS}$ denote the corresponding massless Wilson coefficients and $L_{q}^{\rm NS}$ and 
$\Delta L_{q}^{\rm NS}$ the massive ones, $f_{k(\bar{k})}$ and $\Delta f_{k(\bar{k})}$ are the
unpolarized (polarized) quark and anti-quark distribution functions, and $\mu^2$ is the 
factorization 
scale. Here we follow again the convention used in \cite{Buza:1996wv}, (2.26). The notion `$N_F + 1$' 
in $(\Delta)L_{q}^{\rm NS}$ means that the Wilson coefficient is calculated for $N_F$ massless 
and one massive flavor.\footnote{From 3-loop order onward there are also genuine contributions 
due to two different heavy quarks \cite{TWOH}.} 

For the unpolarized charged current structure functions $F_{1,2,3}(x,Q^2)$ a second Wilson coefficient 
$H_{i,q}^{W^+ - W^-,\rm NS}$  contributes, which in the case of charm describes the flavor 
excitation
\begin{eqnarray}
d \sin^2(\theta_c) + s \cos^2(\theta_c) \rightarrow c
\label{eq:trans}
\end{eqnarray}
in addition to or without heavy flavor pair production and possible virtual heavy quark corrections.
This transition contributes already at tree level. Here $\theta_c$ denotes the Cabibbo-angle 
\cite{Cabibbo:1963yz}. The complete 
corrections to $O(\alpha_s)$ have been calculated in~\cite{Gluck:1997sj,Blumlein:2011zu}\footnote{See 
also \cite{Gottschalk:1980rv} and the discussion in Ref.~\cite{Blumlein:2011zu}.}. 
At $O(\alpha_s^2)$ the asymptotic heavy flavor corrections have been calculated in \cite{Blumlein:2014fqa}
and for $xF_3^{\nu p}$ + $xF_3^{\bar{\nu} p}$ to $O(\alpha_s^3)$ in Ref.~\cite{Behring:2015roa}. Beyond the
terms of $O(\alpha_s)$ we will use the results in the asymptotic case for the numerical illustrations 
given below. Note that the latter contributions are Cabibbo suppressed 
\begin{equation}
\propto |V_{cd}|^2 (d - \overline{d}). 
\end{equation}
Given the present experimental accuracy, this approximation is justified, leaving the 
full calculation for the future.

For the transition (\ref{eq:trans}) the momentum fraction of the massless quarks at tree-level 
changes, 
as well known, to 
\begin{eqnarray}
z = x \left(1 + \frac{m^2}{Q^2}\right) \equiv \frac{x}{\bar{\lambda}},
\end{eqnarray}
because the corresponding Wilson coefficient is a $\delta$-distribution. This is different at higher 
orders, where the Wilson coefficients are given by extended 
distributions,~\cite{Gluck:1997sj,Blumlein:2011zu}. In the asymptotic region $Q^2 \gg m^2$ the following 
representations hold for the Wilson coefficients
$L_{i,q}^{W^+ - W^-,\rm NS}$ and $H_{i,q}^{W^+ - W^-,\rm NS}$~:
\begin{eqnarray}
L_{i,q}^{W^+ - W^-,\rm NS}(N_F+1) &=& a_s^2 \left[ A_{qq,Q}^{(2),\rm NS} + \hat{C}_{i,q}^{(2),W^+ 
- W^-, \rm 
NS}(N_F)\right]  + 
a_s^3 \Biggl[A_{qq,Q}^{(3),\rm NS}  
\nonumber\\  &&
+ A_{qq,Q}^{(2),\rm NS} C_{i,q}^{(1),W^+ - W^-, \rm NS}(N_F+1) + \hat{C}_{i,q}^{(3),W^+ - W^-, \rm 
NS}(N_F)\Biggr]~, 
\label{eq:WIL1}
\\ 
H_{i,q}^{W^+ - W^-,\rm NS}(N_F+1) &=& 
1 + a_s C_{i,q}^{(1),W^+ - W^-, \rm NS}(N_F+1)
\nonumber\\ &&
+ a_s^2 \left[A_{qq,Q}^{(2),\rm NS} + C_{i,q}^{(2),W^+ - W^-, \rm NS}(N_F+1) \right]
\nonumber\\ &&
+ a_s^3 \Biggl[A_{qq,Q}^{(3),\rm NS} 
                + A_{qq,Q}^{(2),\rm NS} C_{i,q}^{(1),W^+ - W^-, \rm NS}(N_F+1) 
\nonumber\\ && 
+ {C}_{i,q}^{(3),W^+ - W^-, \rm NS}(N_F+1)\Biggr] 
\label{eq:WIL2}
\\
&=& 
L_{i,q}^{W^+ - W^-,\rm NS}(N_F+1) + {C}_{i,q}^{W^+ - W^-, \rm 
NS}(N_F),~~~~i = 1,2,3~,
\label{eq:H3}
\end{eqnarray}
with $a_s(\mu^2) = \alpha_s(\mu^2)/(4\pi)$, $A_{qq,Q}^{\rm NS}$ the massive non-singlet operator matrix 
element (OME) 
\cite{Bierenbaum:2009mv,Ablinger:2014vwa} and ${C}_{F_3,q}^{W^+ - W^-, \rm NS}(N_F)$ the massless 
Wilson 
coefficient up to 3-loop order. Here we use the convention
\begin{eqnarray}
\hat{f}(N_F) =  f(N_F+1) - f(N_F)~.
\label{eq:HAT}
\end{eqnarray}
In the following sections we calculate the Wilson coefficients  $(\Delta) L_{i,q}^{\rm NS}$ to
$O(a_s^2)$ in complete form in the deep-inelastic region. In Section~\ref{sec:3} we present
the main details of the calculation, which allows us to focus on the results in the other cases.
\section{\boldmath The polarized non-singlet structure functions}
\label{sec:3}

\vspace*{1mm}
\noindent
The polarized flavor non-singlet neutral current structure functions $g_{1,2}^{\rm NS}$
receive massless and massive QCD corrections, where the latter contribute starting at $O(a_s^2)$. 
In 
the 
following we will give a detailed discussion of the heavy flavor contributions to $g_{1}^{\rm NS}$
as an example. Main aspects of the calculation are given in Appendices~A and B.

\subsection{\boldmath The structure function $g_{1}^{\rm NS}$}
\label{sec:31}

\vspace*{1mm}
\noindent
To $O(a_s^2)$ the non-singlet contribution for $g_1(x,Q^2)$ reads
\begin{eqnarray}
\label{eq:g1NS}
g_1^{\rm NS}(x,Q^2) &=& \left[C_{g_1,q}\left(x,\frac{Q^2}{\mu^2}\right) + 
L_{g_1,q}^{{\rm 
NS},(2)}\left(x,\frac{Q^2}{\mu^2},\frac{m^2}{\mu^2}\right)\right] 
\nonumber\\ &&
\otimes \frac{1}{2} \Bigl[
  \frac{4}{9} \Delta u_v(x,\mu^2) 
+ \frac{1}{9} \Delta d_v(x,\mu^2) 
+ \frac{8}{9} \Delta \bar{u}(x,\mu^2)
+ \frac{2}{9} \left[\Delta \bar{d}(x,\mu^2) + \Delta \bar{s}(x,\mu^2)\right] 
\Bigr]~.
\nonumber\\
\end{eqnarray}
Here $\Delta u_v$ and $\Delta d_v$ denote the polarized valence quark densities, $\Delta \bar{u}, 
\Delta \bar{d}$ and $\Delta 
\bar{d}$ are the polarized sea quark distributions\footnote{For a review on polarized 
deep-inelastic scattering, see \cite{Lampe:1998eu}.}, $\otimes$ denotes the Mellin convolution, 
\begin{eqnarray}
A(x) \otimes B(x) = \int_0^1 dx_1 \int_0^1 dx_2 \delta(x - x_1 x_2) A(x_1) B(x_2)
\end{eqnarray}
and the massless Wilson coefficient is given by
\begin{eqnarray}
C_{g_1,q}\left(x,\frac{Q^2}{\mu^2}\right) = \delta(1-x) + \sum_{k=1}^2 a_s^k C_{g_1,q}^{(k)}\left(x,\frac{Q^2}{\mu^2}\right)~,
\end{eqnarray}
with 
\begin{eqnarray}
C_{g_1,q}^{(1)}\left(x,\frac{Q^2}{\mu^2}\right) &=& P_{qq}^{(0)}(x) \ln\left(\frac{Q^2}{\mu^2}\right) + c_{g_1,q}^{(1)}(x)
\\
C_{g_1,q}^{(2)}\left(x,\frac{Q^2}{\mu^2}\right) &=& \frac{1}{2} \left\{\left[P_{qq}^{(0)} \otimes P_{qq}^{(0)}\right](x) - 
\beta_0
P_{qq}^{(0)}\right\} \ln^2\left(\frac{Q^2}{\mu^2}\right) 
\nonumber\\ &&
+ \left\{P_{qq}^{(1),{\rm NS,-}}(x) + \left[P_{qq}^{(0)} \otimes 
c_{g_1,q}^{(1)}\right](x) - \beta_0 c_{g_1,q}^{(1)}(x) \right\} \ln\left(\frac{Q^2}{\mu^2}\right) + c_{g_1,q}^{(2)}(x)~,
\nonumber\\ 
\end{eqnarray}
cf. e.g. \cite{Zijlstra:1992qd}. Here
$P_{qq}^0$ is the leading order splitting function
\begin{eqnarray}
\label{eq:Pqq0}
P_{qq}^{(0)}(x) = 2 C_F \left(\frac{1+x^2}{1-x}\right)_+,
\end{eqnarray}
with the $+$-prescription being defined by
\begin{eqnarray}
\int_0^1 dx [f(x)]_+ g(x) = \int_0^1 dx [g(x) - g(1)] f(x)~.
\end{eqnarray}
The NLO non-singlet splitting functions $P_{qq}^{(1),{\rm NS, \pm}}(x)$ 
were 
calculated in \cite{PNLO}\footnote{We use the convention $\mu^2 (\partial/\partial \mu^2)$ 
for the scale evolution operator in the renormalization group equation.}, 
the quarkonic one-loop Wilson coefficient $c_{g_1,q}^{(1)}$ 
for the structure function  $g_1$ \cite{Kodaira:1978sh} is given by
\begin{eqnarray}
c_{g_1,q}^{(1)}(z) &=& C_F \Biggl[4 \left(\frac{\ln(1-z)}{1-z}\right)_+ - 
\left(\frac{3}{1-z}\right)_+ - 2(1+z) \ln(1-z)
- 2\frac{1+z^2}{1-z} \ln(z) +4 + 2z 
\nonumber\\ &&
- \delta(1-z)\left[9 + 4 \zeta_2\right] \Biggr]
\end{eqnarray}
and $c_{g_1,q}^{(2)}(z)$ has been calculated in Ref.~\cite{Zijlstra:1993sh}.
The color factors are $C_A = N_c, C_F 
= (N_c^2-1)/(2 N_c), T_F = 1/2$ for $SU(N_c)$ and $N_c = 3$ in Quantum Chromodynamics. 
Here and in the following we set the factorization and renormalization scales both to $\mu$.

\begin{figure}[H]
\centering
\includegraphics[width=0.5\textwidth]{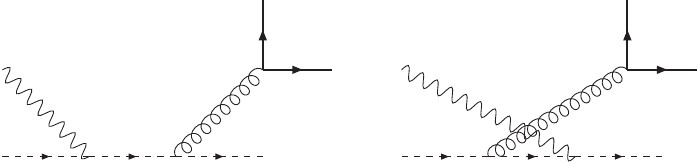}
\caption{\sf \small The $\gamma^* q$ Compton diagrams at 
$O(a_s^2)$. The dashed (full) lines denote massless (massive) quarks, 
respectively.}
\label{QCDCOMP}
\end{figure}

\begin{figure}[H] 
\centering
\includegraphics[width=0.1\textwidth]{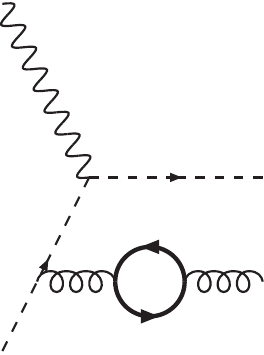} \hspace*{5mm}
\includegraphics[width=0.1\textwidth]{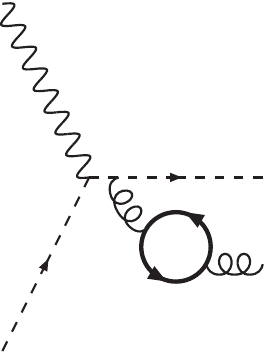} \hspace*{5mm}
\includegraphics[width=0.22\textwidth]{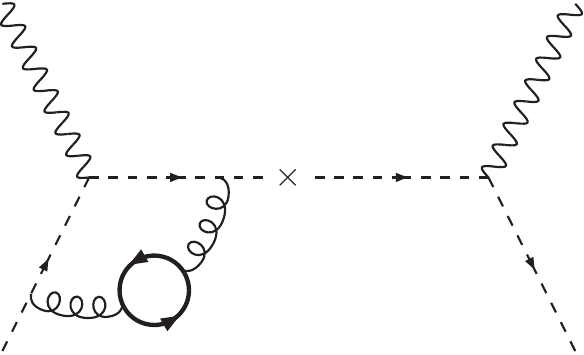} \\
{\footnotesize \sf \hspace*{-0.7cm} (a) \hspace*{2cm} (b) \hspace*{2.7cm} (c)}
\caption{\sf \small The virtual $O(\alpha_s^2)$ heavy flavor corrections. (a) and (b) 
Bremsstrahlung amplitudes; 
(c) interference term of the Born amplitude and the vertex correction. The graphs of 
the self-energy terms contributing to (c) are not shown, but are discussed in Appendix~B.} 
\label{ONELB0Q}
\end{figure}

The $O(a_s^2)$ Wilson coefficient $\Delta L_{g_1,q}^{\rm NS}$ receives contributions from the
Feynman diagrams shown in Figures~\ref{QCDCOMP} and \ref{ONELB0Q}. The diagrams of the Compton process, 
Figure~\ref{QCDCOMP},  describe the real production of a heavy quark pair in the kinematic range
$z \leq Q^2/(Q^2 + 4 m^2)$ of the parton momentum fraction, and contain no singularities, enabling 
their calculation in $d=4$ dimensions.

\vspace*{-5mm}
\noindent
We obtain
\begin{eqnarray}
 \Delta L_{g_1,q}^{\text{NS},(2),C}\left(z,\frac{Q^2}{m^2},\frac{m^2}{\mu^2}\right)
 &=& a_s^2
 C_F T_F \bigg[\bigg\{\frac{4}{3}\frac{1+z^2}{1-z}-\frac{16z}{1-z}\left(\frac{z}{\xi}\right)^2\bigg\}
 \bigg\{\biggl[\ln\left(\frac{1-z}{z^2}\right) 
\nonumber\\ && 
\hspace*{-12mm}
+ \ln\left(\frac{1+\sqrt{1-\frac{4z}{\xi}}}{1-\sqrt{1-\frac{4z}{\xi}}}\right) 
\biggr] \ln\left(\frac{1+\sqrt{1-\frac{4z}{(1-z)\xi}}}{1-\sqrt{1-\frac{4z}{(1-z)\xi}}} \right)
\nonumber\\ &&
\hspace*{-12mm}
+2 \Biggl[
- \Li_2\left(\frac{(1-z)\left(1+\sqrt{1-\frac{4z}{(1-z)\xi}}\right)}{1+\sqrt{1-\frac{4z}{\xi}}} \right)
+ \Li_2\left(\frac{1- \sqrt{1-\frac{4z}{\xi}}}{1 +\sqrt{1-\frac{4z}{(1-z)\xi}}} \right)
\nonumber\\ &&
\hspace*{-12mm}
+ \Li_2\left(\frac{1 - \sqrt{1-\frac{4z}{(1-z)\xi}}}
                  {1 + \sqrt{1-\frac{4z}{\xi}}}\right)
- \Li_2\left(\frac{1 + \sqrt{1-\frac{4z}{(1-z)\xi}}}
                  {1 + \sqrt{1-\frac{4z}{\xi}}}\right)\Biggr]\bigg\}
\nonumber\\ &&
\hspace*{-12mm}
 +\bigg\{-\frac{8}{3}+\frac{4}{1-z}+\left(\frac{z}{(1-z)\xi}\right)^2\left(-16+32z-\frac{8}{1-z}\right)\bigg\}
\nonumber\\ && 
\hspace*{-12mm}
\times \ln\left(\frac{1+\sqrt{1-\frac{4z}{(1-z)\xi}}}{1-\sqrt{1-\frac{4z}{(1-z)\xi}}} \right)
 +\bigg\{\frac{64}{9}+\frac{112}{9}z-\frac{152}{9}\frac{1}{1-z}
+\frac{z}{(1-z)\xi}
\nonumber\\ &&
\hspace*{-12mm}
\times
\bigg[\frac{512}{9}-\frac{128}{3}z
 +\frac{848}{9}z^2\bigg]
+ 
\left(\frac{z}{(1-z)\xi}\right)^2\bigg[-\frac{640}{9}+\frac{1408}{9}z-\frac{2368}{9}z^2
\nonumber\\ &&
\hspace*{-12mm}
+\frac{1600}{9}z^3\bigg]\biggr\}\frac{1}{\sqrt{1-\frac{4z}{\xi}}} \ln\left(\frac{
 \sqrt{1-\frac{4z}{\xi}} + \sqrt{1-\frac{4z}{(1-z)\xi}}}
{\sqrt{1-\frac{4z}{\xi}} - \sqrt{1-\frac{4z}{(1-z)\xi}}}\right)
 +\bigg\{-\frac{188}{27}
\nonumber\\ &&
\hspace*{-12mm}
-\frac{872}{27}z
+\frac{718}{27}\frac{1}{1-z}+\frac{z}{(1-z)\xi}\bigg[-\frac{952}{27}+\frac{1520}{27}z
 -\frac{800}{9}z^2
\nonumber\\
&&
\hspace*{-12mm}
+\frac{20}{27}\frac{1}{1-z}\bigg]\bigg\}
\sqrt{1 - \frac{4z}{(1-z) \xi}}~~
\bigg] 
\theta\left(\frac{\xi}{\xi+4} - z\right),
\label{eq:LNS1C}
\end{eqnarray}
where 
\begin{eqnarray}
\xi = \frac{Q^2}{m^2}. 
\end{eqnarray}
In Appendix~\ref{sec:A} the principal steps of the calculation of (\ref{eq:LNS1C}) are 
outlined. For this contribution to $L_{g_1,q}^{\rm NS}$ we agree with the result given in
\cite{Buza:1996xr}.

The inclusive scattering cross 
section, however, receives also contributions from the virtual corrections shown in Figure~\ref{ONELB0Q}.

\noindent
In the Bremsstrahlung corrections {\sf (a,b)} to these diagrams, the heavy flavor correction is given by the 
one-loop polarization function $\Pi_{QQ}(k^2=0)$. The polarization insertion $\Pi_{QQ}(k^2)$ also appears in the virtual 
correction {\sf (c)}. For technical reasons we decompose $\Pi_{QQ}(k^2) = \Pi_{QQ}(k^2=0) + \left[\Pi_{QQ}(k^2) - 
\Pi_{QQ}(k^2=0)\right]$ and combine the first term with the contributions due to {\sf (a,b)}. This yields the term
\begin{eqnarray}
\label{eq:massless_g1}
\Delta L_{g_1,q}^{{\rm NS},(2), \rm massless}
= - a_s^2 \beta_{0,Q} \ln\left(\frac{m^2}{\mu^2}\right) \left[P_{qq}^{(0)}(z) 
\ln\left(\frac{Q^2}{\mu^2}\right)
+c_{g_1,q}^{(1)}(z) \right],
\end{eqnarray}
with $\beta_{0,Q} = -4 T_F/3$. The term (\ref{eq:massless_g1}) corresponds to a heavy flavor contribution
in the case of massless final states.

There are also self-energy insertions contributing to {\sf (c)}, which, however, vanish for the 
term (\ref{eq:massless_g1}) since the corresponding graphs at 1-loop vanish and $\Pi_{QQ}(k^2=0)$
contributes multiplicatively. For the insertion $\Pi_{QQ}(k^2 \neq 0)$ this is not the case,
cf.~Appendix~B.

The second term $ \left[\Pi_{QQ}(k^2) - \Pi_{QQ}(k^2=0)\right]$ is now used in the interference term calculating the 
form factor. The subtraction term allows to perform the calculation in $d=4$ dimensions, 
\begin{eqnarray}
\label{eq:LqNSg1V}
\Delta L_{g_1,q}^{\text{NS},(2),V}\left(\frac{Q^2}{m^2}\right)
&=&
2 a_s^2 C_F T_F
\Biggl\{\frac{3355}{81}-\frac{952}{9\xi}+\left(\frac{32}{\xi^2}-\frac{16}{3}\right)\zeta_3
+\left(\frac{440}{9\xi}-\frac{530}{27}\right)\ln(\xi)
\nonumber\\
&&
+\tilde{\lambda}\left(\frac{184}{9\xi}-\frac{76}{9}\right)\left[\Li_2\left(\frac{\tilde{\lambda}+1}
{\tilde{\lambda}-1}\right)-\Li_2\left(\frac{\tilde{\lambda}-1}{\tilde{\lambda}+1}\right)\right]
\nonumber\\ 
&&
+\left(\frac{8}{3}-\frac{16}{\xi^2}\right)\left[\Li_3\left(\frac{\tilde{\lambda}-1}
{\tilde{\lambda}+1}\right)+\Li_3\left(\frac{\tilde{\lambda}+1}{\tilde{\lambda}-1}\right)\right]\Biggr\},
\end{eqnarray}
with $\tilde{\lambda} = \sqrt{1 - 4/\xi}$. Details of the calculation are presented in Appendix~B.
\begin{figure}[H] 
\centering 
\includegraphics[scale=0.9]{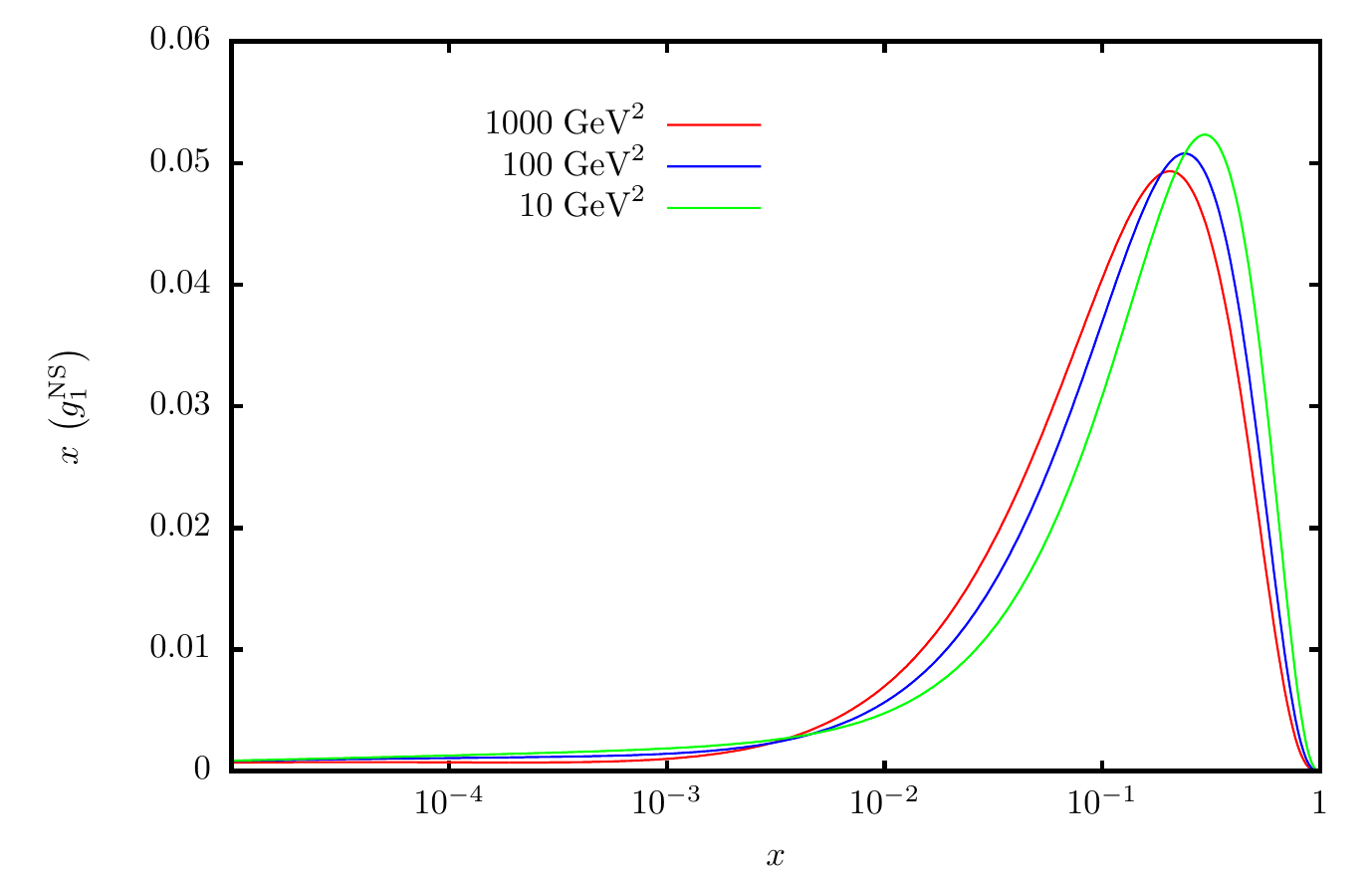} \caption{\sf \small 
The polarized structure function $g_1$ due to photon exchange up to $O(\alpha_s^2)$ including the 
charm and bottom quark corrections in the on-shell scheme with $m_c = 1.59~\GeV$ 
\cite{Alekhin:2012vu} and $m_b = 4.78~\GeV$ \cite{PDG15} using the NLO parton distribution 
functions \cite{Blumlein:2010rn}.} \label{FIG:g1ncall} 
\end{figure} 

\noindent
The massive Wilson coefficient is given by
\begin{eqnarray}
\Delta L_{i,q}^{{\rm NS},(2)}\left(z,\frac{Q^2}{\mu^2},\frac{m^2}{\mu^2}\right)  
&=& \Delta L_{i,q}^{{\rm NS},(2),C}\left(z,\xi\right) + \delta(1-z) L_{i,q}^{{\rm NS},(2),V} 
+ \Delta L_{i,q}^{{\rm NS},(2), \rm massless}\left(z,\frac{Q^2}{\mu^2},\frac{m^2}{\mu^2}\right).
\nonumber\\ 
\end{eqnarray}

\noindent
In the following we will use the values  
\begin{eqnarray}
\label{eq:mass}
m_c = {1.59~\GeV},~~~~~~~~~~~~~~~~~~~~~{m_b = 4.78~\GeV}
\end{eqnarray}
at NNLO in the on-shell scheme \cite{Alekhin:2012vu,PDG15} for all numerical illustrations, both at 
$O(a_s^2)$ and $O(a_s^3)$, 
since we consider the present results as a part of our more general NNLO project, cf.~\cite{Blumlein:2014zxa}, and would like to 
compare with numerical results given at $O(a_s^3)$ in \cite{Ablinger:2014vwa,Ablinger:2014nga,Behring:2015zaa,Behring:2015roa}.
The transformation to the $\overline{\rm MS}$ scheme for the heavy quark masses is well-known 
\cite{MASS}. In fitting the heavy quark masses 
from data one would use the corresponding formula. For the illustration given 
in the following, their {\it equivalent} value in the on-shell scheme has been used for brevity. 
In the numerical results given below, we choose for the factorization and 
renormalization scales $\mu^2 = Q^2$. In the calculation we used the codes {\tt HPLOG, CHAPLIN, 
 HPL} and 
{\tt HarmonicSums}~\cite{Gehrmann:2001pz,Buehler:2011ev,Maitre:2005uu,HARMSUM} at different steps.
\begin{figure}[H] 
\centering 
\includegraphics[scale=0.9]{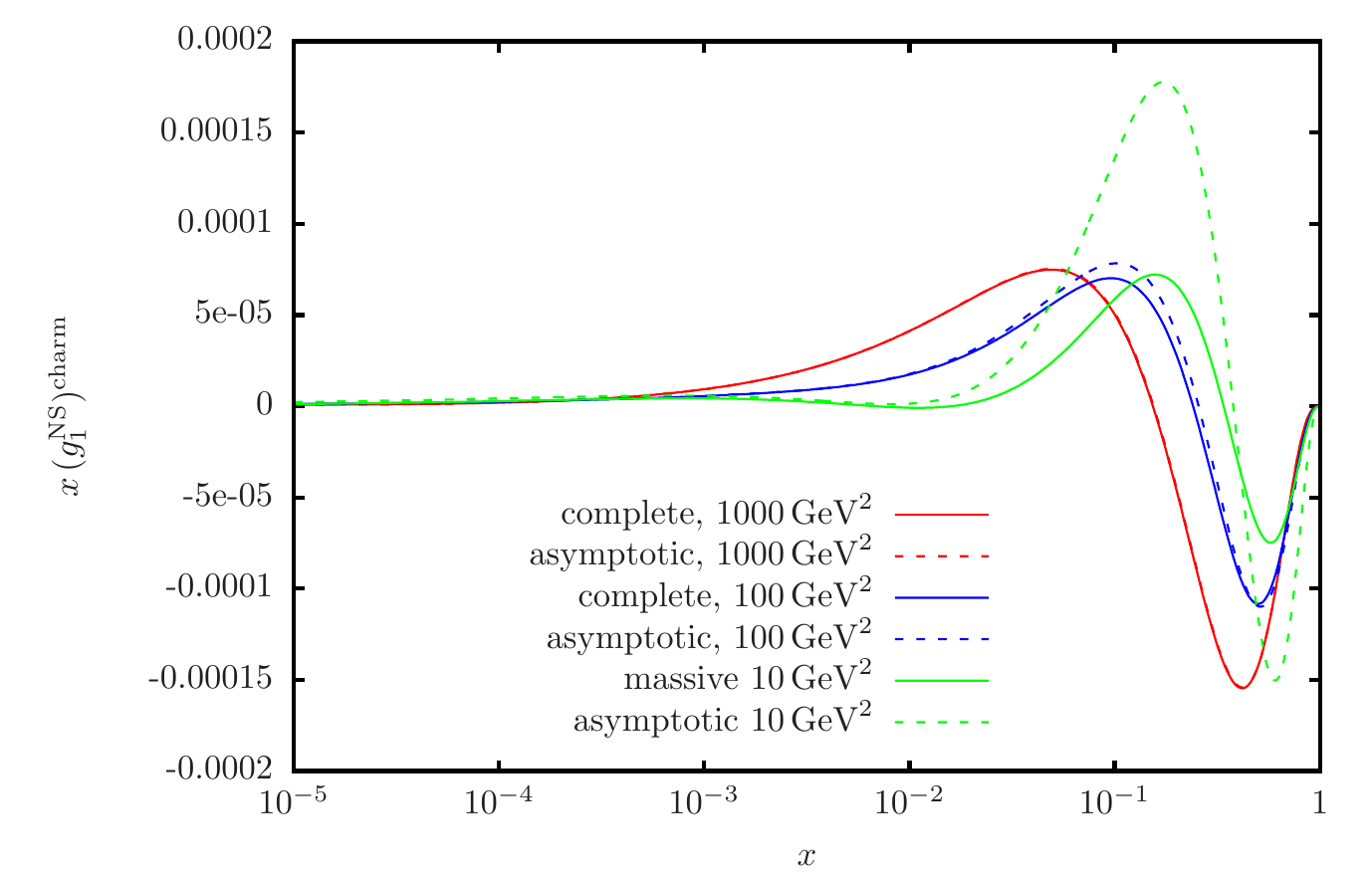} \caption{\sf \small 
The charm quark contribution to the structure function $g_1$ due to photon exchange up to 
$O(\alpha_s^2)$ as a function of $x$ and $Q^2$. The conditions are the same as in 
Figure~\ref{FIG:g1ncall}. Dashed lines: asymptotic representation in $Q^2$ for the heavy flavor 
corrections; full lines: complete heavy flavor contributions. } \label{FIG:g1pow10} 
\end{figure} 
\begin{figure}[H] 
\centering 
\includegraphics[scale=0.8]{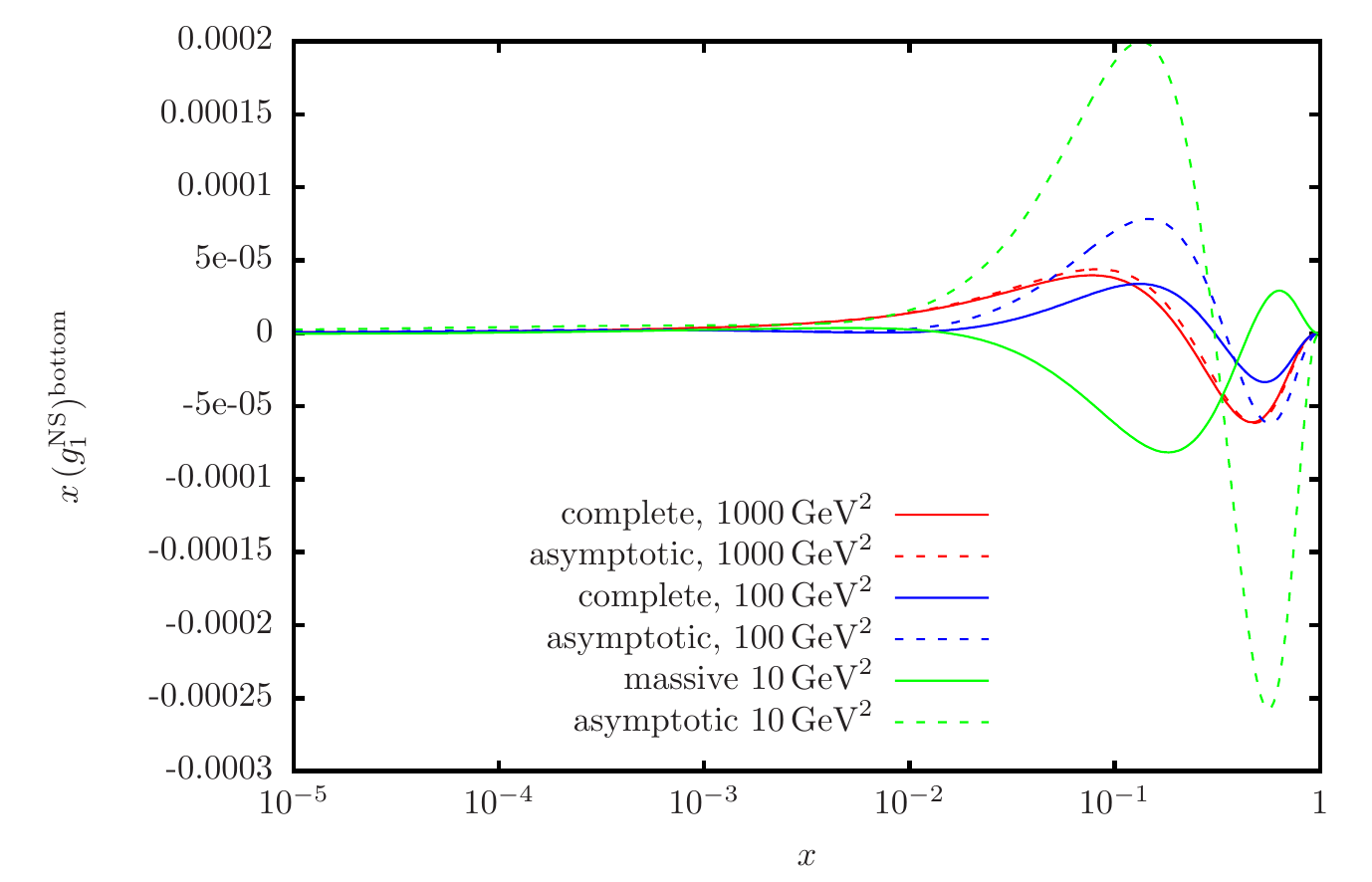} \caption{\sf \small 
The bottom quark contribution to the structure function $g_1$ due to photon exchange up to 
$O(\alpha_s^2)$ as a function of $x$ and $Q^2$. The conditions are the same as in 
Figure~\ref{FIG:g1ncall}. Dashed lines: asymptotic representation in $Q^2$ for the heavy flavor 
corrections; full lines: complete heavy flavor contributions. } 
\label{FIG:g1pow101} 
\end{figure} 
\noindent 
In Figure~\ref{FIG:g1ncall} we illustrate the massless and massive contributions to the 
non-singlet structure function $g_1^{\rm NS}$ to $O(a_s^2)$ as a function of $x$ 
and $Q^2$ using the parton distribution functions \cite{Blumlein:2010rn}. Due to the QCD evolution 
the peak of the function moves towards smaller values of $x$, keeping its valence-like profile. 
The contributions due to charm and bottom are illustrated in Figures~\ref{FIG:g1pow10} and 
\ref{FIG:g1pow101}. Here we also compare the asymptotic expressions with the complete results, 
which show differences for $Q^2 \sim 10~\GeV^2$ and become very close for $Q^2 = 100$ and 
$1000~\GeV^2$ for charm and at higher scales also for bottom.

\begin{figure}[H] 
\centering 
\includegraphics[scale=0.8]{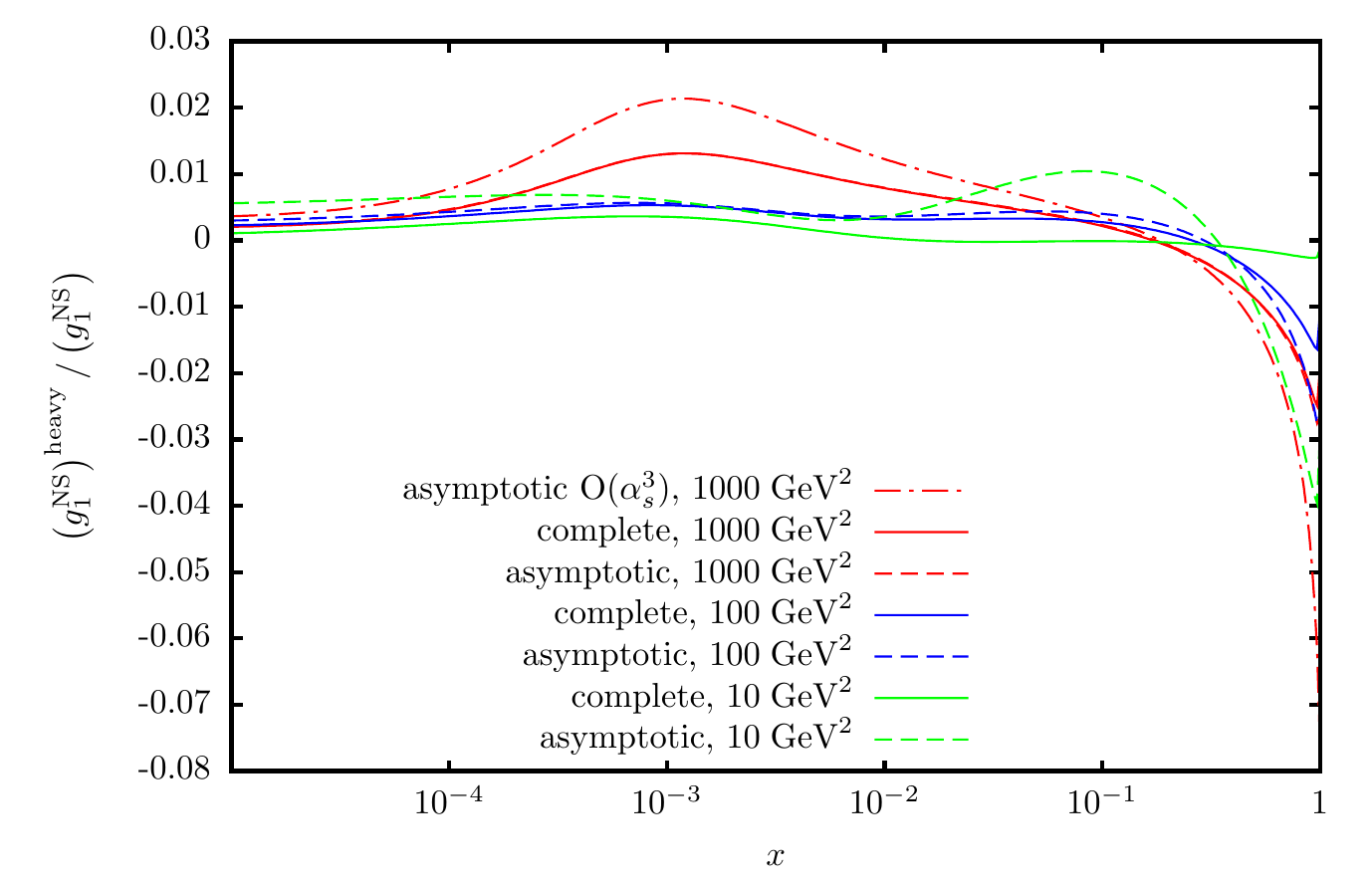} \caption{\sf 
\small The ratio of the heavy flavor non-singlet contributions to the structure function $g_1$ due 
to photon exchange to the complete structure function up to $O(\alpha_s^2)$ as a function of $x$ 
and $Q^2$. The conditions are the same as in Figure~\ref{FIG:g1ncall}. Dashed lines: asymptotic 
representation in $Q^2$ for the heavy flavor corrections; full lines: complete heavy flavor 
contributions. The dash-dotted line shows the asymptotic result at $O(\alpha_s^3)$ for $Q^2 = 
1000~\GeV^2$.} 
\label{FIG:g1ncratios} 
\end{figure}

The ratio of the heavy quark contributions to the complete structure function are 
illustrated in Figure~\ref{FIG:g1ncratios}. In the range of smaller values of $x$ 
the fraction amounts to $< +1.2\%$, while at larger 
values of $x$ the corrections become negative amounting to $-3$. The asymptotic 
3-loop corrections \cite{Behring:2015zaa} at $Q^2 = 1000~\GeV^2$ are even larger 
and contribute to $O(2\%)$ at lower values of $x$ and amount to $O(-6\%)$ at large 
$x$.

Here and in the following we often will make the observation that the asymptotic expressions 
tend to agree better in the region of small $x$ even at lower values of $Q^2$, where this is not expected
a priori. The reason for this is that the relevant effective scale, inside the corresponding 
integrals, is the hadronic mass squared $W^2$, rather than $Q^2$ itself. 
\subsection{\boldmath The structure function $g_2^{\rm NS}$}
\label{sec:g2}

\vspace*{1mm}
\noindent
At leading twist, the structure function $g_2(x,Q^2)$ is obtained through the Wandzura-Wilczek relation  
\begin{eqnarray}
g_2(x,Q^2) = - g_1(x,Q^2) + \int_x^1 \frac{dy}{y} g_1(y,Q^2)~.
\end{eqnarray}
Here $g_2(x,Q^2)$ denotes the non-singlet distribution, calculated using
$g_1(x,Q^2) \equiv g_1^{\rm NS}(x,Q^2)$, Eq.~(\ref{eq:g1NS}).
The Wandzura-Wilczek relation has been derived for massless quarks in \cite{Wandzura:1977qf}, see also 
\cite{Blumlein:1996tp,Blumlein:1996vs}, but 
possesses a much wider validity as has been shown in later years. It also holds for 
scattering off massive quarks \cite{Blumlein:1998nv} 
and for the target mass corrections \cite{Piccione:1997zh,Blumlein:1998nv}, as well as for non-forward 
\cite{Blumlein:1999sc,Blumlein:2000cx,Geyer:2004bx} and diffractive scattering \cite{Blumlein:2002fw,Blumlein:2008di} and heavy 
flavor production in photon-gluon fusion  \cite{Blumlein:2003wk}. At leading twist the structure functions $g_1$ and $g_2$ are 
connected by an operator relation, cf.~\cite{Blumlein:1999sc}. Representations in the covariant 
parton model were given in 
Refs.~\cite{Jackson:1989ph,Roberts:1996ub,Blumlein:1996tp,Blumlein:2003wk}.
\begin{figure}[H]
\centering
\includegraphics[scale=0.9]{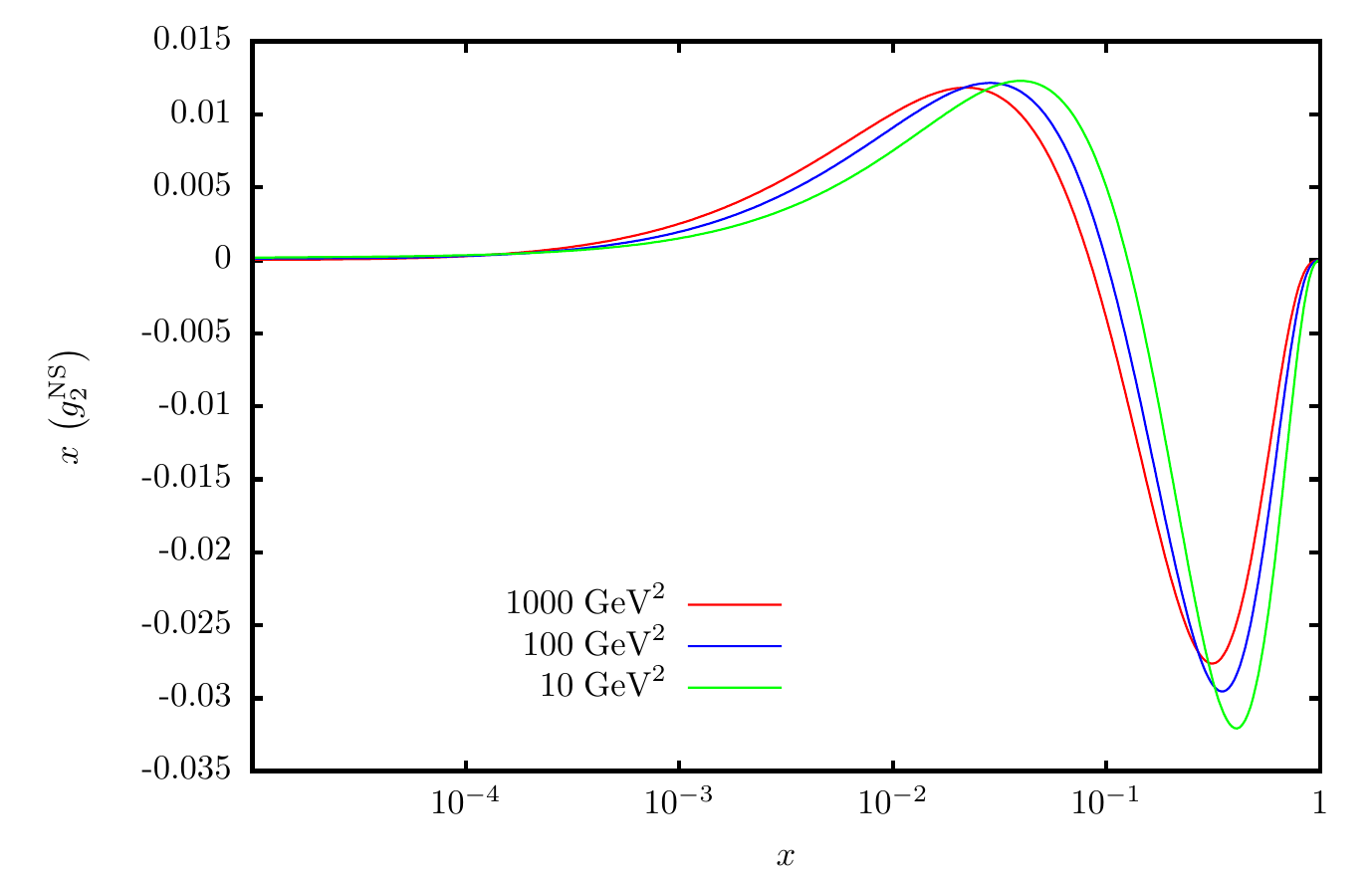}
\caption{\sf \small 
The polarized structure function $g_2$ due to photon exchange up to $O(\alpha_s^2)$ including the charm and  bottom quark 
corrections in
the on-shell scheme with $m_c = 1.59~\GeV$ \cite{Alekhin:2012vu} and $m_b = 4.78~\GeV$ \cite{PDG15} using the NLO parton distribution
functions \cite{Blumlein:2010rn}.}
\label{FIG:g2ncall}
\end{figure}
\begin{figure}[H]
\centering
\includegraphics[scale=0.9]{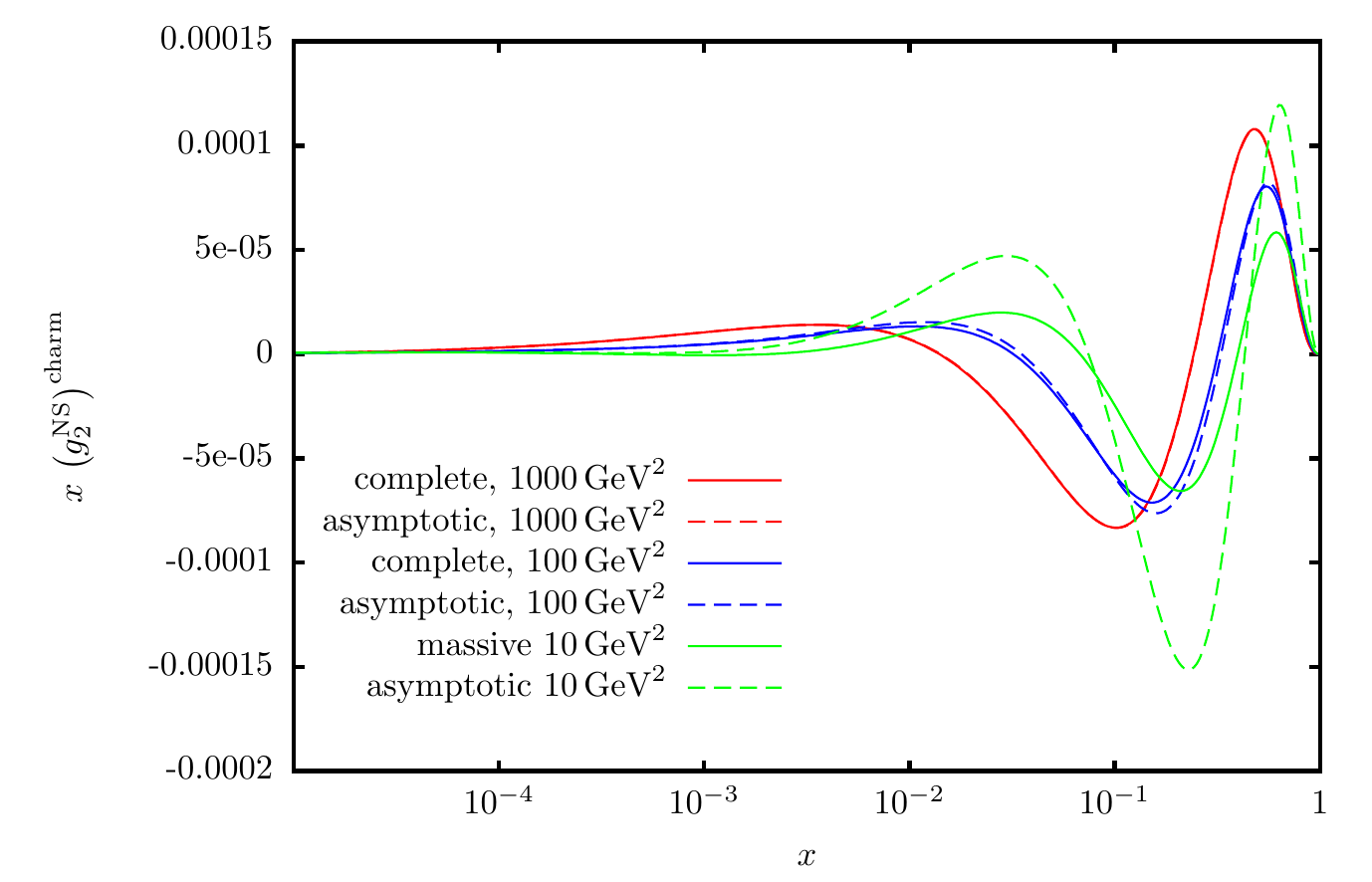}
\caption{\sf \small 
The charm contribution to the structure function $g_2$ due to photon exchange up to $O(\alpha_s^2)$ 
as a function of $x$ and $Q^2$. The conditions are the same as in Figure~\ref{FIG:g2ncall}. 
Dashed lines: 
asymptotic representation in $Q^2$ for the heavy flavor corrections; full lines: complete heavy flavor contributions.
}
\label{FIG:g2heavy}
\end{figure}

\begin{figure}[H]
\centering
\includegraphics[scale=0.9]{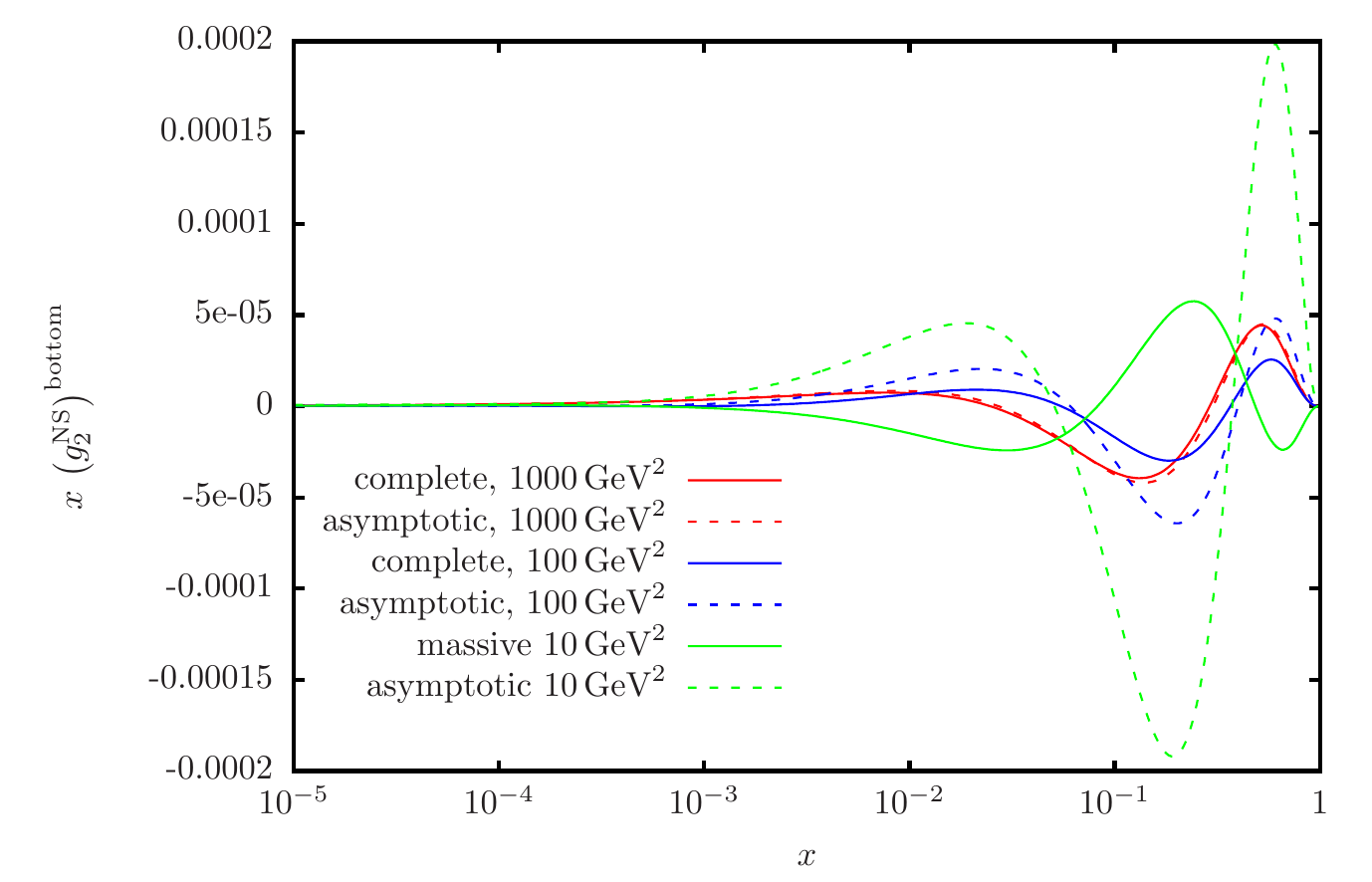}
\caption{\sf \small
The bottom  contribution to the structure function $g_2$ due to photon exchange up to $O(\alpha_s^2)$ 
as a function of $x$ and $Q^2$. The conditions are the same as in Figure~\ref{FIG:g2ncall}. 
Dashed lines: 
asymptotic representation in $Q^2$ for the heavy flavor corrections; full lines: complete heavy flavor contributions.
}
\label{FIG:g2heavy1}
\end{figure}

In Figure~\ref{FIG:g2ncall} we illustrate the flavor non-singlet contribution at twist 2 to the structure function $xg_2(x,Q^2)$ for 
pure photon exchange up to $O(\alpha_s^2)$. It takes values in the range $+0.01$ to $-0.03$, with 
only mild scaling violations 
varying $Q^2$ from $10~{\GeV^2}$ to $1000~{\GeV^2}$. In Figures~\ref{FIG:g2heavy} and \ref{FIG:g2heavy1} we illustrate the heavy 
flavor corrections due to charm and bottom, respectively.
The effect is of $O(1\%)$ in the case of charm. We also compare the exact results with those using the 
asymptotic
\begin{figure}[H]
\centering
\includegraphics[scale=0.9]{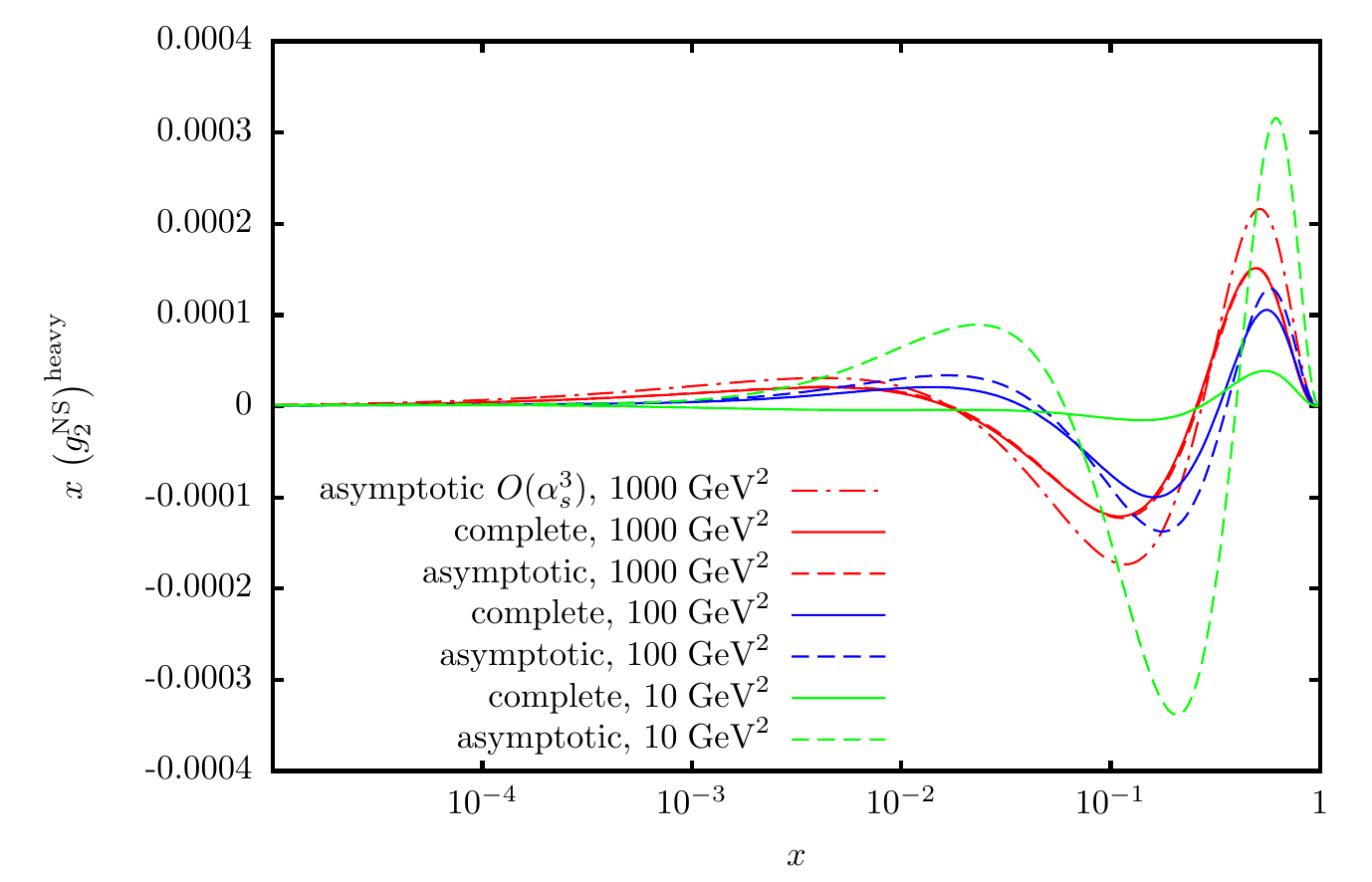}
\caption{\sf \small
The complete heavy flavor contributions to the non-singlet structure function $g_2$ due to photon 
exchange up to $O(\alpha_s^2)$ 
as a function of $x$ and $Q^2$. Full lines: $O(a_s^2)$ contributions; dashed lines asymptotic $O(a_s^2)$ contributions. The 
dash-dotted line for $Q^2 = 1000~\GeV^2$ corresponds to all contributions including also the asymptotic $O(a_s^3)$ term.
The conditions are the same as in Figure~\ref{FIG:g2ncall}.}
\label{FIG:g2heavy2}
\end{figure}

\noindent
representation, in which the power corrections are disregarded, cf.~\cite{Behring:2015zaa}. The effect is clearly visible at 
lower scales, and fully disappears at $Q^2 \sim 100~\GeV^2$ in the case of charm.

In Figure~\ref{FIG:g2heavy2} we illustrate the combined heavy flavor effect and also show the asymptotic 3-loop corrections, 
which turn out to be larger than the exact corrections.

The structure of the Wandzura-Wilczek relation implies that the associated sum rule for the first moment yields  
zero. However, this is not a prediction which derives from the light-cone expansion \cite{Blumlein:1996vs}, since
the corresponding moment does not contribute to it as a term. Rather the Wandzura-Wilczek relation, as an 
analytic 
continuation,
is compatible with the result, which is also called (flavor non-singlet) Burkhardt-Cottingham sum rule
\cite{Burkhardt:1970ti}. It results from the fact that the imaginary part of $g_2(\vec{q}^2,q_0)$ obeys 
a superconvergence relation. Unlike a series of other sum rules, it cannot be expressed as an expectation value of
(axial)vector operators \cite{Ravindran:2001dk}.
\section{The unpolarized non-singlet structure functions \boldmath 
$F_{1,2}^{\rm NS}$}
\label{sec:4}

\vspace*{1mm}
\noindent
In the case of pure photon exchange, the unpolarized neutral current scattering cross section is 
parameterized by
two deep-inelastic structure functions $F_{1,2}(x,Q^2)$ which obey
\begin{eqnarray}
2xF_1(x,Q^2) = F_2(x,Q^2) - F_L(x,Q^2),
\end{eqnarray}
in the absence of target mass corrections \cite{Georgi:1976ve}. Here $F_L(x,Q^2)$ denotes the longitudinal structure function.
In the following we will refer to the structure functions $F_2$ and $F_L$. The calculation proceeds in a similar way to
that outlined in Section~\ref{sec:3}.

\subsection{\boldmath 
$F_{L}^{\rm NS}$}
\label{sec:4.1}

\vspace*{1mm}
\noindent
In the case of the structure function $F_L$, the Compton contribution is given by
\begin{eqnarray}
 L_{F_L,q}^{\text{NS},(2),C}\left(z,\frac{Q^2}{m^2},\frac{m^2}{\mu^2}\right)
 &=&
 a_s^2 C_F T_F \Biggl\{
 96 \frac{z^3}{\xi^2} \Biggl[\ln\left(\frac{1+\sqrt{1-\frac{4z}{(1-z)\xi}}}{1-\sqrt{1-\frac{4z}{(1-z)\xi}}}\right)
 \ln\left(\frac{1+\sqrt{1-\frac{4z}{\xi}}}{1-\sqrt{1-\frac{4z}{\xi}}}\right)
\nonumber\\ &&
+2 \Biggl[
- \Li_2\left(\frac{(1-z)\left(1+\sqrt{1-\frac{4z}{(1-z)\xi}}\right)}{1+\sqrt{1-\frac{4z}{\xi}}} \right)
+ \Li_2\left(\frac{1- \sqrt{1-\frac{4z}{\xi}}}{1 +\sqrt{1-\frac{4z}{(1-z)\xi}}} \right)
\nonumber\\ &&
+ \Li_2\left(\frac{1 - \sqrt{1-\frac{4z}{(1-z)\xi}}}
                  {1 + \sqrt{1-\frac{4z}{\xi}}}\right)
- \Li_2\left(\frac{1 + \sqrt{1-\frac{4z}{(1-z)\xi}}}
                  {1 + \sqrt{1-\frac{4z}{\xi}}}\right)
 \Biggr] \Biggr]
\nonumber\\ &&
- \frac{16}{3 \xi} (22 z^2 - z\xi) \sqrt{1 - \frac{4z}{
\xi}}
\ln\left(\frac{
 \sqrt{1-\frac{4z}{\xi}} + \sqrt{1-\frac{4z}{(1-z)\xi}}}
{\sqrt{1-\frac{4z}{\xi}} - \sqrt{1-\frac{4z}{(1-z)\xi}}}\right)
\nonumber\\ &&
-\frac{8}{9(1-z) \xi} \sqrt{1-\frac{4z}{(1-z)\xi}}\left[60 z -478 z^2 +372 z^3 \right. 
\nonumber\\ && 
\left.
- \xi (6 - 31 z + 25 z^2)\right]
+\ln\left(\frac{1+\sqrt{1-\frac{4z}{(1-z)\xi}}}{1-\sqrt{1-\frac{4z}{(1-z)\xi}}} \right)
\nonumber\\ &&
\times
\left[ \frac{32(6 z^2 - 9z +2) z^2}{(1-z)^2 \xi^2} 
+ 96 \frac{z^3}{\xi^2} \ln \left(\frac{1-z}{z^2}\right)\right]
\Biggr\}
\theta\left(\frac{\xi}{\xi+4} - z\right)
\label{eq:FLR}
\end{eqnarray}
and we confirm the result given in \cite{Buza:1995ie}.
The virtual contribution vanishes and the contribution corresponding to massless final states reads
\begin{eqnarray}
L_{F_L,q}^{{\rm NS},(2), \rm massless}\left(z,\frac{Q^2}{\mu^2},\frac{m^2}{\mu^2}\right) =
- a_s^2 \beta_{0,Q} \ln\left(\frac{m^2}{\mu^2}\right) c_{F_L,q}^{(1)}(z),
\end{eqnarray}
with \cite{Zee:1974du}
\begin{eqnarray}
c_{F_L,q}^{(1)}(z) = 4 C_F z~.
\end{eqnarray}
Expanding $L_{L,q}^{{\rm NS},(2)}$ for large values of $\xi$ leads to $\hat{C}_{F_L,q}^{(2)}$, the corresponding massless
2-loop Wilson coefficient \cite{FL:2L,Moch:1999eb}, as predicted by  renormalization in Ref.~\cite{Bierenbaum:2009mv,Behring:2014eya}
with no logarithmic term $\sim \ln(\xi)$ left, unlike the case where we take just the term 
$L_{q,L}^{\text{NS},(2),C}$ into account, cf. \cite{Buza:1995ie,Blumlein:2006mh}, where 
\cite{Zijlstra:1992qd}
\begin{eqnarray}
\hat{C}_{F_L,q}^{{\rm NS},(2)}(z) = - \beta_{0,Q} c_{L,q}^{(1)} \ln\left(\frac{Q^2}{\mu^2}\right) + \hat{c}_{F_L,q}^{{\rm 
NS},(2)}(z),
\end{eqnarray}
and
\begin{eqnarray}
\hat{c}_{F_L,q}^{{\rm NS},(2)}(z) = C_F T_F \Biggl\{\frac{16}{3} - \frac{200}{9} z - \frac{16}{3} z \left[2 \ln(z) - 
\ln(1-z)\right]\Biggr\}.
\end{eqnarray}

\noindent
The non-singlet structure function for $F_L$ reads
\begin{eqnarray}
F_L^{\rm NS}(x,Q^2)&=& x \left[C_{F_L,q}\left(x,\frac{Q^2}{\mu^2}\right) + 
L_{F_L,q}^{{\rm 
NS},(2)}\left(x,\frac{Q^2}{\mu^2},\frac{m^2}{\mu^2}\right)\right] 
\nonumber\\ &&
\otimes \Bigl[
 \frac{4}{9} u_v(x,\mu^2)
+ \frac{1}{9} d_v(x,\mu^2)
+ \frac{8}{9} \bar{u}(x,\mu^2)
+ \frac{2}{9} \left[\bar{d}(x,\mu^2) + \bar{s}(x,\mu^2)\right]
\Bigr]~,
\end{eqnarray}
\begin{figure}[H]
\centering
\includegraphics[scale=0.9]{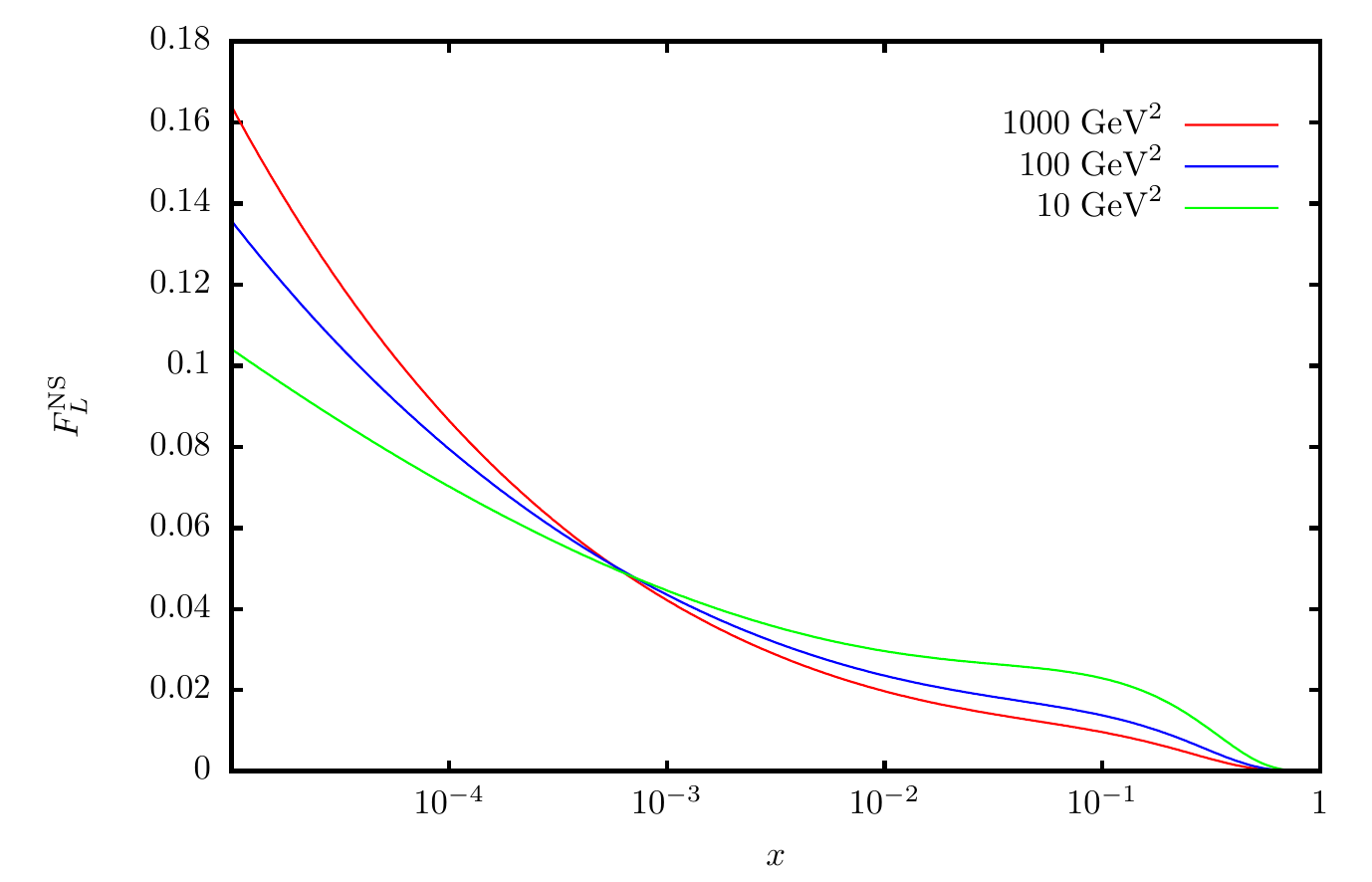}
\caption{\sf \small The structure function $F_L$ due to photon exchange up to $O(\alpha_s^2)$ 
including the charm and  bottom quark 
corrections in
the on-shell scheme with $m_c = 1.59~\GeV$ \cite{Alekhin:2012vu} 
and $m_b = 4.78~\GeV$ \cite{PDG15} using the NNLO parton distribution
functions \cite{Alekhin:2013nda}.}
\label{FIG:FLncall}
\end{figure}
\noindent
where $u_v$ and $d_v$ are the unpolarized valence quark densities and $\bar{u}$ and $\bar{d}$ the sea quark densities,
\begin{figure}[H]
\centering
\includegraphics[scale=0.9]{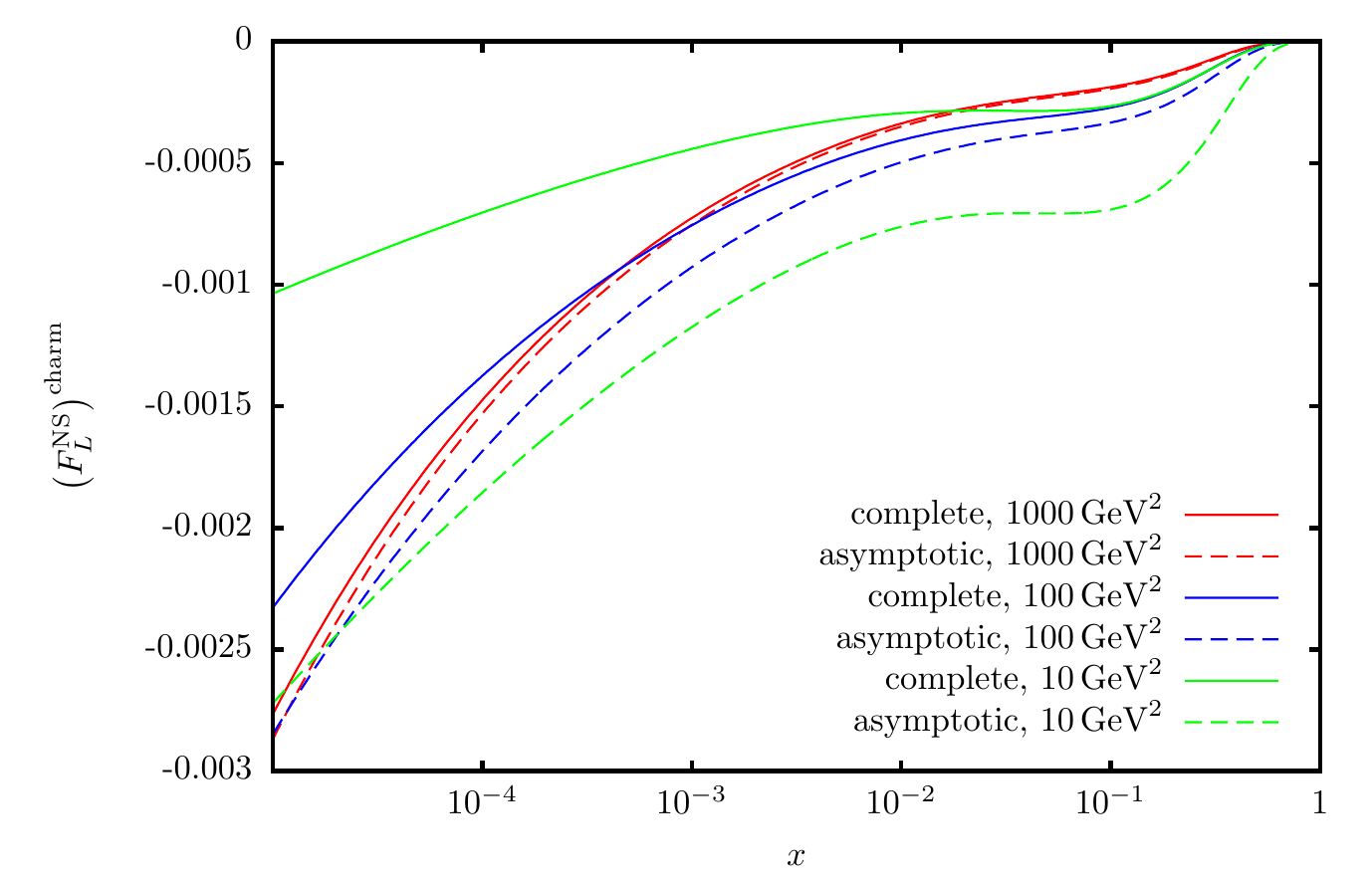}
\caption{\sf \small The charm quark contribution to the structure function $F_L$ due to photon 
exchange up to $O(\alpha_s^2)$ 
as a function of $x$ and $Q^2$. The conditions are the same as in Figure~\ref{FIG:FLncall}. 
Dashed lines: 
asymptotic representation in $Q^2$ for the heavy flavor corrections; full lines: complete heavy flavor contributions.
}
\label{FIG:FLLvA}
\end{figure}
\begin{figure}[H]
\centering
\includegraphics[scale=0.9]{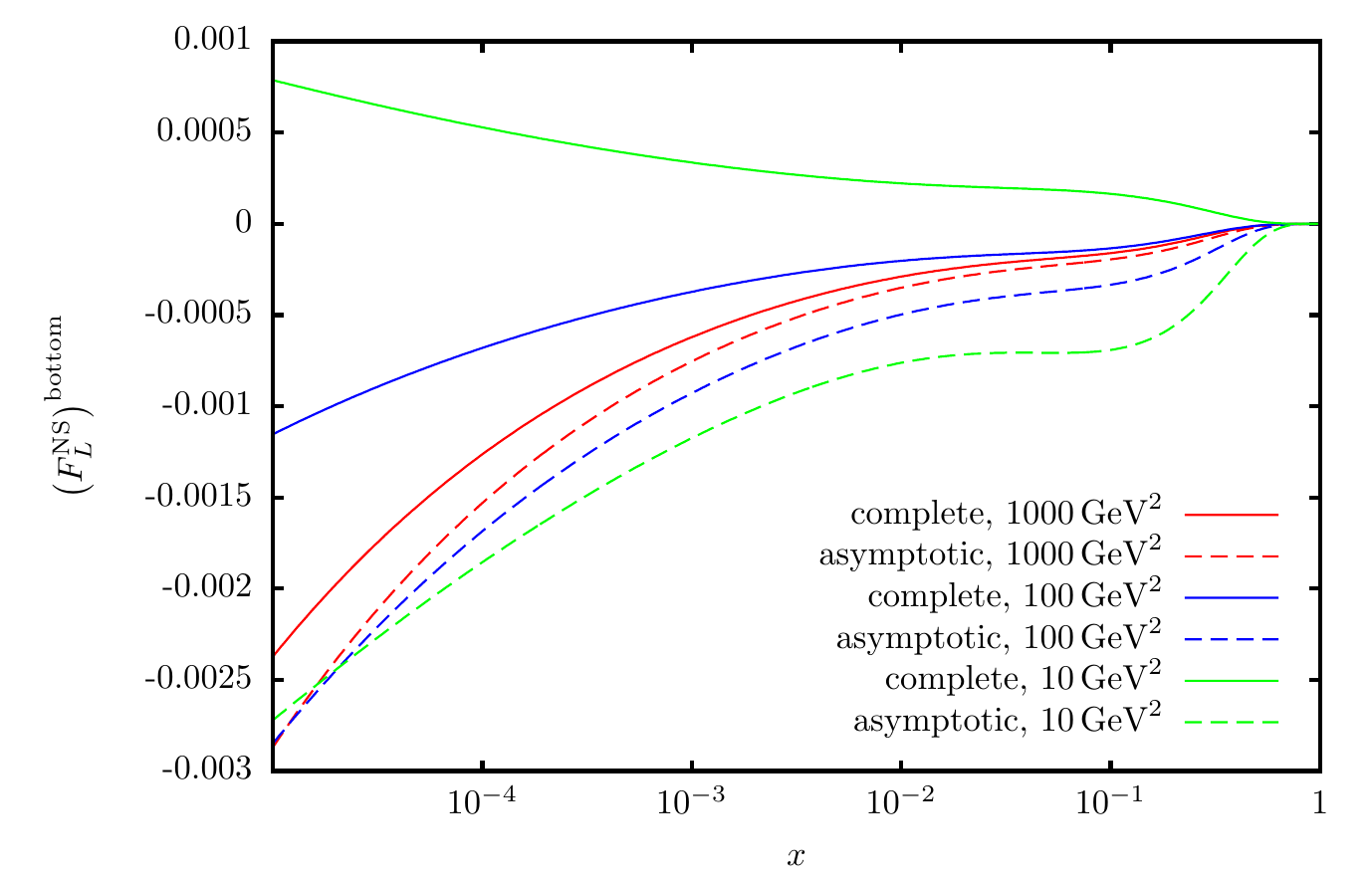}
\caption{\sf \small The bottom quark  contribution to the structure function $F_L$ due to photon 
exchange up to $O(\alpha_s^2)$ 
as a function of $x$ and $Q^2$. The conditions are the same as in Figure~\ref{FIG:FLncall}. 
Dashed lines: 
asymptotic representation in $Q^2$ for the heavy flavor corrections; full lines: complete heavy flavor contributions.
}
\label{FIG:FLLvA1}
\end{figure}
\noindent
and the massless Wilson coefficient is given by
\begin{eqnarray}
C_{F_L,q}\left(x,\frac{Q^2}{\mu^2}\right) = \sum_{k=1}^2 a_s^k C_{F_L,q}^{(k)}\left(x,\frac{Q^2}{\mu^2}\right)~,
\end{eqnarray}
with \cite{Zijlstra:1992qd}
\begin{eqnarray}
C_{F_L,q}^{(1)}\left(x,\frac{Q^2}{\mu^2}\right) &=& c_{F_L,q}^{(1)}(x)
\\
C_{F_L,q}^{(2)}\left(x,\frac{Q^2}{\mu^2}\right) &=& 
\Biggl\{\left[P_{qq}^{(0)} \otimes c_{F_L,q}^{(1)}\right](x) - \beta_0 c_{F_L,q}^{(1)}(x)\Biggr\} 
\ln\left(\frac{Q^2}{\mu^2}\right)
+ c_{F_L,q}^{(2)}(x)~.
\nonumber\\ 
\end{eqnarray}

\begin{figure}[H]
\centering
\includegraphics[scale=0.9]{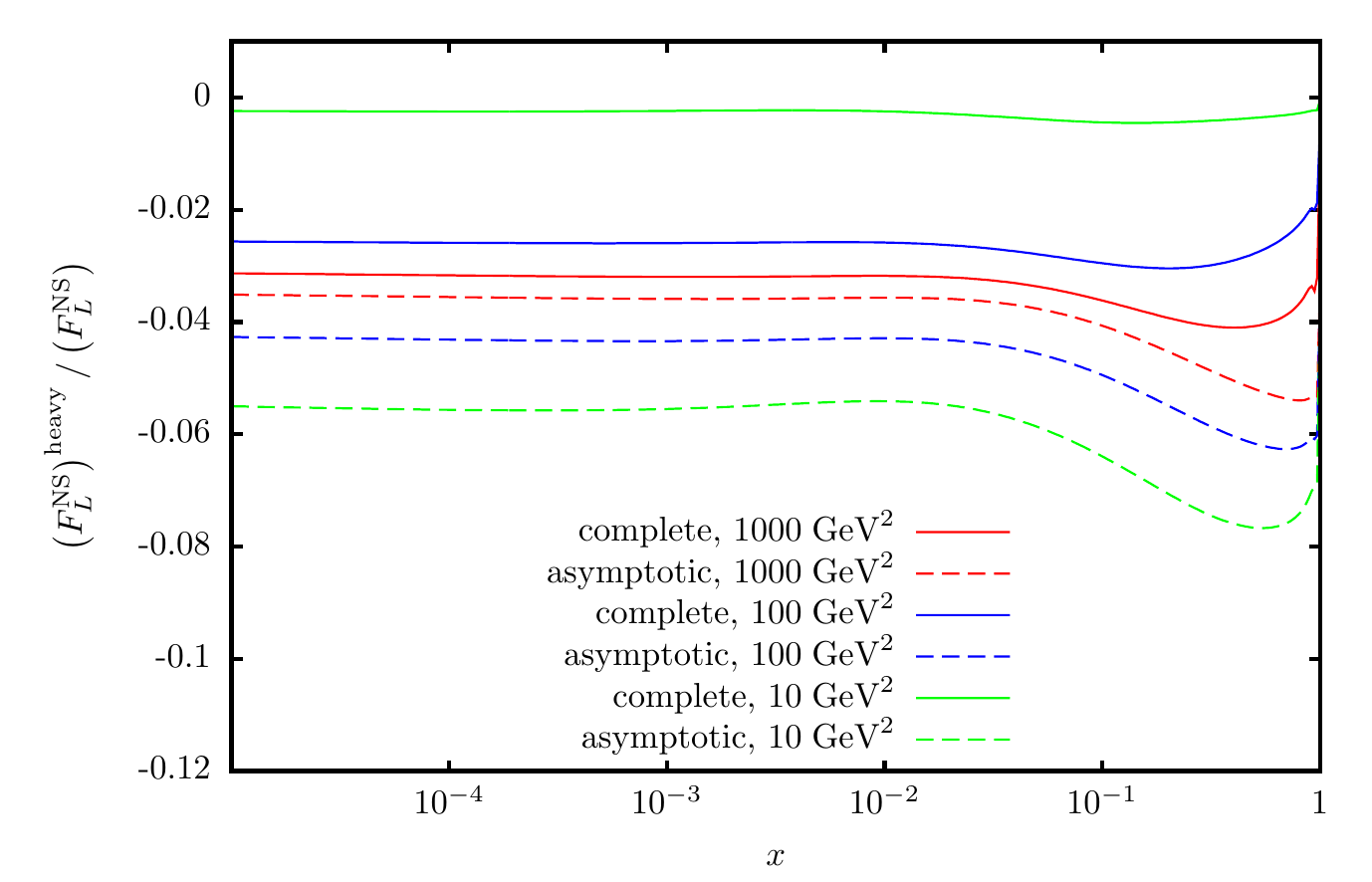}
\caption{\sf \small The ratio of the heavy flavor contributions to the structure function $F_L$ 
due to photon exchange to the complete
structure function up to $O(\alpha_s^2)$ as a function of $x$ and $Q^2$. The conditions are the same as in Figure~\ref{FIG:FLncall}.
Dashed lines: 
asymptotic representation in $Q^2$ for the heavy flavor corrections; full lines: complete heavy flavor contributions.}
\label{FIG:FLncratios}
\end{figure}

In Figure~\ref{FIG:FLncall} we show the $O(a_s^2)$ corrections to the non-singlet structure function 
$F_L^{\rm NS}$,
including the complete charm and bottom quark corrections. During evolution this structure function grows towards small values of 
$x$. 
The absolute charm and bottom quark contributions are illustrated in Figures~\ref{FIG:FLLvA}, \ref{FIG:FLLvA1}.
In the present case, the corrections in the asymptotic limit are sufficiently close to the complete corrections only for
$Q^2 \gsim 1000~\GeV^2$ in the case of charm. It is well known that for  $F_L$ the asymptotic 
representation holds at very high scales only, which also applies to the non-singlet case. 
For the charm quark corrections the asymptotic 
representation holds at $Q^2 \sim 1000~\GeV^2$. Below there are significant differences. The situation is 
correspondingly worse
for the bottom quark corrections shown in Figure~\ref{FIG:FLLvA1}. In general the asymptotic corrections give larger negative
corrections than found in the complete calculation. The relative heavy flavor corrections for $F_L^{\rm NS}$ are shown in 
Figure~\ref{FIG:FLncratios}. They behave nearly constant in 
the small $x$ region, amounting to $-0.3$ to $-4\%$ in the region $Q^2 = 10$ to $1000~\GeV^2$, with larger asymptotic 
corrections.
\subsection{\boldmath 
$F_{2}^{\rm NS}$}
\label{sec:4.2}

\vspace*{1mm}
\noindent
For the structure function $F_2$, we obtain the following Compton contribution 
\begin{eqnarray}
L_{F_2,q}^{\text{NS},(2),C}\left(z,\frac{Q^2}{m^2},\frac{m^2}{\mu^2}\right)
 &=&
 a_s^2C_F T_F \Biggl\{
 \left[-\frac{16 z^2}{\xi^2(1-z)} \left(1-9z+9z^2\right) + \frac{4}{3} \frac{1+z^2}{1-z} \right]
\nonumber\\ &&
\times  \Biggl[ \Biggl[
 \ln\left(\frac{1+\sqrt{1-\frac{4z}{\xi}}}{1-\sqrt{1-\frac{4z}{\xi}}}\right)
+ \ln\left(\frac{1-z}{z^2}\right) \Biggr]
 \ln\left(\frac{1+\sqrt{1-\frac{4z}{(1-z)\xi}}}{1-\sqrt{1-\frac{4z}{(1-z)\xi}}}\right)
\nonumber\\ &&
+2 \Biggl[
- \Li_2\left(\frac{(1-z)\left(1+\sqrt{1-\frac{4z}{(1-z)\xi}}\right)}{1+\sqrt{1-\frac{4z}{\xi}}} \right)
+ \Li_2\left(\frac{1- \sqrt{1-\frac{4z}{\xi}}}{1 +\sqrt{1-\frac{4z}{(1-z)\xi}}} \right)
\nonumber\\ &&
+ \Li_2\left(\frac{1 - \sqrt{1-\frac{4z}{(1-z)\xi}}}
                  {1 + \sqrt{1-\frac{4z}{\xi}}}\right)
- \Li_2\left(\frac{1 + \sqrt{1-\frac{4z}{(1-z)\xi}}}
                  {1 + \sqrt{1-\frac{4z}{\xi}}}\right)
 \Biggr] \Biggr]
\nonumber\\ &&
+ \frac{8}{9(1-z)\xi}\left[26 z -168 z^2 +188 z^3 - (8 - 6z+17z^2) \xi\right]
\sqrt{1-\frac{4z}{\xi}}
\nonumber\\ && \times
\ln\left(\frac{
 \sqrt{1-\frac{4z}{\xi}} + \sqrt{1-\frac{4z}{(1-z)\xi}}}
{\sqrt{1-\frac{4z}{\xi}} - \sqrt{1-\frac{4z}{(1-z)\xi}}}\right)
+\frac{2}{27 (1-z)^2 \xi} \sqrt{1-\frac{4z}{(1-z) \xi}}
\nonumber\\ && \times
\left[ -1702 z +9516 z^2 -14260 z^3
+6456 z^4  \right.
\nonumber\\ && \left.
+(223-709 z +1108 z^2 -622 z^3) \xi\right]\nonumber\\ &&
-\frac{4}{3(1-z)^3\xi^2}\left[6z^2(-15+70z-90z^2+36z^3)-\xi^2(1-3z^2+2z^3)\right]\nonumber\\ &&
\times\ln\left(\frac{1+\sqrt{1-\frac{4z}{(1-z)\xi}}}{1-\sqrt{1-\frac{4z}{(1-z)\xi}}}\right)
\Biggr\}\theta\left(\frac{\xi}{\xi+4} - z\right).
\end{eqnarray}
This expression agrees with a result given in \cite{Buza:1995ie}.
The virtual correction is the same as in the case of the structure function $g_1^{\rm NS}$, 
Eq.~(\ref{eq:LqNSg1V}), and 
the contribution with massless final states is given by:
\begin{eqnarray}
L_{F_2,q}^{{\rm NS},(2), \rm massless}\left(z,\frac{Q^2}{\mu^2},\frac{m^2}{\mu^2}\right)
= - a_s^2 \beta_{0,Q} \ln\left(\frac{m^2}{\mu^2}\right) \left[P_{qq}^{(0)}(z) 
\ln\left(\frac{Q^2}{\mu^2}\right)
+c_{F_2,q}^{(1)}(z) \right],
\end{eqnarray}
with  \cite{Furmanski:1981cw}
\begin{eqnarray}
\label{eq:CF21}
c_{F_2,q}^{(1)}(z)  &=& 
C_F \Biggr\{
4 \left(\frac{\ln(1-z)}{1-z}\right)_+ 
- \left(\frac{3}{(1-x)}\right)_+
+ 6 + 4 z - 2(1+z) \ln(1-z) 
\nonumber\\ && 
- 2 \left(\frac{1+z^2}{1-z}\right) \ln(z) 
-(9 + 4\zeta_2) \delta(1-z)\Biggr\}~. 
\end{eqnarray}
Up to $O(a_s^2)$ the non-singlet structure function  $F_2(x,Q^2)$ reads
\begin{eqnarray}
F_2^{\rm NS}(x,Q^2) &=& x 
\Biggr\{\left[C_{F_2,q}\left(x,\frac{Q^2}{\mu^2}\right) + L_{F_2,q}^{{\rm 
NS},(2)}\left(x,\frac{Q^2}{\mu^2},\frac{m^2}{\mu^2}\right)\right] 
\nonumber\\ &&
\otimes \Bigl[
 \frac{4}{9} u_v(x,\mu^2)
+ \frac{1}{9} d_v(x,\mu^2)
+ \frac{8}{9} \bar{u}(x,\mu^2)
+ \frac{2}{9} \left[\bar{d}(x,\mu^2) + \bar{s}(x,\mu^2)\right]
\Bigr]\Biggr\}~,\nonumber\\
\end{eqnarray}
and the massless Wilson coefficient is given by
\begin{eqnarray}
C_{F_2,q}\left(x,\frac{Q^2}{\mu^2}\right) = \delta(1-x) + \sum_{k=1}^2 a_s^k C_{F_2,q}^{(k)}\left(x,\frac{Q^2}{\mu^2}\right)~,
\end{eqnarray}
with \cite{Zijlstra:1992qd}
\begin{eqnarray}
C_{F_2,q}^{(1)}\left(x,\frac{Q^2}{\mu^2}\right) &=& P_{qq}^{(0)}(x) \ln\left(\frac{Q^2}{\mu^2}\right) + c_{F_2,q}^{(1)}(x)
\\
C_{F_2,q}^{(2)}\left(x,\frac{Q^2}{\mu^2}\right) &=& \frac{1}{2} \left\{\left[P_{qq}^{(0)} \otimes P_{qq}^{(0)}\right](x) - 
\beta_0
P_{qq}^{(0)}\right\} \ln^2\left(\frac{Q^2}{\mu^2}\right) 
\nonumber\\ &&
+ \left\{P_{qq}^{(1),{\rm NS,+}}(x) + \left[P_{qq}^{(0)} \otimes 
c_{F_2,q}^{(1)}\right](x) - \beta_0 c_{F_2,q}^{(1)}(x) \right\} \ln\left(\frac{Q^2}{\mu^2}\right) + c_{F_2,q}^{(2)}(x)~.
\nonumber\\ 
\end{eqnarray}

\begin{figure}[H]
\centering
\includegraphics[scale=0.9]{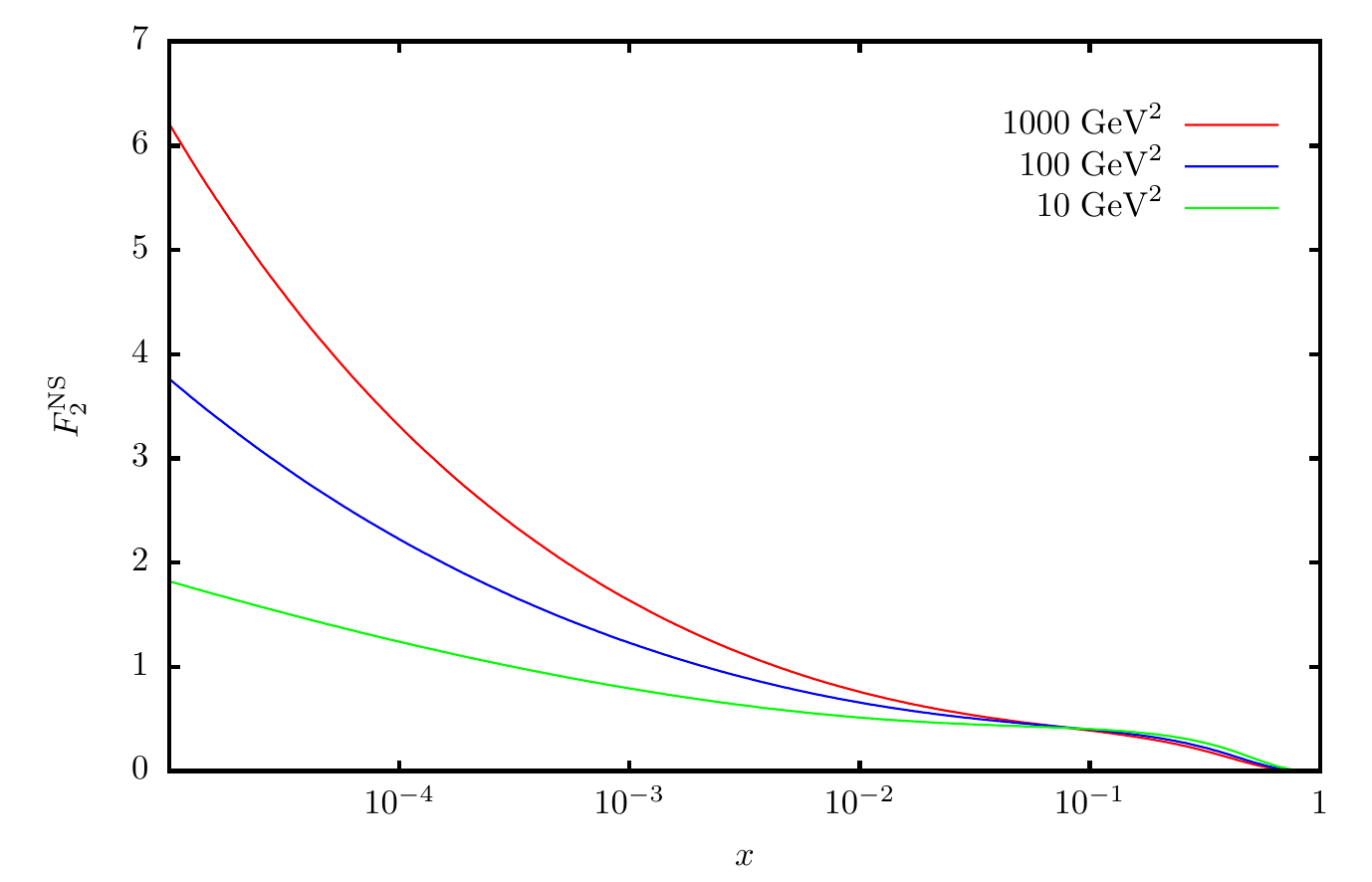}
\caption{\sf \small The structure function $F_2$ due to photon exchange up to $O(\alpha_s^2)$ 
including the charm and  bottom quark corrections in
the on-shell scheme with $m_c = 1.59~\GeV$ \cite{Alekhin:2012vu} and $m_b = 4.78~\GeV$ \cite{PDG15} 
using the NNLO parton distribution
functions \cite{Alekhin:2013nda}.}
\label{FIG:F2ncall}
\end{figure}

\begin{figure}[H]
\centering
\includegraphics[scale=0.9]{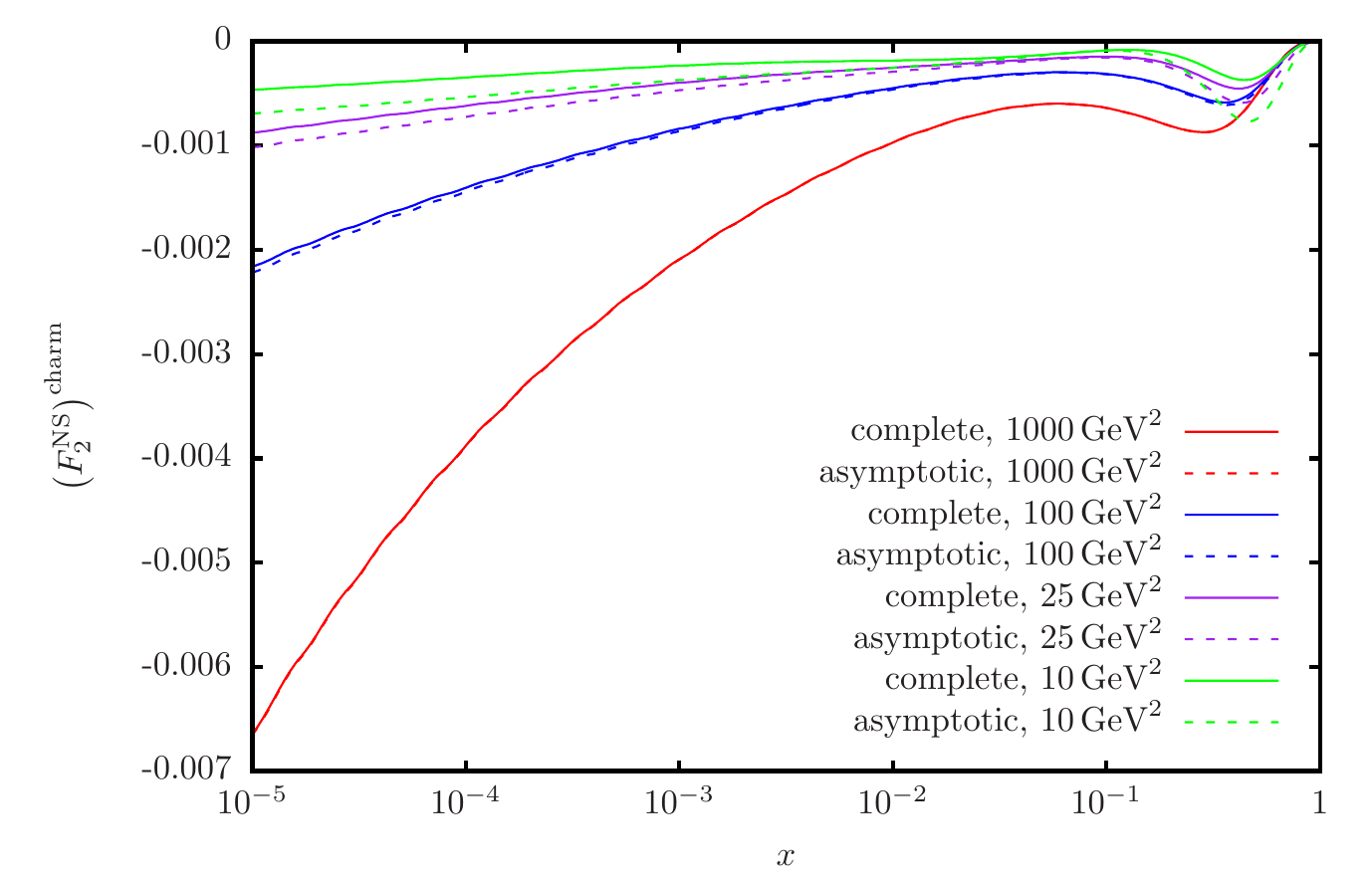}
\caption{\sf \small The charm quark contribution to the structure function $F_2$ due to photon 
exchange up to $O(\alpha_s^2)$ 
as a function of $x$ and $Q^2$. The conditions are the same as in Figure~\ref{FIG:F2ncall}. 
Dashed lines: 
asymptotic representation in $Q^2$ for the heavy flavor corrections; full lines: complete heavy flavor contributions.
}
\label{FIG:f2pow10}
\end{figure}

\begin{figure}[H]
\centering
\includegraphics[scale=0.9]{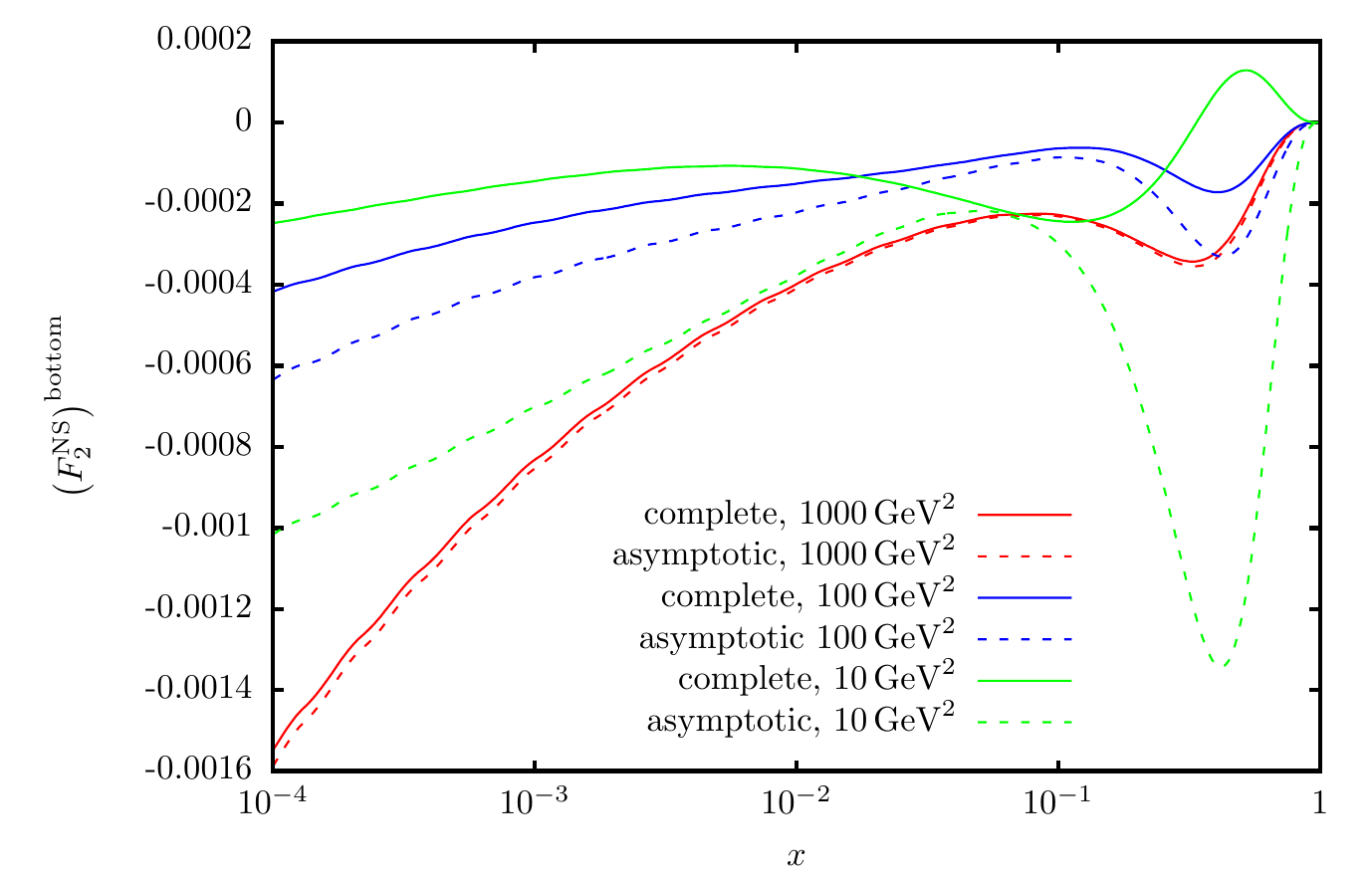}
\caption{\sf \small The bottom quark contribution to the structure function $F_2$ due to photon 
exchange up to $O(\alpha_s^2)$ 
as a function of $x$ and $Q^2$. The conditions are the same as in Figure~\ref{FIG:F2ncall}. 
Dashed lines: 
asymptotic representation in $Q^2$ for the heavy flavor corrections; full lines: complete heavy flavor contributions.
}
\label{FIG:f2pow101}
\end{figure}

\vspace*{1mm}
In Figure~\ref{FIG:F2ncall} we show $F_2^{\rm NS}$ up to $O(a_s^2)$ with the complete charm and 
bottom 
corrections. It rises for growing values of $Q^2$ for small values of $x$.
\noindent
The absolute charm and bottom quark corrections are illustrated in Figures~\ref{FIG:f2pow10} and 
\ref{FIG:f2pow101}, 
illustrating as well the effect of the asymptotic results. They get close to the exact ones much earlier than in the case of 
$F_L^{\rm NS}$.

\noindent
The bottom quark contributions shown in Figure~\ref{FIG:f2pow101} are about one order of magnitude smaller 
than those
for charm quarks, still with clear differences between the exact and asymptotic result at $Q^2 \sim 
100~\GeV^2$.  

\begin{figure}[H]
\centering
\includegraphics[scale=0.9]{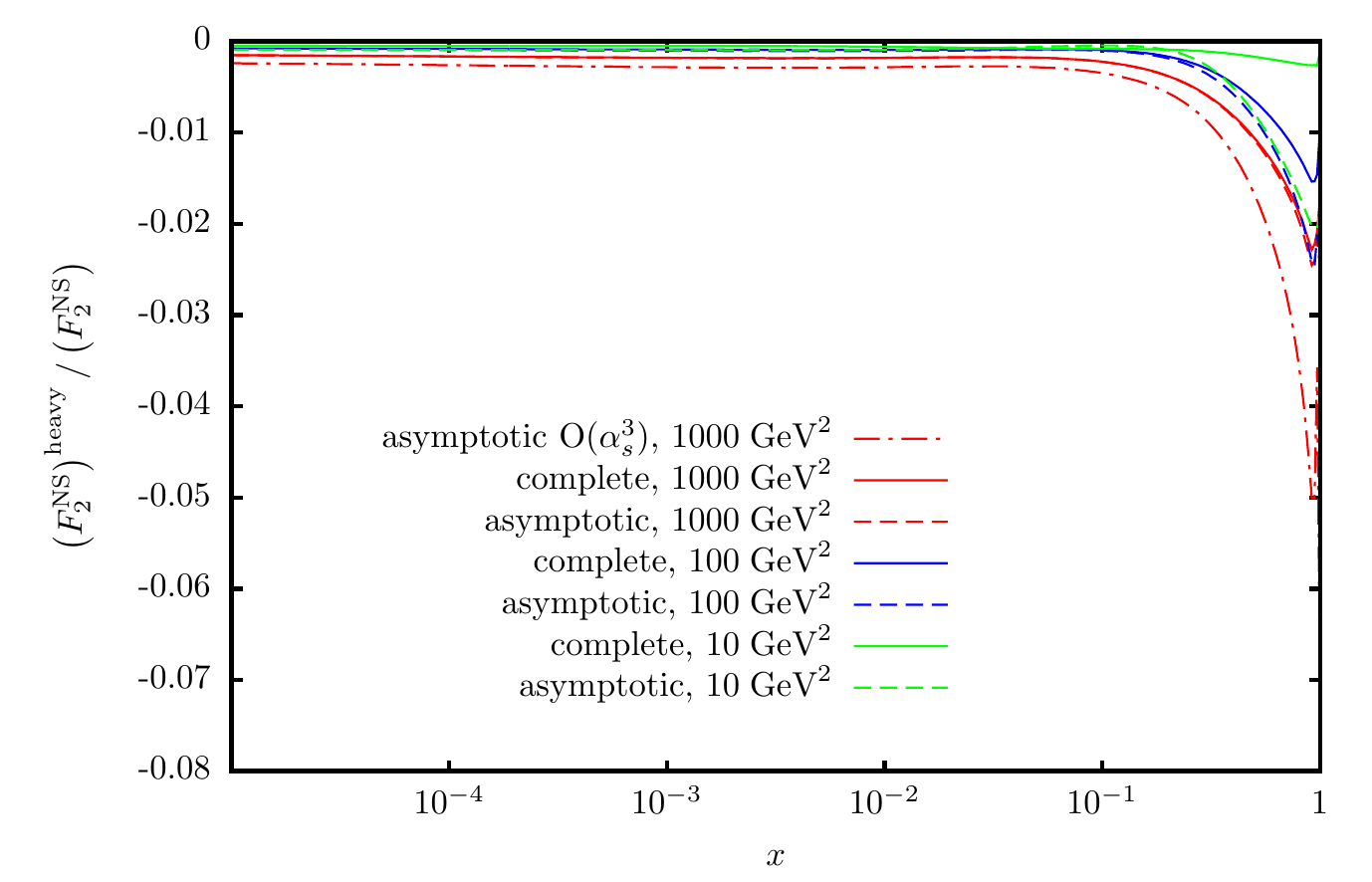}
\caption{\sf \small The ratio of the heavy flavor contributions to the structure function $F_2$ 
due to photon exchange to the complete
structure function up to $O(\alpha_s^2)$ as a function of $x$ and $Q^2$. The conditions are the same as in Figure~\ref{FIG:F2ncall}.
Dashed lines: 
asymptotic representation in $Q^2$ for the heavy flavor corrections; full lines: complete heavy flavor contributions.}
\label{FIG:F2ncratios}
\end{figure}

Figure~\ref{FIG:F2ncratios} illustrates the relative contribution of the heavy flavor corrections up to $O(a_s^2)$. 
The corrections are rather flat in the small $x$ region and amount to $-0.1$ to $-0.6~\%$ for $x < 0.1$ growing towards
$-2.5~\%$ at large $x$ from $Q^2 = 10~\GeV^2$ to $1000~\GeV^2$.
\section{The unpolarized non-singlet charged current structure functions}
\label{sec:5}

\vspace*{1mm}
\noindent
In the non-singlet charged current case we have to distinguish transitions between light flavors accompanied with
heavy flavor production and the excitation of charm from massless down-type quarks. Due to the smallness of 
the corresponding 
CKM-matrix element \cite{PDG15} we will not consider the excitation of bottom quarks from the massless quarks. The current 
values of the contributing CKM-matrix elements are 
\begin{eqnarray}
\begin{array}{ll}
|V_{ud}| = 0.97425,  &~~~~~~~~~~~~|V_{us}| = 0.2253 \\
|V_{cd}| = 0.225,    &~~~~~~~~~~~~|V_{cs}| = 0.986. \\
\end{array}
\end{eqnarray}
The corresponding flavor non-singlet combinations are given by
\begin{eqnarray}
\label{eq:F1CC}
F_1^{\bar{\nu}p}(x,Q^2) - F_1^{{\nu}p}(x,Q^2) &=& 
\left[C_{F_1,q}^{\rm NS}\left(x,\frac{Q^2}{\mu^2}\right) +
               L_{F_1,q}^{\rm NS}\left(x,\frac{Q^2}{\mu^2},\frac{m^2}{\mu^2}\right) \right]
\otimes
\bigl[
\left(|V_{du}|^2 
+ |V_{su}|^2\right) 
\nonumber\\ &&
\times u_v(x,\mu^2) 
- |V_{du}|^2 d_v(x,\mu^2) \bigr] 
\nonumber\\   &&
- H_{F_1,q}^{\rm NS}\left(x,\frac{Q^2}{\mu^2},\frac{m^2}{\mu^2}\right) \otimes 
|V_{cd}|^2  d_v(x,\mu^2),
\\
\label{eq:F2CC}
F_2^{\bar{\nu}p}(x,Q^2) - F_2^{{\nu}p}(x,Q^2) &=& 
2x \Biggl\{\left[C_{F_2,q}^{\rm NS}\left(x,\frac{Q^2}{\mu^2}\right) +
               L_{F_2,q}^{\rm NS}\left(x,\frac{Q^2}{\mu^2},\frac{m^2}{\mu^2}\right) \right]
\otimes
\bigl[
 \left(|V_{du}|^2 
+ |V_{su}|^2\right) 
\nonumber\\ &&
\times u_v(x,\mu^2) 
- |V_{du}|^2 d_v(x,\mu^2) 
\bigr] 
\nonumber\\ &&
- H_{F_2,q}^{\rm NS}\left(x,\frac{Q^2}{\mu^2},\frac{m^2}{\mu^2}\right) \otimes 
|V_{cd}|^2  d_v(x,\mu^2) \Biggr\},
\\
\label{eq:F3CC}
F_3^{\bar{\nu}p}(x,Q^2) + F_3^{{\nu}p}(x,Q^2) &=& 
2 \Biggl\{\left[C_{F_3,q}^{\rm NS}\left(x,\frac{Q^2}{\mu^2}\right) +
               L_{F_3,q}^{\rm NS}\left(x,\frac{Q^2}{\mu^2},\frac{m^2}{\mu^2}\right) \right]
\otimes
\bigl[|V_{du}|^2 d_v(x,\mu^2) 
\nonumber\\ &&
+ \left(|V_{du}|^2 
+ |V_{su}|^2\right) u_v(x,\mu^2) \bigr] 
\nonumber\\ &&
+ H_{F_3,q}^{\rm NS}\left(x,\frac{Q^2}{\mu^2},\frac{m^2}{\mu^2}\right) \otimes 
|V_{cd}|^2  d_v(x,\mu^2) \Biggr\}.
\end{eqnarray}
Here, $C_{F_i,q}^{\rm NS}, L_{F_i,q}^{\rm NS}$, and $H_{F_i,q}^{\rm NS}$ denote the massless $(C)$ 
and 
massive  
Wilson coefficients $(L,H)$ for the coupling of the weak bosons to only massless quarks $(C,L)$ and for 
charm excitation $(H)$. We assume that the sea quark distributions obey
\begin{eqnarray}
u_s(x,\mu^2) = \bar{u}(x,Q^2),~~~~~d_s(x,\mu^2) = \bar{d}(x,Q^2),~~~~~s(x,\mu^2) = \bar{s}(x,Q^2)~.
\end{eqnarray}
The contributions due to the Wilson coefficients $H_{F_i,q}^{\rm NS}, i=1,2,3$ are Cabbibo 
suppressed. 

The combinations (\ref{eq:F1CC}--\ref{eq:F3CC})
are related to the unpolarized Bjorken sum rule \cite{Bjorken:1967px}, the Adler sum rule 
\cite{Adler:1965ty}, and the Gross-Llewellyn Smith sum rule \cite{Gross:1969jf}, respectively, by their 
first  moments. First we consider these combinations themselves and turn to the sum rules later. 
Up to $O(\alpha_s)$ the single heavy 
quark excitations have been calculated in Refs.~\cite{Gluck:1997sj,Blumlein:2011zu} correcting 
results in \cite{Gottschalk:1980rv}. 
\begin{figure}[H]
\centering
\includegraphics[scale=0.9]{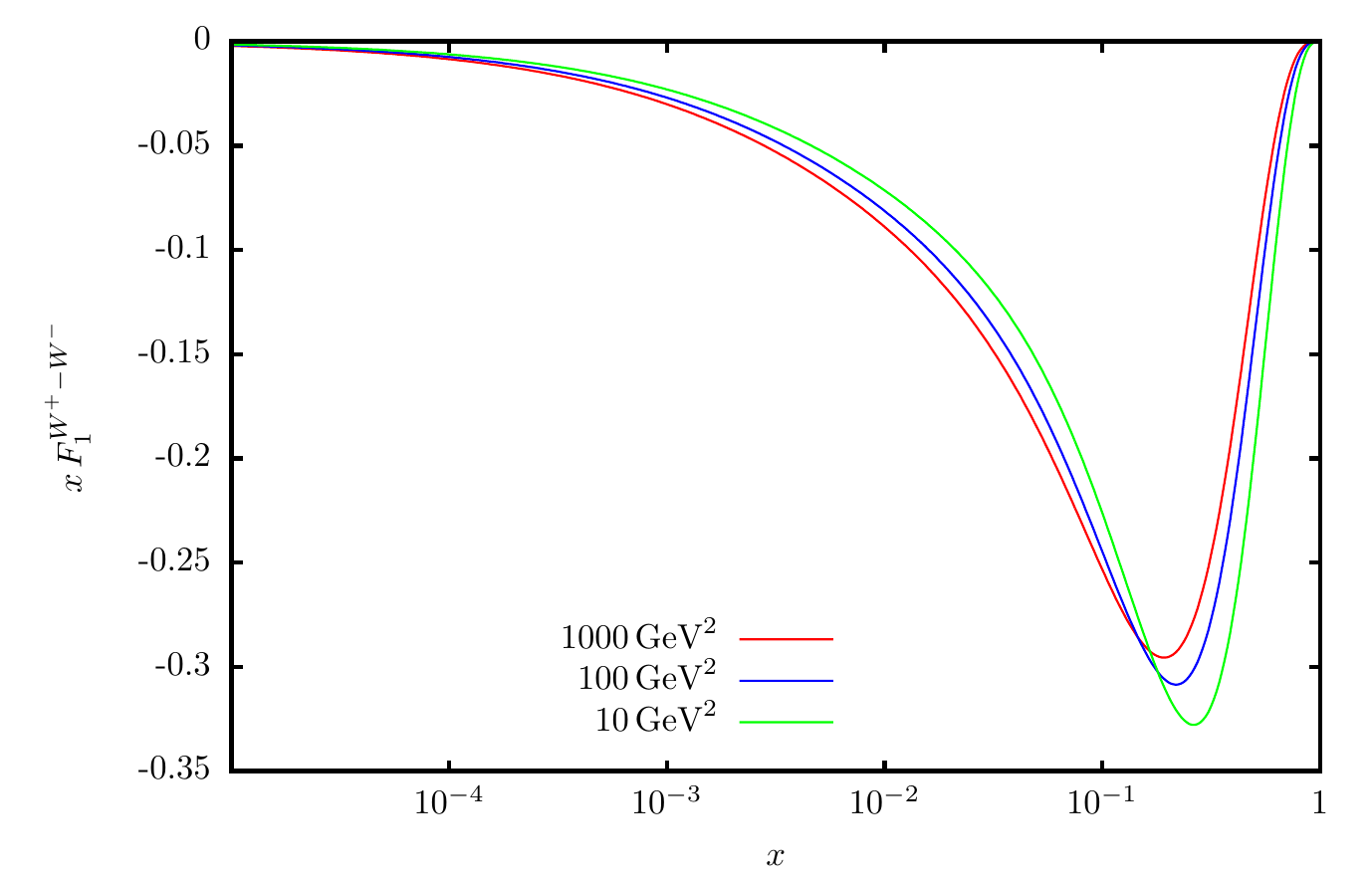}
\caption{\sf \small The charged current structure function $xF_1^{W^+ - W^-}$ up to 
$O(\alpha_s^2)$ including 
the charm quark corrections in the on-shell scheme with $m_c = 1.59~\GeV$ \cite{Alekhin:2012vu} and 
using the NNLO parton distribution functions \cite{Alekhin:2013nda}.}
\label{FIG:F1rec}
\end{figure}
\noindent
At two-loop order, only the asymptotic results for $Q^2 \gg m^2$
are available \cite{Blumlein:2014fqa}\footnote{A sign error in \cite{Buza:1997mg} has been corrected.},
to which we refer in the following. We will limit our considerations to the case of the charm contributions.

\begin{figure}[H]
\centering
\includegraphics[scale=0.9]{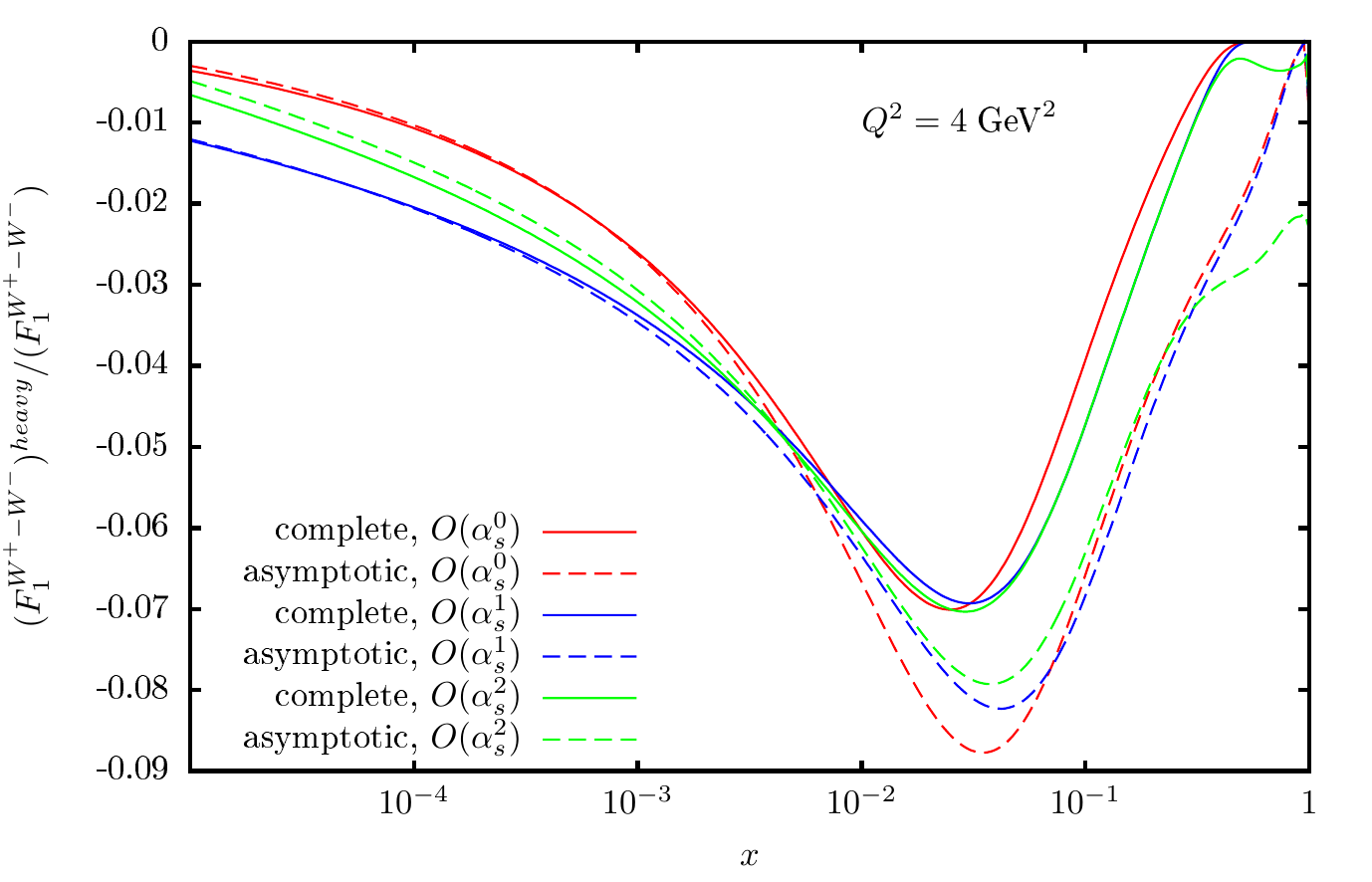}
\caption{\sf \small The ratio of the charm quark contributions to the charged current structure 
function $xF_1^{W^+ - W^-}$ 
up to $O(\alpha_s^2)$ to the full corrections at $Q^2 = 4~\GeV^2$. The other conditions are the same as in Figure~\ref{FIG:F1rec}.}
\label{FIG:F1ratios4}
\end{figure}

\begin{figure}[H]
\centering
\includegraphics[scale=0.9]{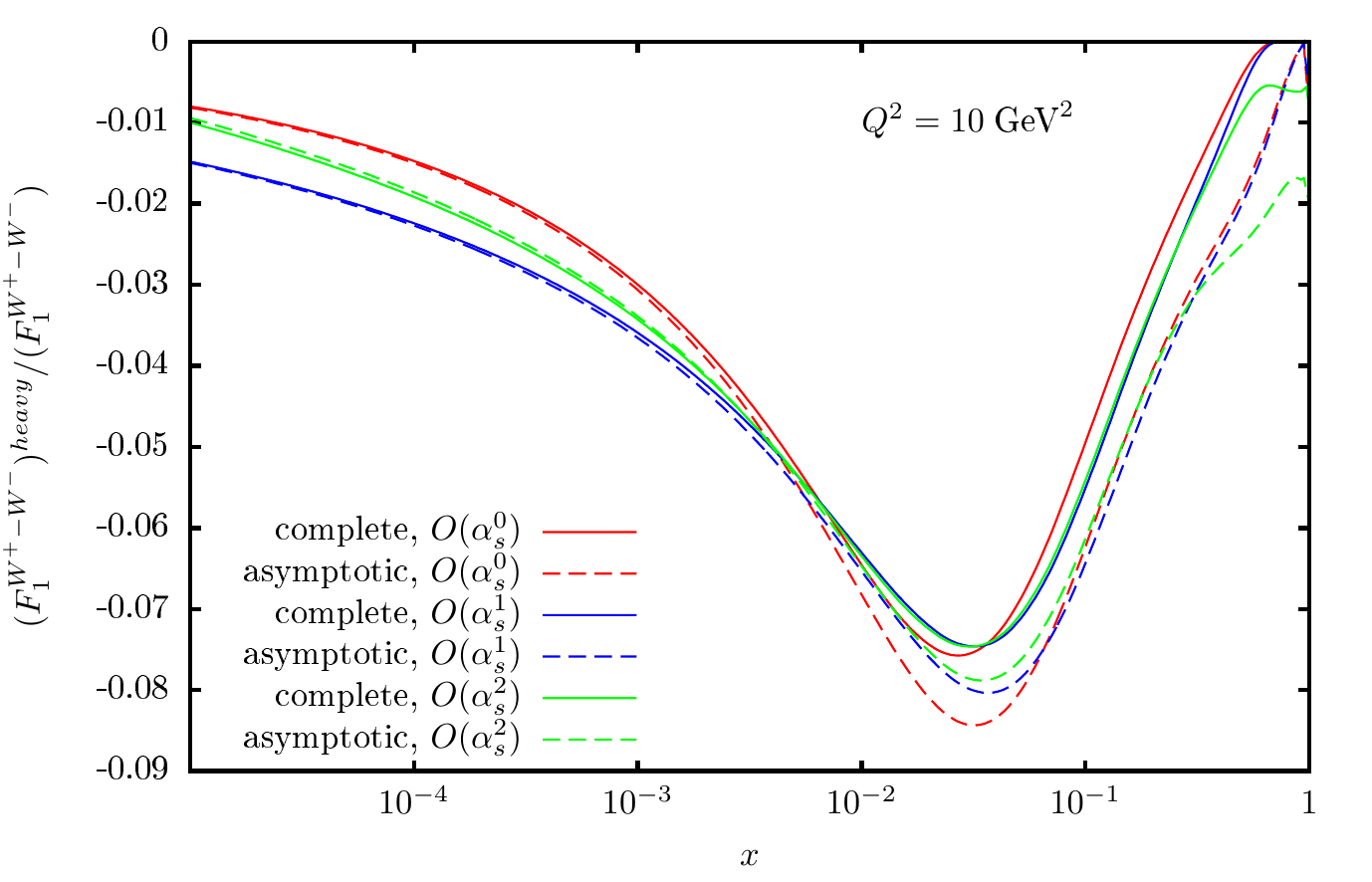}
\caption{\sf \small The ratio of the charm quark contributions to the charged current structure 
function $xF_1^{W^+ - W^-}$ 
up to $O(\alpha_s^2)$ to the full corrections at $Q^2 = 10~\GeV^2$. The other conditions are the same as in Figure~\ref{FIG:F1rec}.}
\label{FIG:F1ratios10}
\end{figure}

\begin{figure}[H]
\centering
\includegraphics[scale=0.9]{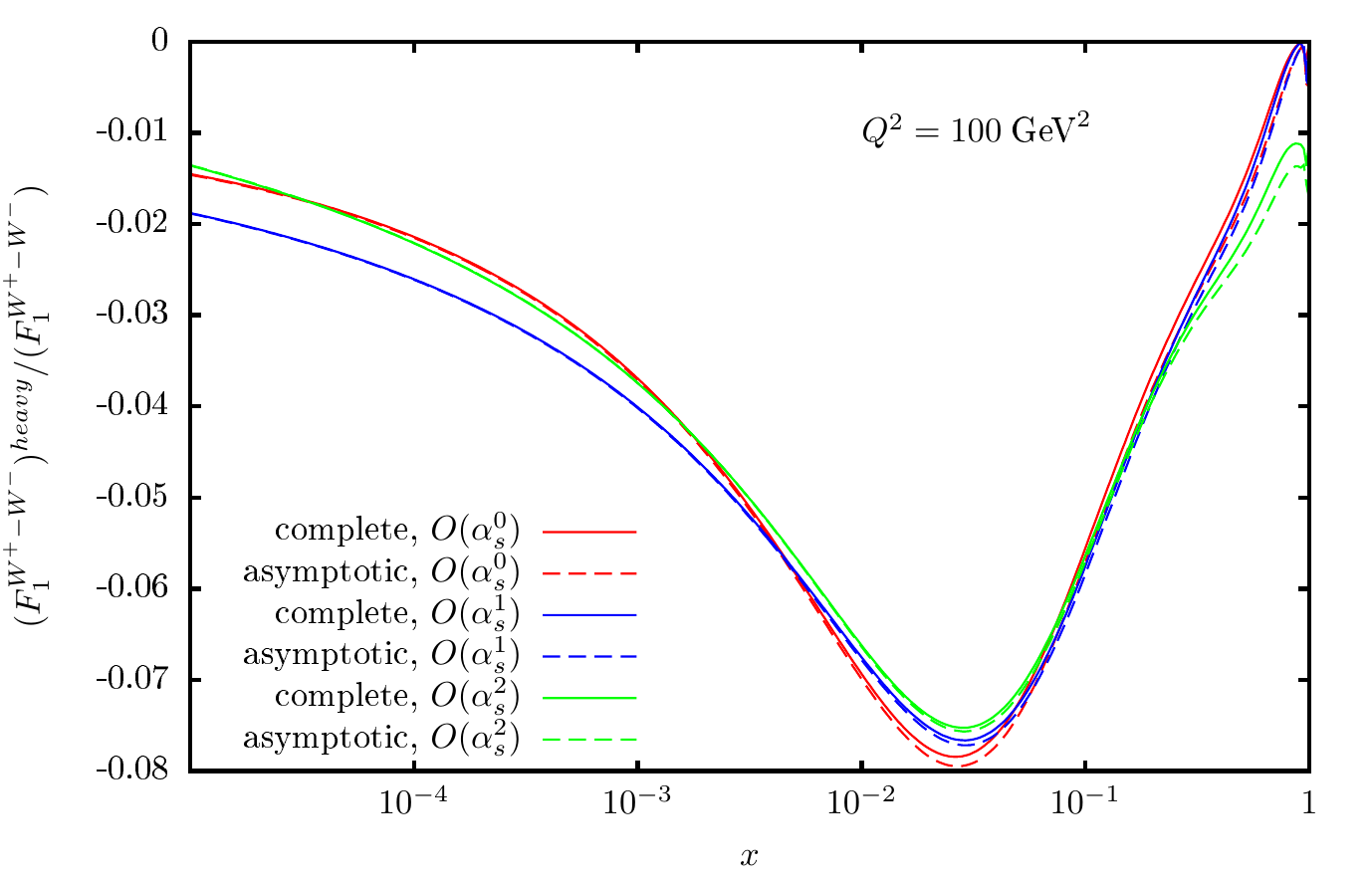}
\caption{\sf \small The ratio of the charm quark contributions to the charged current structure 
function $xF_1^{W^+ - W^-}$ up to $O(\alpha_s^2)$ to the full corrections at $Q^2 = 100~\GeV^2$. The other conditions are the 
same as in Figure~\ref{FIG:F1rec}.}
\label{FIG:F1ratios}
\end{figure}

In Figure~\ref{FIG:F1rec} we illustrate the non-singlet structure function $xF_1^{W^+ - W^-}$ up to
$O(\alpha_s^2)$, showing its scaling violations in the range $Q^2 = 10$ to $1000~\GeV^2$.
Its charm quark corrections up to $O(\alpha_s^2)$, using the asymptotic corrections for the
$O(\alpha_s^2)$ term of the flavor excitation contributions as a first approximation, are 
illustrated in
Figures~\ref{FIG:F1ratios4}--\ref{FIG:F1ratios} for virtualities $Q^2 = 4, 10$ and $100~\GeV^2$.
\noindent
The higher order results lower the corrections in the small $x$ region and enlarge it at large $x$.
The asymptotic corrections work well in the whole region $Q^2 \in [4, 100]~\GeV^2$ for small values of $x$
and in the whole region at $Q^2 = 100~\GeV^2$. The relative corrections amount to values between $0$ and $-8\%$.
\begin{figure}[H]
\centering
\includegraphics[scale=0.9]{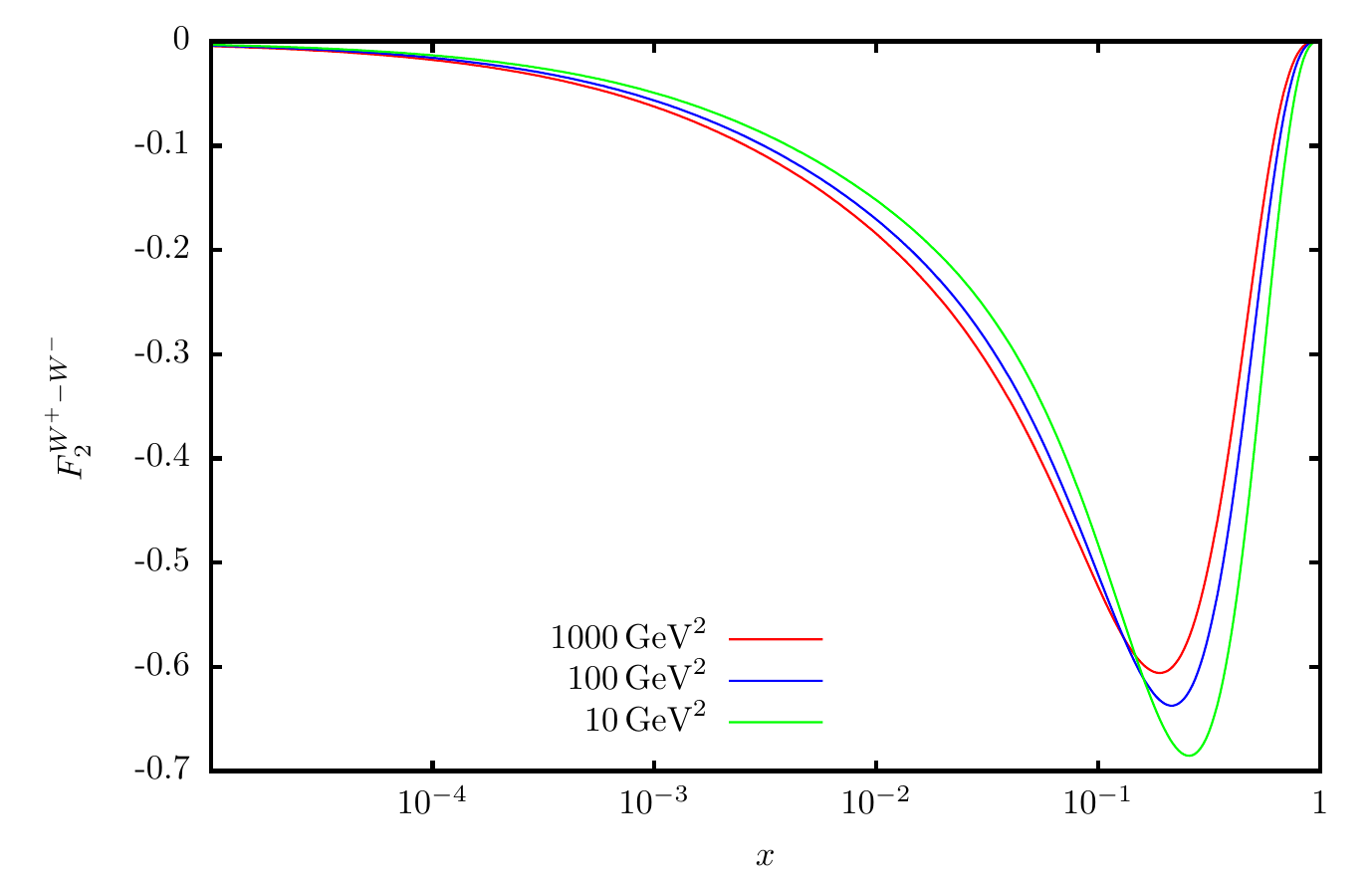}
\caption{\sf \small The charged current structure function $F_2^{W^+ - W^-}$ up to $O(\alpha_s^2)$ including the 
charm quark corrections in the on-shell scheme with $m_c = 1.59~\GeV$ \cite{Alekhin:2012vu} and using the NNLO 
parton distribution functions \cite{Alekhin:2013nda}.}
\label{FIG:F2rec}
\end{figure}

\begin{figure}[H]
\centering
\includegraphics[scale=0.9]{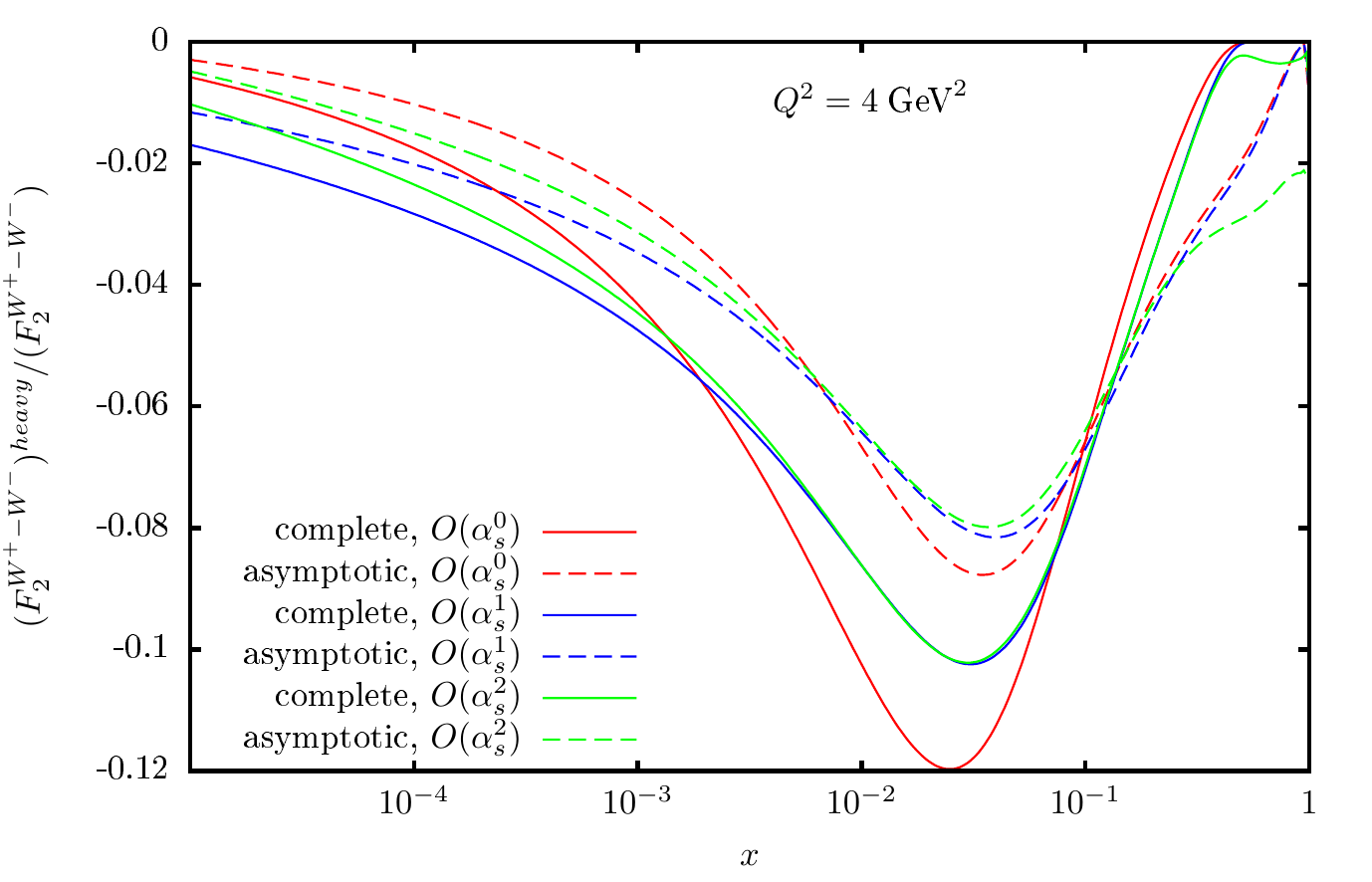}
\caption{\sf \small The ratio of the charm quark contributions to the charged current structure 
function $F_2^{W^+ - W^-}$ 
up to $O(\alpha_s^2)$ to the full corrections at $Q^2 = 4~\GeV^2$. The other conditions are the same as in 
Figure~\ref{FIG:F2rec}.}
\label{FIG:F2ratios4}
\end{figure}

\begin{figure}[H]
\centering
\includegraphics[scale=0.9]{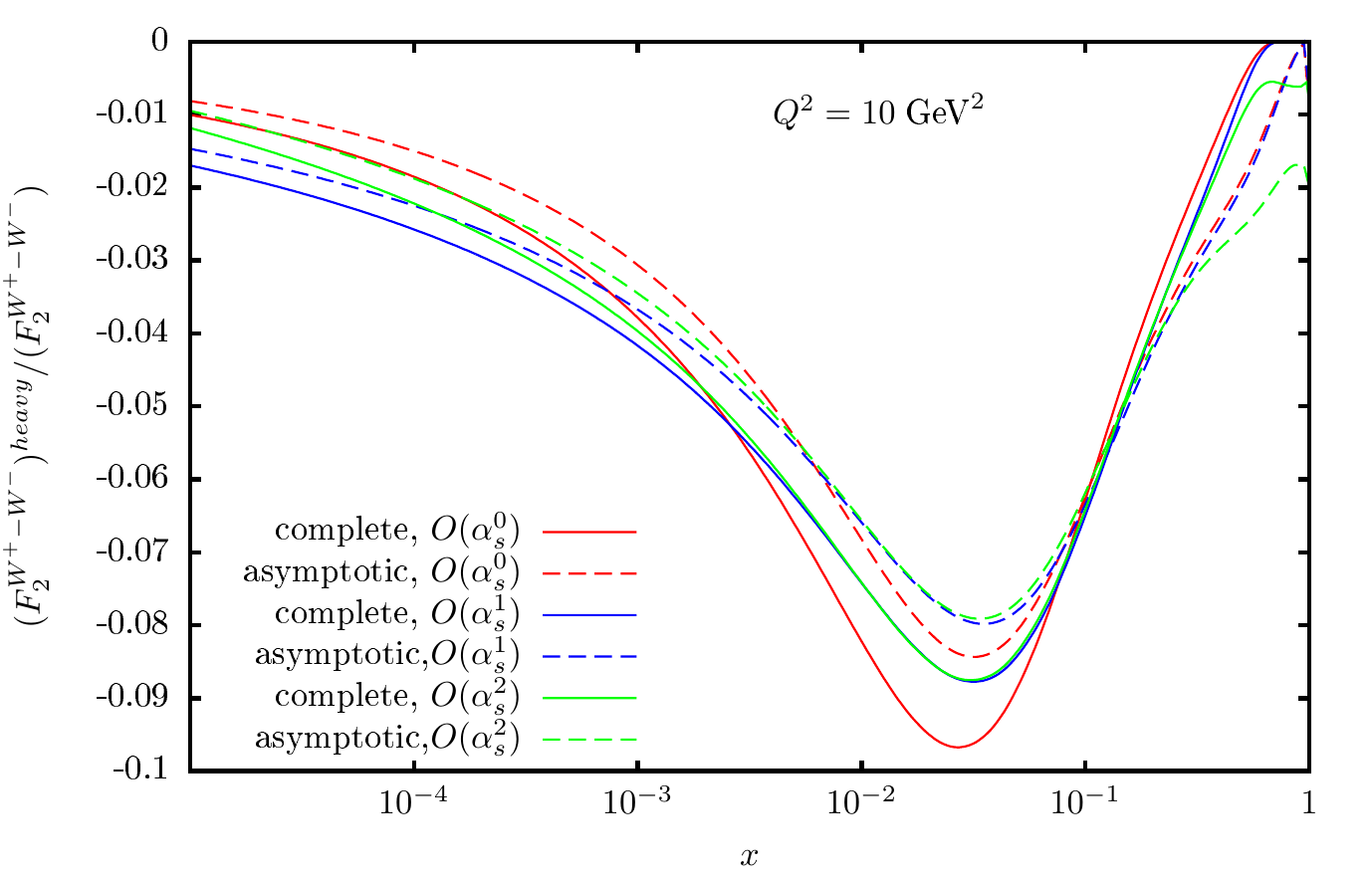}
\caption{\sf \small The ratio of the charm quark contributions to the charged current structure 
function $F_2^{W^+ - W^-}$
up to $O(\alpha_s^2)$ to the full corrections at $Q^2 = 10~\GeV^2$. The other conditions are the same as in 
Figure~\ref{FIG:F2rec}.}
\label{FIG:F2ratios10}
\end{figure}

\begin{figure}[H]
\centering
\includegraphics[scale=0.9]{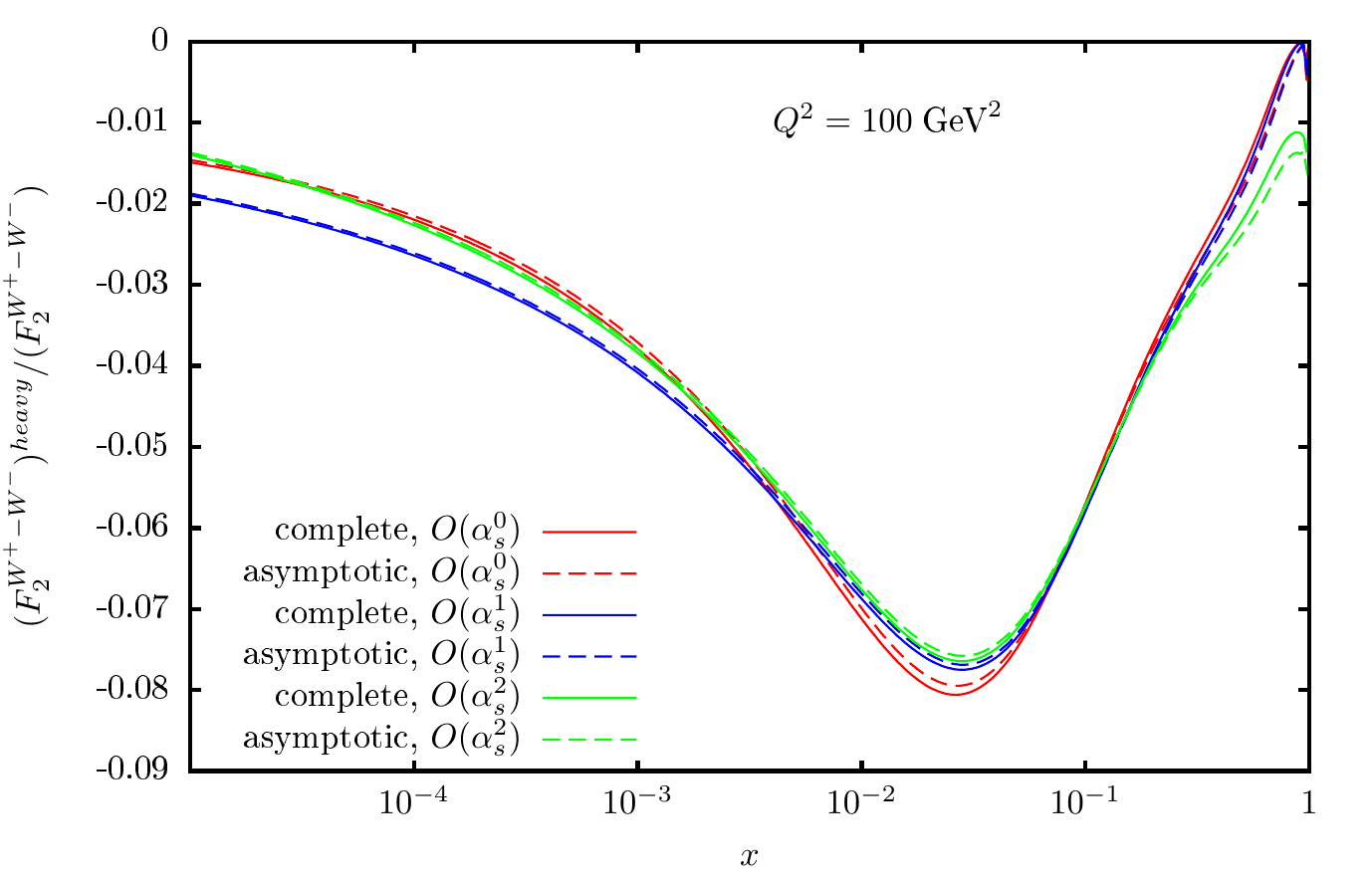}
\caption{\sf \small The ratio of the charm quark contributions to the charged current structure 
function $F_2^{W^+ - W^-}$ 
up to $O(\alpha_s^2)$ to the full corrections at $Q^2 = 100~\GeV^2$. The other conditions are the same as in 
Figure~\ref{FIG:F2rec}.}
\label{FIG:F2ratios}
\end{figure}

In Figure~\ref{FIG:F2rec} the charged current structure function $F_2^{W^+ - W^-}$ is shown, including the charm
quark corrections. Again QCD-evolution moves the profile towards smaller values of $x$.
The relative corrections due to charm are shown in Figures~\ref{FIG:F2ratios4}--\ref{FIG:F2ratios} for the scales
$Q^2 = 4, 10$ and $100~\GeV^2$.

\noindent
Here we also compare the asymptotic result against the complete ones.
The charm contribution is found in the range of $0$ to $\sim -12\%$ at $Q^2 = 4~\GeV^2$ to  $0$ to $\sim -8\%$, 
at $Q^2 = 100~\GeV^2$ peaking around $x \sim 0.03$.

\begin{figure}[H]
\centering
\includegraphics[scale=0.9]{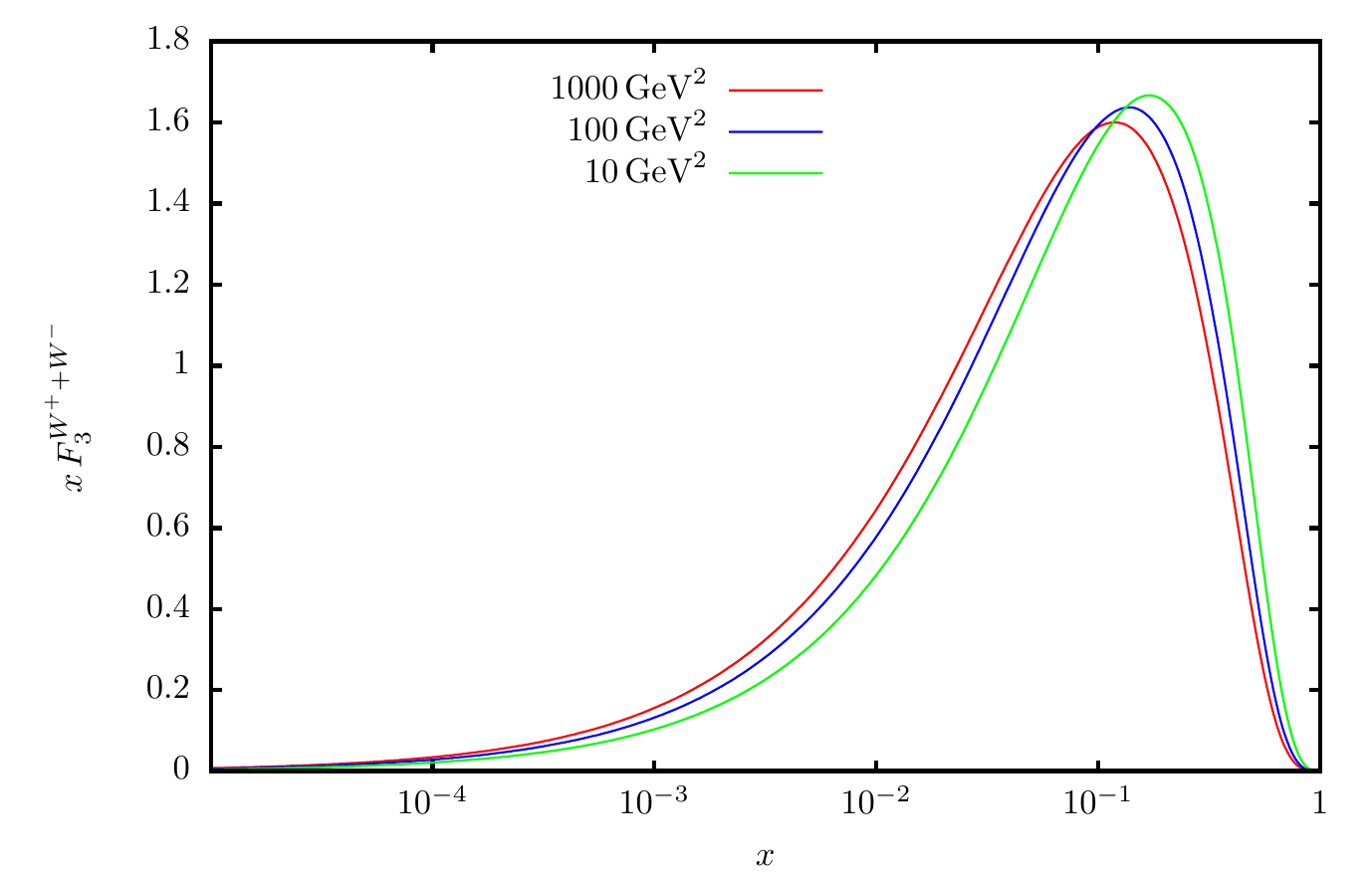}
\caption{\sf \small The charged current structure function $xF_3^{W^+ + W^-}$ up to 
$O(\alpha_s^2)$ including the 
charm quark corrections in the on-shell scheme with $m_c = 1.59~\GeV$ \cite{Alekhin:2012vu} and 
using the NNLO parton distribution functions \cite{Alekhin:2013nda}.}
\label{FIG:F3rec}
\end{figure}

\begin{figure}[H]
\centering
\includegraphics[scale=0.9]{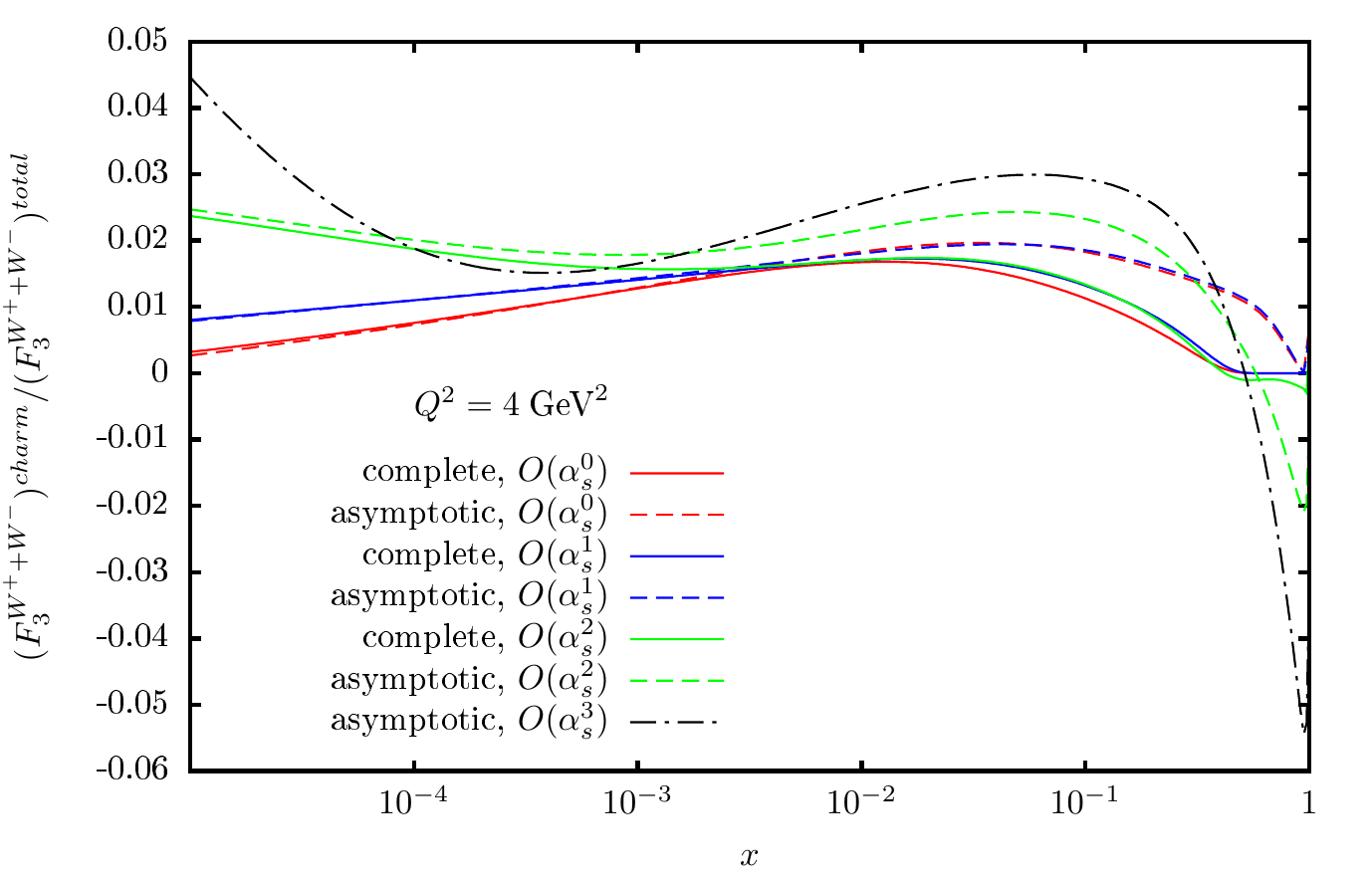}
\caption{\sf small The ratio of the charm quark contributions to the charged current structure 
function $xF_3^{W^+ + W^-}$ 
up to $O(\alpha_s^2)$ to the full corrections at $Q^2 = 4~\GeV^2$. 
Dash-dotted line: corrections to  $O(\alpha_s^3)$ in the asymptotic case.
The other conditions are the same as in Figure~\ref{FIG:F2rec}.}
\label{FIG:F3ratios4}
\end{figure}

\begin{figure}[H]
\centering
\includegraphics[scale=0.9]{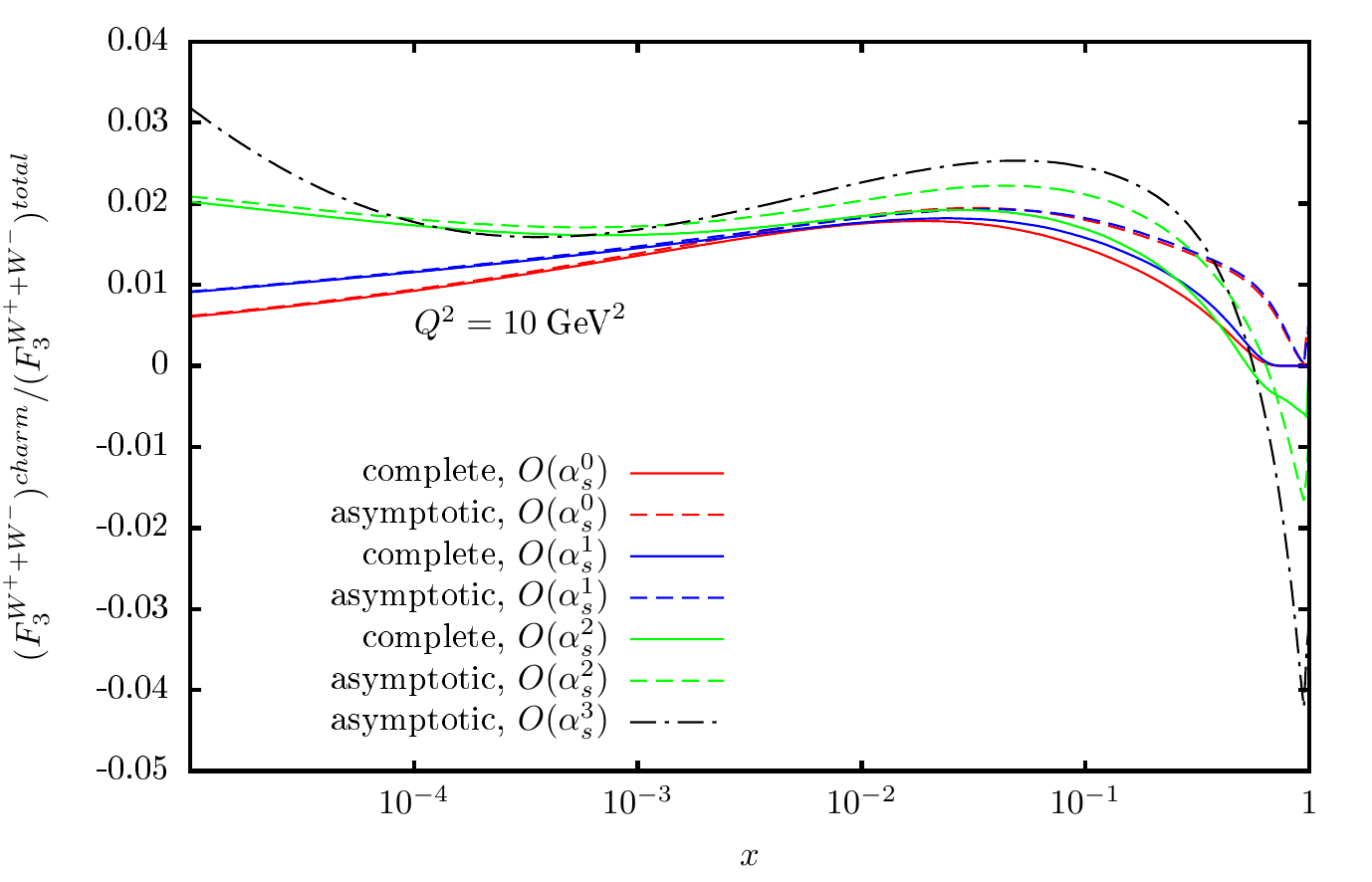}
\caption{\sf \small The ratio of the charm quark contributions to the charged current structure 
function $xF_3^{W^+ + W^-}$ 
up to $O(\alpha_s^2)$ to the full corrections at $Q^2 = 10~\GeV^2$. Dash-dotted line: corrections to  $O(\alpha_s^3)$ in 
the asymptotic case. The other conditions are the same as in 
Figure~\ref{FIG:F2rec}.}
\label{FIG:F3ratios10}
\end{figure}

\begin{figure}[H]
\centering
\includegraphics[scale=0.9]{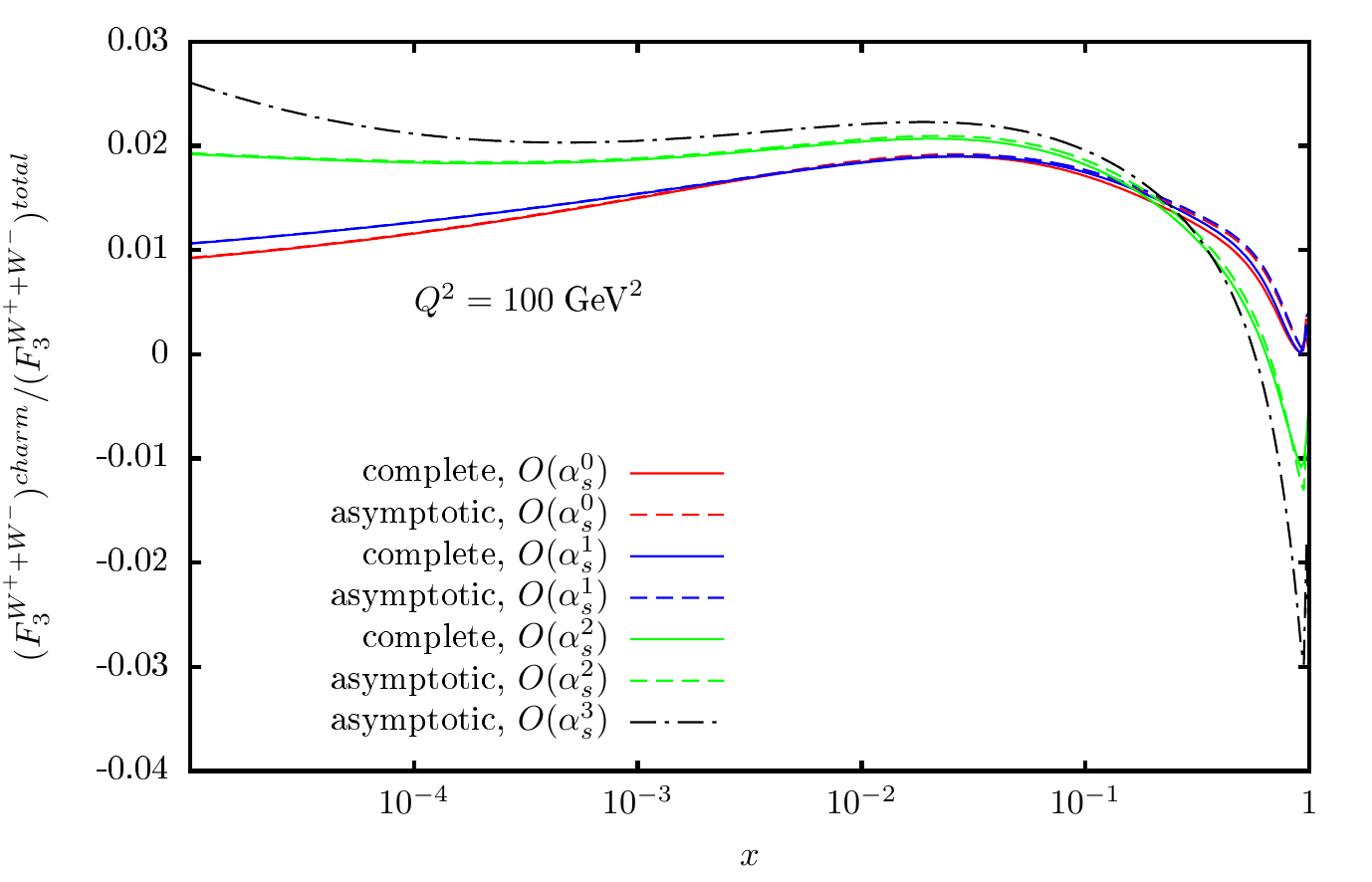}
\caption{\sf \small The ratio of the charm quark contributions to the charged current structure function $xF_3^{W^+ + W^-}$ 
up to $O(\alpha_s^2)$ to the full corrections at $Q^2 = 100~\GeV^2$. Dash-dotted line: corrections to  $O(\alpha_s^3)$ in 
the asymptotic case. The other conditions are the same as in 
Figure~\ref{FIG:F2rec}.}
\label{FIG:F3ratios}
\end{figure}

In Figure~\ref{FIG:F3rec} the charged current structure function $xF_3^{W^+ + W^-}$ including the charm quark corrections
are shown. In this case also the asymptotic 3-loop corrections have been calculated \cite{Behring:2015roa}. As 
shown in Figures~\ref{FIG:F3ratios4}--\ref{FIG:F3ratios}, the charm quark corrections vary form $\sim +2~\%$ 
to $-2~\%$
from small to large $x$. With rising values of $Q^2$ the corrections become more pronounced at large values of $x$.
Note that the asymptotic $O(\alpha_s^3)$ corrections yield significant contributions both at small and large values of $x$.

\section{The Sum Rules}
\label{sec:6}

\vspace*{1mm}
\noindent
In the following we discuss the corrections to the Adler sum rule \cite{Adler:1965ty}, which have to vanish, 
and calculate the corrections to the polarized Bjorken sum rule \cite{Bjorken:1969mm}, 
the unpolarized Bjorken sum rule \cite{Bjorken:1967px}, and the Gross-Llewellyn Smith sum rule \cite{Gross:1969jf}, 
which are obtained as the first moments of the massive Wilson coefficients calculated in the previous sections. The
combination of the parton distributions is partly different, as here differences between structure functions 
in the neutral current case are considered. But this affects only the normalization factor of the sum rules, which are 
known constants.
As has been outlined in Refs.~\cite{Behring:2015zaa,Behring:2015roa} up to 3-loop order, in the asymptotic region $Q^2 
\gg m^2$ the sum rules only modify the massless approximation by replacing the number of massless flavors 
from $N_F \rightarrow N_F+1$. Given the factorization of the massive Wilson coefficients 
\cite{Buza:1995ie,Buza:1996wv}, this holds for all orders in the coupling constant, since the first moment 
of the massive non-singlet OMEs vanish order by order in the coupling constant due to fermion number 
conservation.
The 4-loop corrections to these sum rules have been calculated in 
Refs.~\cite{Baikov:2010je,Chetyrkin:14,Baikov:2012zn}. Earlier Pad\'e estimates were given in \cite{Kataev:1995vh}.

We emphasize that
in the present paper the inclusive Wilson coefficients are calculated for deep-inelastic scattering, but not those in the 
flavor tagged case. The relations obtained do not smoothly transform into the photo-production limit $Q^2 \approx 0$, 
both for the Wilson coefficients and
the parton distribution functions, setting $\mu^2 = Q^2$. They are valid only up to a 
lower scale $Q_0^2$, which usually 
should be at least of $O(m_c^2)$ or larger, also to stay outside the region of higher twist corrections. In the case 
of the sum rules
discussed below, in the limit $Q^2/m^2 \rightarrow 0$ logarithmic contributions survive, while this is the 
not the case
in the limit of large virtualities  $m^2/Q^2 \rightarrow 0$. The photo-production region for the corresponding structure
functions needs a separate treatment.

In the following we will discuss the complete massive corrections to the four sum rules in the deep-inelastic 
region. The power corrections of a single heavy quark {\sf c} or {\sf b} will be shown to basically 
interpolate between $N_F$ and $N_F +1$ massless flavors in the limit $m^2/Q^2 \rightarrow 0$, while at lower 
scales $Q^2$, partly negative virtual corrections are possible. The sum rules are observables and we represent them 
choosing the factorization scale $\mu^2 = Q^2$. The scale matching 
can be performed analytically in Mellin space up to the respective order in $a_s$ in which the quantity is 
calculated, cf.~\cite{Blumlein:2006be}.

To get closer to the unitary representation for the CKM matrix elements, we calculate the functions 
$H_{F_i,q}$, 
(\ref{eq:WIL2}, \ref{eq:H3}), allowing for massive charm quarks to be pair produced also for this Cabbibo suppressed 
term, but referring to massless $s \rightarrow c$ charged current transitions for the real 
and virtual corrections.

Finally, we also consider the target mass corrections to the deep-inelastic sum rules, as they are of relevance in the
region of lower values of $Q^2$.
\subsection{The Adler sum rule}
\label{sec:61}

\vspace*{1mm}
\noindent
The Adler sum rule \cite{Adler:1965ty} states
\begin{eqnarray}
\label{eq:ADLER}
\int_0^1 \frac{dx}{x} \left[F_2^{\overline{\nu}p}(x,Q^2) - F_2^{{\nu}p}(x,Q^2)\right] = 2 [1 + \sin^2(\theta_c)]
\end{eqnarray}
for three massless flavors. Here $\theta_c$ denotes the Cabibbo angle \cite{Cabibbo:1963yz}. The integral 
(\ref{eq:ADLER}) neither receives QCD nor quark- or target mass corrections, cf. also 
\cite{Ravindran:2001dk,Adler:2009dw}. 

The Compton contribution yields
\begin{eqnarray}
\int_0^z dz L_{F_2,q}^{{\rm NS},(2),C}(z) &=& 
a_s^2 C_F T_F \Biggl\{
-\frac{2426}{81} - \frac{476}{9} \tilde{\lambda}^2
+ \left(\frac{400}{27} + \frac{220}{9} \tilde{\lambda}^2\right) \ln(\xi)
\nonumber\\ &&
- \left[\frac{20}{3} \tilde{\lambda}  + \frac{92}{9} \tilde{\lambda}^3 \right] 
\left[
  \Li_2\left(-\frac{1-\tilde{\lambda}}{1+\tilde{\lambda}}\right)
- \Li_2\left(-\frac{1+\tilde{\lambda}}{1-\tilde{\lambda}}\right) \right]
\nonumber\\ &&
- \left[\frac{10}{3} + 4 \tilde{\lambda}^2 -  2 \tilde{\lambda}^4\right] \left[ 
  \Li_3\left(-\frac{1-\tilde{\lambda}}{1+\tilde{\lambda}}\right)
+ \Li_3\left(-\frac{1+\tilde{\lambda}}{1-\tilde{\lambda}}\right) - 2 \zeta_3 \right]\Biggr\},
\nonumber\\
\end{eqnarray}
which is canceled by the virtual correction $\int_0^1 dz L_{F_2,q}^{{\rm NS},(2),V}(z)$, 
(\ref{eq:LqNSg1V}). 
The first 
moment of the contribution with massless final states 
\begin{eqnarray}
L_{F_2,q}^{{\rm NS},(2),\rm massive}\left(\frac{Q^2}{m^2},\frac{m^2}{\mu^2}z\right) = -a_s^2 
\beta_{0,Q} 
\ln\left(\frac{m^2}{\mu^2}\right)
\left[P_{qq}^{(0)}(z) \ln\left(\frac{Q^2}{\mu^2}\right) + c_{F_2,q}^{(1)}(z) \right],
\end{eqnarray}
also vanishes, cf.~(\ref{eq:CF21}, \ref{eq:Pqq0}).

For the charged current flavor excitation slow rescaling at tree level yields
\begin{eqnarray}
\int_0^1 dx \frac{F_2^{\bar{\nu} p}(x,Q^2) - F_2^{{\nu} p}(x,Q^2)}{x} 
&=& \int_0^{\tfrac{1+\xi}{\xi}}
\frac{dz}{z} \left[F_2^{\bar{\nu} p}(z,Q^2) - F_2^{{\nu} p}(z,Q^2) \right]
\nonumber\\ 
&=& \int_0^1
\frac{dz}{z} \left[F_2^{\bar{\nu} p}(z,Q^2) - F_2^{{\nu} p}(z,Q^2) \right],
\end{eqnarray}
since the support of $F_2(z,Q^2)$ is $z \in [0,1]$.
For the first order massive QCD corrections given in \cite{Gluck:1997sj,Blumlein:2011zu} 
the first moment (\ref{eq:ADLER}) vanishes. The corresponding $O(a_s^2)$ corrections have only been studied 
in the asymptotic case 
\cite{Blumlein:2014fqa} and vanish. For massless quarks, the Adler sum rule has been checked at $O(\alpha_s^3)$ in
\cite{Moch:2007gx}. It seems that in the case of massless  4-loop corrections, the validity  of the sum rule 
has not 
yet been 
checked perturbatively \cite{ADLER4}. The target mass corrections are studied in 
Section~\ref{sec:65}.

In contrast, the QCD-, quark mass- and target mass corrections to the first moments of the structure functions 
$g_1, F_1$ and $F_3$ do not vanish. 

\subsection{The polarized Bjorken sum rule}
\label{sec:62}

\vspace*{1mm}
\noindent
The polarized  Bjorken sum rule \cite{Bjorken:1969mm} refers to the first moment of the flavor non-singlet combination
\begin{eqnarray}
\label{eq:PBJ}
\int_0^1 {dx} \left[g_1^{ep}(x,Q^2) - g_1^{en}(x,Q^2) \right] = \frac{1}{6}\left|\frac{g_A}{g_V}\right|
C_{\rm pBJ}(\hat{a}_s),
\end{eqnarray}
with $g_{A,V}$ the neutron decay constants, $g_A/g_V \approx -1.2767 \pm 0.0016$ \cite{Mund:2012fq} and 
\begin{eqnarray}
\hat{a}_s = \frac{\alpha_s}{\pi}~. 
\end{eqnarray}
The 1- \cite{Kodaira:1978sh}, 2-\cite{Gorishnii:1985xm}, 3- \cite{Larin:1991tj} and 4-loop QCD corrections 
\cite{Baikov:2010je} 
in the massless case are given by
\begin{eqnarray}
\label{eq:BJSR}
C_{\rm pBJ}(\hat{a}_s),
&=&
1 -  \hat{a}_s
+ \hat{a}_s^2 (-4.58333 + 0.33333 N_F)
+ \hat{a}_s^3 (-41.4399 + 7.60729 N_F - 0.17747 N_F^2)
\nonumber\\ &&
+ \left. \hat{a}_s^4 (-479.448 + 123.391 N_F - 7.69747 N_F^2 + 0.10374 N_F^3)\right|_{\rm NS}
\nonumber\\ &&
+ \left. \hat{a}_s^4 (12.2222-0.740741 N_F)\right|_{\rm SI} \sum_{k=1}^{N_F} e_k
\end{eqnarray}
choosing the renormalization scale $\mu^2 = Q^2$, cf.~\cite{Zijlstra:1993sh} for $SU(3)_c$. Here $N_F$ denotes the number of
active light flavors and the labels {\rm NS} and {\rm SI} refer to the genuine `non-singlet' and `singlet' contributions, 
respectively.  The expression for general color factors was given in Ref.~\cite{Baikov:2010je,Baikov:2015tea}.\footnote{
An estimate of the singlet contribution has been made in Ref.~\cite{Larin:2013yba}. We refer to the the result of the 
calculation in Ref.~\cite{Baikov:2015tea}.} The massless corrections for $N_F = 3$ and $N_F = 4$ are
\begin{eqnarray}
\label{eq:BJSR3}
C_{\rm pBJ}(N_F=3) &=& 1 -  \hat{a}_s -3.58334 \hat{a}_s^2 - 20.2153 \hat{a}_s^3 - 175.781
\hat{a}_s^4~~~~~\text{and}
\\
\label{eq:BJSR4}
C_{\rm pBJ}(N_F=4) &=& 1 -  \hat{a}_s - 3.25001 \hat{a}_s^2 - 13.8503 \hat{a}_s^3 - 98.2889
 \hat{a}_s^4~.
\end{eqnarray}

For the asymptotic massive corrections (\ref{eq:WIL1}--\ref{eq:WIL2}) only the first moments of the massless Wilson coefficients
$\hat{C}_{g_1,q}^{(2,3),\rm NS}(N_F)$ contribute, since the first moments of the massive 
non-singlet OMEs 
vanish due to fermion 
number conservation, a property holding even at higher order. Therefore, any new heavy quark changes 
Eq.~(\ref{eq:BJSR}) by a shift in $N_F \rightarrow N_F +1$ only, for the asymptotic corrections. 

We turn now to the heavy quark corrections, which are given by
\begin{eqnarray}
C_{\rm pBJ}^{{\sf massive},(2)} &=& 3 C_F T_F \Biggl\{
        \frac{6 \xi ^2+2735 \xi +11724}{5040 \xi } -\frac{\sqrt{\xi +4}}{\xi^{3/2}} 
        \frac{\big(3 \xi ^3+106 \xi ^2+1054 \xi +4812\big)}{5040}
\nonumber\\ && \times
        \ln \left[\frac{\sqrt{1+\frac{4}{\xi }}+1}{\sqrt{1+\frac{4}{\xi }}-1}\right]
        -\frac{1}{\xi^2} \frac{5}{12} \ln^2\left[\frac{\sqrt{1+\frac{4}{\xi }}+1}{\sqrt{1+\frac{4}{\xi }}-1}\right]
        +\frac{\big(3 \xi ^2+112 \xi +1260\big)}{5040} \ln(\xi) \Biggr\},
\nonumber\\
\end{eqnarray}
see Appendix~C. In the asymptotic region $\xi \gg 1$, $C_{\rm pBJ}^{{\sf 
massive},(2)}$ behaves like
\begin{eqnarray}
C_{\rm pBJ}^{{\sf massive},(2)} &\propto& 3 C_F T_F \Biggl\{
\frac{1}{2}-\frac{5}{12 \xi^2} \ln^2(\xi) - \frac{4}{3 \xi} \ln(\xi) + \frac{17}{9 \xi} 
+ O\left(\frac{\ln(\xi)}{\xi^2}\right) \Biggr\}~.
\end{eqnarray}
\begin{figure}[H]
\centering
\includegraphics[scale=0.7]{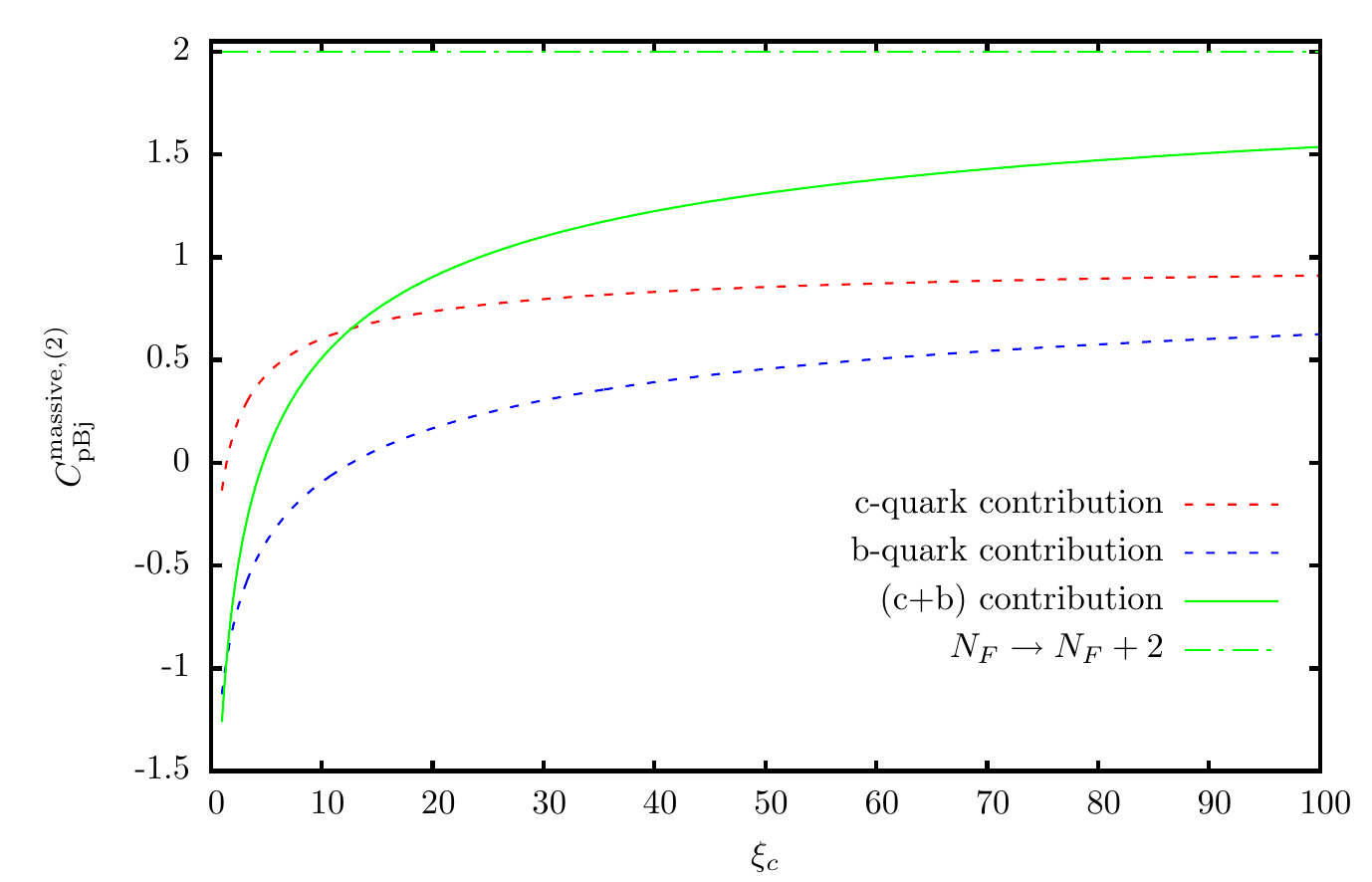}
\caption{\sf \small The $O(a_s^2)$ coefficients $C_{\rm pBJ}^{{\sf massive},(2)}$ for charm and 
bottom quarks as a function of $\xi_c$.}
\label{FIG:pBJ1}
\end{figure}

\noindent
Up to 2-loop order the massless and the massive corrections to the polarized Bjorken sum rule are given by
\begin{eqnarray}
C_{\rm pBJ}(\xi_c) = 1 - \hat{a}_s - \hat{a}_s^2 \left\{-\frac{55}{12} + \frac{1}{3} \left[N_F 
+ C_{\rm pBJ}^{{\sf massive},(2)}(\xi_c)
+ C_{\rm pBJ}^{{\sf massive},(2)}\left(\xi_c \frac{m_c^2}{m_b^2}\right)\right]
\right\} + O(\hat{a}_s^3),
\nonumber\\
\end{eqnarray}
accounting for the charm and bottom quark contributions, with $\xi_c = Q^2/m_c^2$. In Figure~\ref{FIG:pBJ1}
the effect of the heavy flavor Wilson coefficients $C_{\rm pBJ}^{{\sf massive},(2)}$ for charm and bottom
are illustrated as a function of $\xi_c$.
\begin{table}[H]
\centering
\begin{tabular}{|lr|r|}
\hline
$\xi$ =&   1 & -4.003 \\
$N_F =3$&    & -3.583 \\
$\xi$ =&   5 & -3.569 \\
$\xi$ =&  10 & -3.413 \\
$\xi$ =&  24 & -3.251 \\
$N_F =4$&    & -3.250 \\
$\xi$ =&  50 & -3.146 \\
$\xi$ =& 100 & -3.071 \\
$\xi$ =& 500 & -2.970 \\
$N_F =5$&    & -2.917 \\
\hline
\end{tabular}
\caption[]{\label{Tab1}
\sf \small The massless and massive 2-loop corrections to the Bjorken sum rule as a function of 
$\xi_c$.
We also indicated the respective purely massless results for $N_F = 3,4$ and 5.} 
\end{table}
At low scales the corrections are negative and the interpolation to the asymptotic value 2 for $N_F \rightarrow N_F +2$
in $\xi_c$ proceeds very slowly. In Table~\ref{Tab1} we illustrate the mass effects for the 2-loop terms. The massless
prediction is only reached for considerably large values of $Q^2$, namely for $\xi_c \sim 24$ in the case of $N_F=4$ 
and
$\xi_c \gg 500$ for $N_F = 5$.
\begin{table}[H]
\centering
\begin{tabular}{|r|r|r|r|r|}
\hline
$Q^2/\GeV^2$ & $O(\hat{a}_s^4)$ massless  & $O(\hat{a}_s^3)$ massless  & $\Delta_{4-3}$ & 
massive $O(\hat{a}_s^2)$\\ 
\hline
30           & 0.9180 & 0.9205 & -0.0025  & -0.0008 \\
100          & 0.9321 & 0.9335 & -0.0014  & -0.0011 \\
10000        & 0.9587 & 0.9590 & -0.0003  & -0.0008 \\ 
\hline
\end{tabular}
\caption[]{\label{Tab2}
\sf \small Comparison of $C_{\rm pBJ}$ in the massless approximation to $O(\hat{a}_s^4)$, 
$O(\hat{a}_s^3)$ for $N_F=3$ massless flavors, 
and the $O(\hat{a}_s^2)$ contributions due to charm and bottom.}
\end{table}
In Table~\ref{Tab2} we compare the values of the polarized Bjorken sum rule for different values of 
$Q^2$ to illustrate the effect of the heavy flavor contribution. The massive contribution turns out 
to be comparable in size to the massless 4-loop contribution. Due to surviving logarithms 
in $\xi$ in the large $\xi$ region in the tagged flavor case, different results are obtained 
\cite{Blumlein:1998sh,vanNeerven:1999ec}. However, the corresponding quantity does not describe the 
heavy flavor contributions to the structure functions, which are inclusive quantities.
\subsection{The unpolarized Bjorken sum rule}
\label{sec:64}

\begin{figure}[H]
\centering
\includegraphics[scale=0.9]{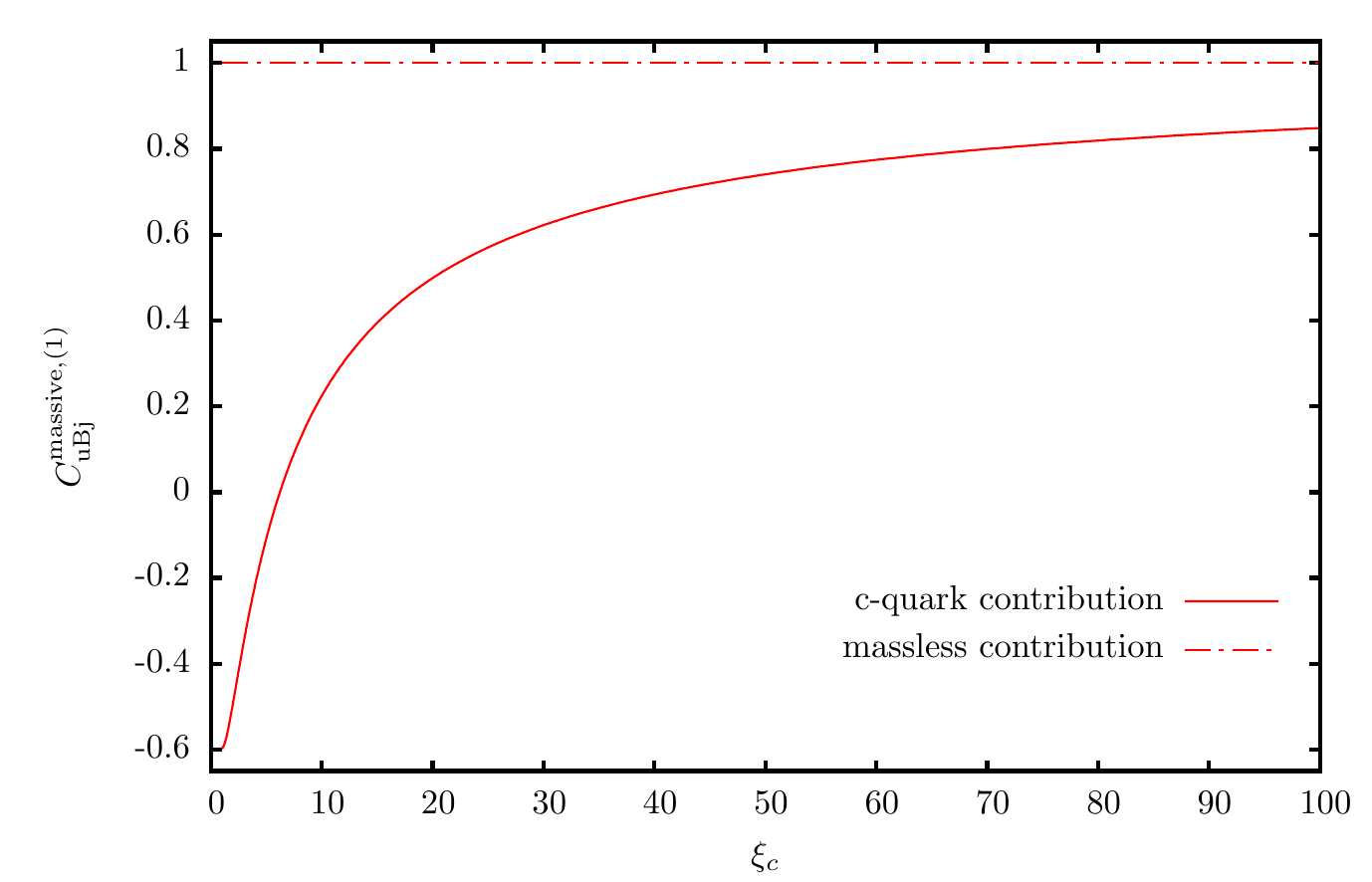}
\caption{\sf \small The $O(a_s)$ coefficient $C_{\rm uBJ}^{{\sf massive},(1)}$ as a function of 
$\xi_c$, normalized to 1 in the limit $\xi_c \rightarrow \infty$.}
\label{FIG:UBJ1}
\end{figure}

\begin{figure}[H]
\centering
\includegraphics[scale=0.9]{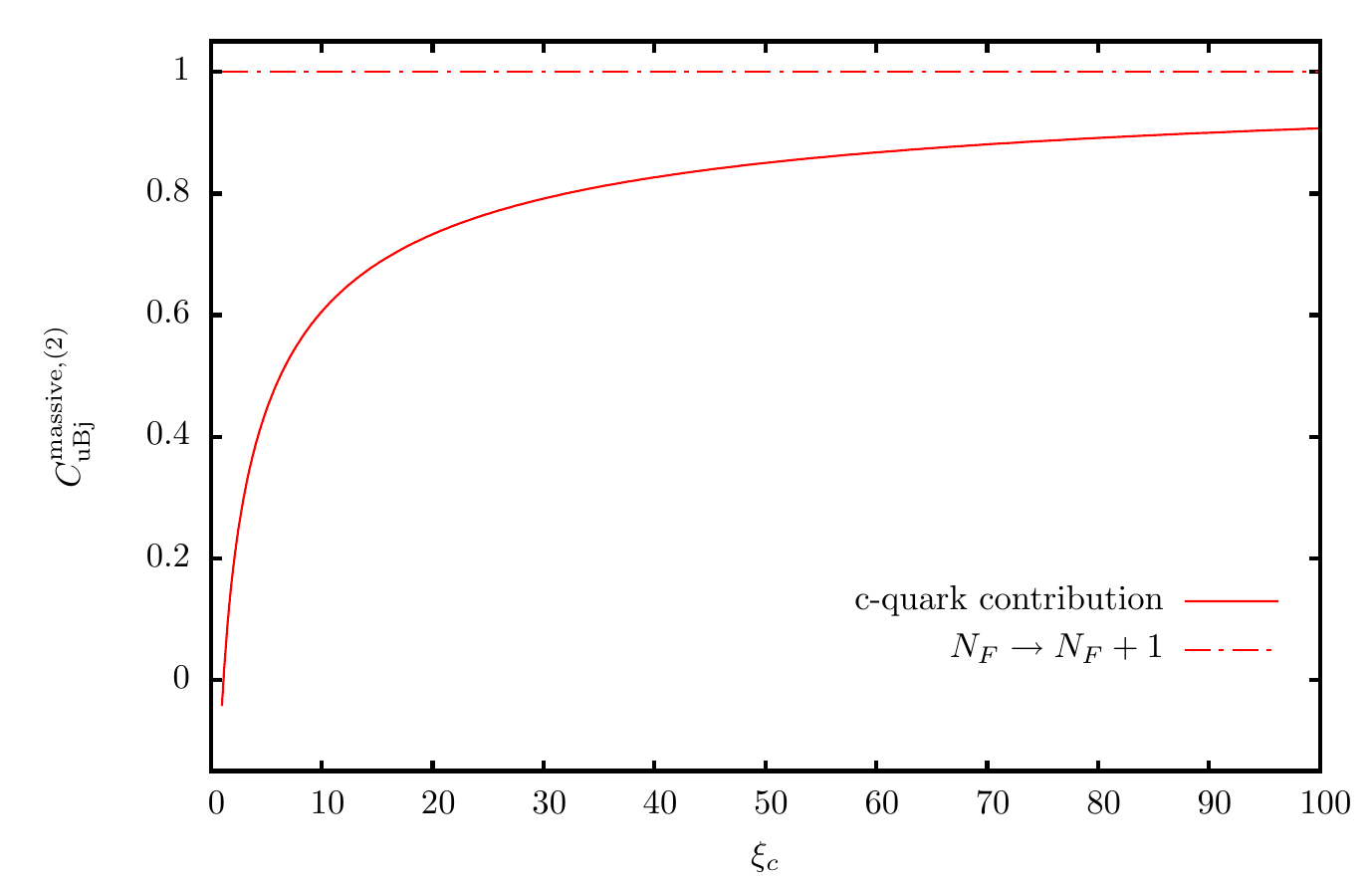}
\caption{\sf \small The $O(a_s^2)$ coefficient $C_{\rm uBJ}^{{\sf massive},(2)}$ as a function of 
$\xi$, normalized to 1 in the limit $\xi_c \rightarrow \infty$.}
\label{FIG:CuBJ2}
\end{figure}

\vspace*{1mm}
\noindent
The unpolarized Bjorken sum rule \cite{Bjorken:1967px} is given by
\begin{eqnarray}
\label{eq:UPBJ}
\int_0^1 {dx} \left[F_1^{\bar{\nu} p}(x,Q^2) - F_1^{{\nu} p}(x,Q^2)\right] 
= C_{\rm uBJ}(\hat{a}_s). 
\end{eqnarray}
The massless 1- 
\cite{Bardeen:1978yd,Altarelli:1978id,Humpert:1980uv,Furmanski:1981cw}, 2-loop \cite{Gorishnii:1985xm}, 3-loop 
\cite{Larin:1990zw} and 4-loop 
\cite{Chetyrkin:14} QCD corrections have been calculated 
\begin{eqnarray}
\label{eq:uBJSR}
C_{\rm uBJ}(\hat{a}_s),
&=& 
1 -  0.66667 \hat{a}_s 
+ \hat{a}_s^2 (-3.83333 + 0.29630 N_F)
\nonumber\\ &&
+ \hat{a}_s^3 (-36.1549 + 6.33125 N_F - 0.15947 N_F^2) 
\nonumber\\ &&
+ \hat{a}_s^4 (-436.768 + 111.873 N_F - 7.11450 N_F^2 + 0.10174 N_F^3)~,
\end{eqnarray} 
setting $\mu^2 = Q^2$ for $SU(3)_c$.
For $N_F = 3, 4$ the massless QCD corrections are given by
\begin{eqnarray}
\label{eq:uBJSRM0}
C_{\rm uBJ}(\hat{a}_s, N_F=3)
&=& 
1 -  0.66667 \hat{a}_s  - 2.94444 \hat{a}_s^2 - 18.5963 \hat{a}_s^3 - 162.436 \hat{a}_s^4 
\\
C_{\rm uBJ}(\hat{a}_s, N_F=4) &=&
1 -  0.66667 \hat{a}_s  - 2.64815 \hat{a}_s^2 - 13.3813 \hat{a}_s^3 - 96.6032 \hat{a}_s^4~. 
\end{eqnarray} 
The massive corrections start at $O(\alpha_s^0)$ with the $s' \rightarrow c$ transitions 
\cite{Gluck:1997sj,Blumlein:2011zu}
\begin{eqnarray}
C_{\rm uBJ}^{{\sf massive},(0)}(\xi) &=& \frac{\xi}{1+\xi} 
\\ 
C_{\rm uBJ}^{{\sf massive},(1)}(\xi) &=& 
C_F \frac{1}{4} \Biggl\{\frac{2 + \xi - 2 \xi^2}{\xi(1+\xi)}
- 2 \frac{1 + \xi - 3 \xi^2}{\xi^2 (1+ \xi)} \ln(1+\xi) \Biggr\}.
\end{eqnarray} 
$C_{\rm uBJ}^{{\sf massive},(1)}(\xi)$ approaches the asymptotic value of $-2/3$ given in 
(\ref{eq:uBJSR}). Its behaviour as a function of $\xi_c$ is shown in 
Figure~\ref{FIG:UBJ1}.
The massive 2-loop corrections are given by
\begin{eqnarray}
\label{eq:uBJSRM1}
C_{\rm uBJ}^{{\sf massive},(2)}(\xi) &=& \hat{a}_s^2 C_F T_F \frac{1}{16}\Biggl\{
-\frac{8}{\xi^2} 
\ln^2\left(\frac{\sqrt{1+\frac{4}{\xi}}+1}{\sqrt{1+\frac{4}{\xi}}-1}\right)
+\ln\left(\frac{\sqrt{1+\frac{4}{\xi}}+1}{\sqrt{1+\frac{4}{\xi}}-1}\right) 
\nonumber\\ && \times
\left[
-\frac{344}{21 \xi}
-\frac{268}{105}
-\frac{4 \xi}{105}
+\frac{2 \xi^2}{105} \right] 
\sqrt{1+\frac{4}{\xi}} 
\nonumber\\ && 
+\left(\frac{8}{3} - \frac{2 \xi^2}{105}\right) \ln(\xi) 
+ \frac{856}{21 \xi} 
+\frac{2258}{315} 
- \frac{4\xi}{105}
\Biggr\}~. 
\end{eqnarray} 
In the region $\xi \gg 1$ one obtains
\begin{eqnarray}
\label{eq:uBJSRM1a}
C_{\rm uBJ}^{{\sf massive},(2)}(\xi) &\approx&  \hat{a}_s^2 C_F T_F \Biggl\{
\frac{4}{9} + \frac{1}{\xi} \left[ \frac{20}{9} - \frac{4}{3} \ln(\xi)\right]\Biggr\} + 
O\left(\frac{\ln^2(\xi)}{\xi^2}\right)~.
\end{eqnarray} 
In Figure~\ref{FIG:CuBJ2},  $C_{\rm uBJ}^{{\sf massive},(2)}(\xi)$ is shown as a function of $\xi$.

\noindent
To $O(\hat{a}_s^2)$ the unpolarized Bjorken sum rule reads
\begin{eqnarray} 
C_{\rm uBj}(\xi) &=& 
\left[1 - |V_{cd}|^2\left[C_{\rm uBj}^{{\sf charm},(0)}(\xi) -1 \right]\right] - \hat{a}_s 
\left\{\frac{2}{3} + |V_{cd}|^2 \left[
C_{\rm uBj}^{{\sf charm},(1)}(\xi) + \frac{2}{3} \right]\right\}
\nonumber\\ &&
+ \hat{a}^2_s \left[-\frac{23}{6} + \frac{8}{27} N_F + C_{\rm uBJ}^{{\sf charm},(2)}(\xi)\right] + 
O(\hat{a}_s^3)~. 
\end{eqnarray} 

In Table~\ref{Tab3} we compare the values of the unpolarized Bjorken sum rule for different values 
of $Q^2$
to illustrate the effect of the heavy flavor contribution.
\begin{table}[H]
\centering
\begin{tabular}{|r|r|r|r|r|}
\hline
$Q^2/\GeV^2$ & $O(\hat{a}_s^4)$ massless  & $O(\hat{a}_s^3)$ massless  & $\Delta_{4-3}$ & massive 
\\ 
\hline
30           & 0.9414 & 0.9437 & -0.0023  & 0.0032 \\
100          & 0.9520 & 0.9533 & -0.0013  & 0.0014 \\
10000        & 0.9714 & 0.9717 & -0.0003  & 0.0004 \\ 
\hline
\end{tabular}
\caption[]{\label{Tab3}
\sf \small Comparison of $C_{\rm uBJ}$ in the massless approximation to $O(\hat{a}_s^4)$, 
$O(\hat{a}_s^2)$ for $N_F=3$ massless flavors, 
and the $O(\hat{a}_s^2)$ contributions due to charm.}
\end{table}
\noindent
The charm corrections at $O(\hat{a}_s^2)$ are of the same size as the massless $O(\hat{a}_s^4)$ corrections.
\subsection{The Gross-Llewellyn Smith sum rule}
\label{sec:63}

\vspace*{1mm}
\noindent
The Gross-Llewellyn Smith sum rule \cite{Gross:1969jf} refers to the first moment of the flavor non-singlet combination
\begin{eqnarray}
\label{eq:GLS0}
\int_0^1 {dx} \left[F_3^{\bar{\nu} p}(x,Q^2) + F_3^{{\nu} p}(x,Q^2)\right] 
= 6 C_{\rm GLS}(\hat{a}_s),
\end{eqnarray}
assuming idealized CKM mixing. The 1-loop 
\cite{Bardeen:1978yd,Altarelli:1978id,Humpert:1980uv,Furmanski:1981cw}, 2-loop 
\cite{Gorishnii:1985xm}, 3-loop \cite{Larin:1991tj} and 4-loop QCD corrections 
\cite{Baikov:2012zn,Baikov:2010je} in the 
massless case are given by
\begin{eqnarray}
\label{eq:GLSSR}
C_{\rm GLS}(\hat{a}_s)
&=& 
1 -  \hat{a}_s 
+ \hat{a}_s^2 (-4.58333 + 0.33333 N_F)
+ \hat{a}_s^3 (-41.4399 + 8.02047 N_F - 0.17747 N_F^2) 
\nonumber\\ &&
+ \hat{a}_s^4 (-479.448 + 129.193 N_F - 7.93065 N_F^2 + 0.10374 N_F^3)~,
\end{eqnarray} 
choosing the renormalization scale $\mu^2 = Q^2$ for $SU(3)_c$. The expression for general color 
factors was given in Ref.~\cite{Baikov:2012zn,
Baikov:2010je}. Note that the QCD corrections to the 
Gross-Llewellyn Smith sum rule and to the polarized Bjorken sum rule \cite{Bjorken:1969mm} 
are identical up to $O(\hat{a}_s^2)$.

The excitation of charm basically interpolates between
\begin{eqnarray}
\label{eq:GLS3}
C_{\rm GLS}(\hat{a}_s, N_F=3)
&=& 
1 -  \hat{a}_s - 3.58334 \hat{a}_s^2 - 18.9757 \hat{a}_s^3 - 160.444 \hat{a}_s^4~~~~~\text{and}
\nonumber\\
\label{eq:GLS4}
C_{\rm GLS}(\hat{a}_s, N_F=4)
&=& 
1 -  \hat{a}_s - 3.25001 \hat{a}_s^2 - 12.1975 \hat{a}_s^3 - 82.9270 \hat{a}_s^4~.
\end{eqnarray} 
The charm corrections at lowest and first order are given by
\begin{eqnarray}
C_{\rm GLS}^{{\sf charm},(0)}(\xi) &=& \frac{\xi}{1 + \xi}\\
C_{\rm GLS}^{{\sf charm},(1)}(\xi)  &=& C_F \Biggl\{-\frac{3 \xi}{4(1+\xi)} + \frac{3}{2} \frac{\ln(1+\xi)}{1+\xi} 
\Biggr\},
\end{eqnarray} 
while at $O(a_s^2)$ the contributions to $L_{F_3,q}$ are given by
\begin{eqnarray}
C_{\rm GLS}^{{\sf charm},(2)}(\xi) &=& \frac{1}{3} C_{\rm pBJ}^{{\sf charm},(2)}(\xi)~. 
\end{eqnarray} 
The heavy flavor corrections up to $O(\hat{a}_s^2)$ are given by
\begin{eqnarray}
C_{\rm GLS}(\xi) &=& 1 + |V_{cd}|^2 \left[C_{\rm GLS}^{{\sf charm},(0)}(\xi) - 1\right] 
+ \hat{a}_s \left\{-1+ 
\frac{|V_{cd}|^2}{6}\left[C_{\rm 
GLS}^{{\sf charm},(1)}(\xi) + 1\right]\right\}
\nonumber\\ && + \hat{a}_s^2\left\{-\frac{55}{12} 
+ \frac{1}{3} N_F + C_{\rm GLS}^{{\sf charm},(2)}(\xi)\right\} + O(\hat{a}_s^3).
\end{eqnarray} 
\begin{figure}[H]
\centering
\includegraphics[scale=0.9]{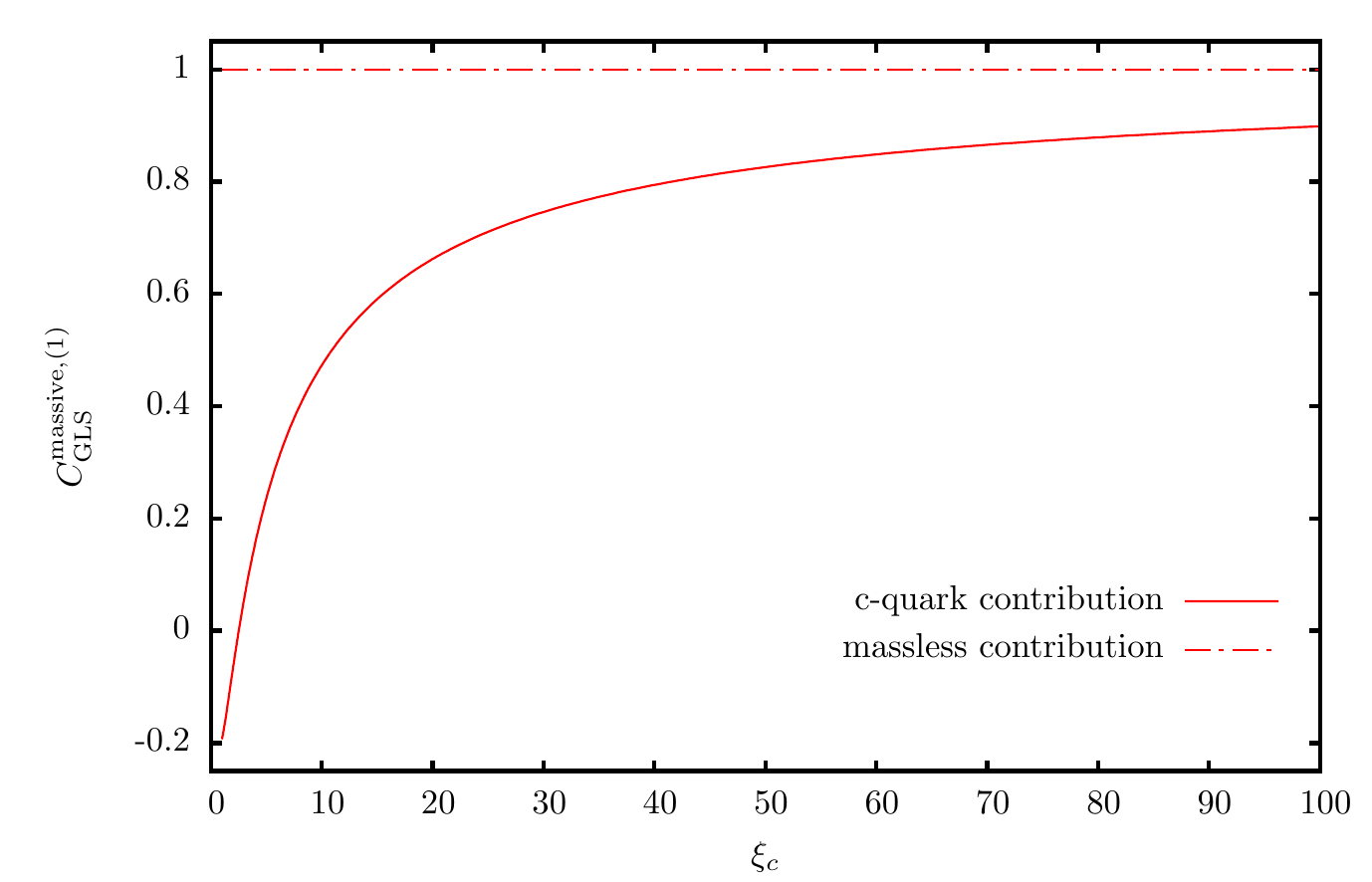}
\caption{\sf \small The $O(a_s)$ coefficient $C_{\rm GLS}^{{\sf massive},(1)}$ as a function of 
$\xi$, normalized to 1 in the limit $\xi \rightarrow  \infty$.}
\label{FIG:GLS1}
\end{figure}

\noindent
In Table~\ref{Tab2b} we compare the values of the Gross-Llewellyn Smith sum rule for different 
values of $Q^2$
to illustrate the effect of the heavy flavor contribution.
\begin{table}[H]
\centering
\begin{tabular}{|r|r|r|r|r|}
\hline
$Q^2/\GeV^2$ & $O(\hat{a}_s^4)$ massless  & $O(\hat{a}_s^3)$ massless  & $\Delta_{4-3}$ & massive 
\\ 
\hline
30           & 0.9185 & 0.9207 & -0.0022  &  0.0024
\\
100          & 0.9324 & 0.9337 & -0.0013  &  0.0013\\
10000        & 0.9588 & 0.9590 & -0.0002  &  0.0004\\ 
\hline
\end{tabular}
\caption[]{\label{Tab2b}
\sf \small Comparison of $C_{\rm pBJ}$ in the massless approximation to $O(\hat{a}_s^4)$, 
$O(\hat{a}_s^2)$ for $N_F=3$ massless flavors, 
and the $O(\hat{a}_s^2)$ contributions due to charm.}
\end{table}
\noindent
As in the case of the other sum rules, the charm corrections at $O(\hat{a}_s^2)$ 
turn out to be of the same size as the massless $O(\hat{a}_s^4)$ corrections.
\subsection{The Target Mass Corrections to The Sum Rules}
\label{sec:65}

\vspace*{1mm}
\noindent
For the target mass corrections it has been shown \cite{Georgi:1976ve} that the correction factor
to the massless structure function $F_2(N,Q^2)$ in Mellin space is given by
\begin{eqnarray}
F_2^{\sf TM}(N,Q^2) &=& \sum_{j=0}^\infty \left(\frac{M^2}{Q^2}\right)^j \binom{N+j}{j} \frac{N (N-1)}{(N+2j)(N+2j-1)}
\frac{C_2^{N+2j} a_{N+2j}^{(2)}}
     {C_2^{1+2j} a_{1+2j}^{(2)}}
\\
a_k^{(2)}      &=& \int_0^1 dx x^{k-1} \left[u_v(x,Q^2) - d_v(x,Q^2)\right], \text{with}~~~~a_1^{(2)} = 1 \\
C_{2}^k        &=& \int_0^1 dx x^{k-1} C_{2}(x,Q^2),~~~~~C_2^{1} = 1~,
\end{eqnarray}
with $M$ the nucleon mass, $(\Delta)a_{N+2j}^{(2)}$ the (non-perturbative) moments of the massless 
PDFs and $C_2$ the moments of the Wilson coefficient 
contributing to $F_2$. Here we consider the flavor-non singlet contribution  $(F_2^{\bar{\nu} p} - F_2^{{\nu} p})/x$ which 
is relevant for the Adler-sum rule. 
Note that the first moment of $C_2$, except for the tree-level contribution, vanishes, as has been 
proven to 3-loop order for the massless and massive 
Wilson coefficients (in the asymptotic region) by explicit calculations 
\cite{Furmanski:1981cw,vanNeerven:1991nn,Moch:2007gx,Ablinger:2014vwa} and above for the massive 
contributions to the complete corrections at 2-loop order. One obtains

\begin{eqnarray}
\lim_{N \rightarrow 1} F_2^{\sf TM}(N,Q^2) = 0.
\end{eqnarray}
In contrast, the first moments of the structure functions $F_1$ and $F_3$ do not vanish 
at higher orders in QCD both in the massless and massive 
cases \cite{Blumlein:2012bf,Steffens:2012jx}.
Moreover, both in the unpolarized \cite{Blumlein:1998nv,Ravindran:2001dk,Steffens:2012jx} 
and in the polarized cases, the target
mass corrections are different for different structure functions, which are usually associated to other ones 
by current conservation, as
in the case of $F_4(x,Q^2)$ and $F_5(x,Q^2)$.

In the case of the unpolarized Bjorken sum rule, the target mass correction factor is given by 
\cite{Blumlein:2012bf,Steffens:2012jx}
\begin{eqnarray}
F_1^{\sf TM}(N=1,Q^2) &=& \sum_{j=0}^\infty \left(\frac{M^2}{Q^2}\right)^j 
(1+j) \frac{\left[C_1^{1+2j} +\frac{1}{1+2j} C_2^{1+2j}\right] a_{1+2j}}{C_1^{1} a_1}~. 
\end{eqnarray}
Here $C_1^{(k)}$ denotes the $k$th moment of the Wilson coefficient contributing to the structure function
$F_1^{\bar{\nu} p} - F_1^{{\nu} p}$.

The target mass corrections to the polarized Bjorken sum rule are given by
\cite{Piccione:1997zh,Blumlein:1998nv}
\begin{eqnarray}
g_1^{\sf TM}(N=1,Q^2) &=& \sum_{j=0}^\infty \left(\frac{M^2}{Q^2}\right)^j 
\frac{(1+j)}{(1+2j)^2} \frac{\Delta C_1^{1+2j} \Delta a_{1+2j}}{\Delta C_1^{1} \Delta a_1} \\
\Delta{a}_{k} &=& \frac{1}{6} \int_0^1 dx~x^{k-1}~\left[ \Delta u_v(x,Q^2) - \Delta d_v(x,Q^2) 
+ 2 \left( \Delta \overline{u}(x,Q^2) -  \Delta \overline{d}(x,Q^2) \right) \right] 
\nonumber
\\
\\
\Delta C_{1}^k      &=& \int_0^1 dx x^{k-1} \Delta C_{1}(x,Q^2),
\end{eqnarray}
where $\Delta C_{1}$ is the polarized flavor-non singlet Wilson coefficient corresponding to the structure 
function 
$g_1^{ep} - g_1^{en}$.

For the target mass corrections to the Gross-Llewellyn Smith sum rule one obtains 
\cite{Blumlein:2012bf,Steffens:2012jx}
\begin{eqnarray}
F_3^{\sf TM}(N=1,Q^2) &=& \sum_{j=0}^\infty \left(\frac{M^2}{Q^2}\right)^j 
\frac{1+j}{1+2j} 
\frac{C_3^{1+2j} a_{1+2j}^{(3)}}{C_3^{1} a_{1}^{(3)}},\\
a_k^{(3)}      &=& 2 \int_0^1 dx x^{k-1} \left[u_v(x,Q^2) + d_v(x,Q^2)\right], \text{with}~~~~a_1^{(3)} = 6, 
\end{eqnarray}
and $C_3^{k}$ are the moments of the Wilson coefficient contributing
to the flavor non-singlet combination  $F_3^{\bar{\nu} p} + F_3^{{\nu} p}$.

\begin{figure}[H]
\centering
\includegraphics[scale=0.9]{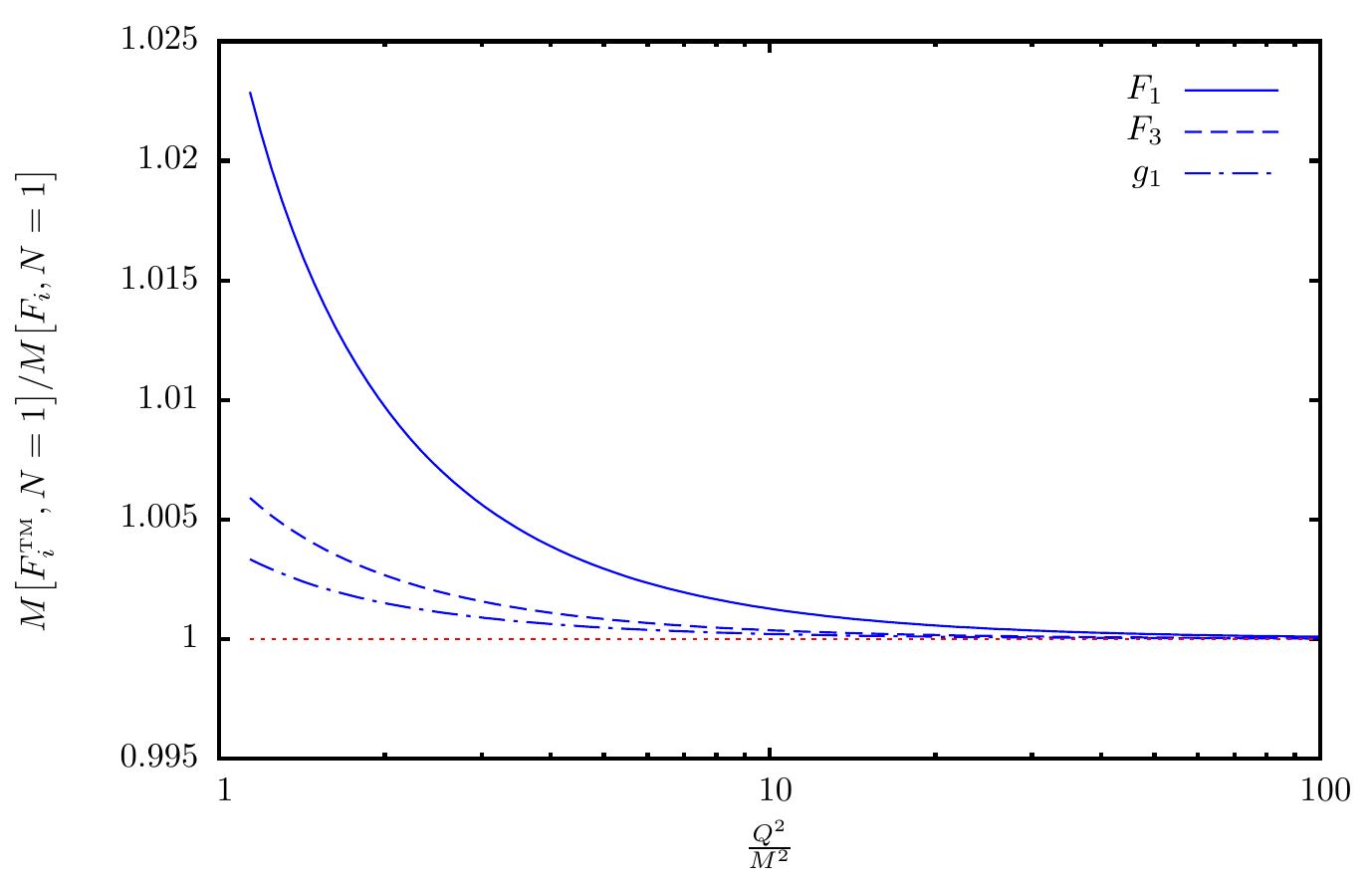}
\caption{\sf \small The target mass corrections, normalized to the massless case, to the 
unpolarized ($F_1$) and polarized Bjorken sum rule 
($g_1$) and the Gross-Llewellyn Smith sum tule ($F_3$).}
\label{FIG:TM}
\end{figure}

In Figure~\ref{FIG:TM} we illustrate the effect of the target mass corrections to the unpolarized and polarized Bjorken sum rule
as well as the Gross-Llewellyn Smith sum rule, as a function of $Q^2/M^2$, accounting only for the operator 
matrix elements
$(\Delta) a_k^{(1,3)}$. We refer to the unpolarized PDFs \cite{Alekhin:2013nda} at NNLO and polarized PDFs \cite{Blumlein:2010rn} at NLO, 
and $\alpha_s$ at NNLO, to allow for a comparison of the different contributions up to NNLO.
The corrections diminish towards large virtualities $Q^2$. At $Q^2 \sim 1~\GeV^2$ they amount $+2.3 \%, 
0.60~\%$ and
$0.33 \%$ for the unpolarized Bjorken sum rule, the Gross-Llewellyn Smith sum rule and the polarized Bjorken 
sum rule, respectively.

\section{Conclusions}
\label{sec:7}

\vspace*{1mm}
\noindent
We have calculated the complete heavy flavor corrections to the flavor non-singlet deep-inelastic 
structure functions $F_{1,2}$ and $g_{1,2}$ in the neutral current case, and to $F_{1,2}^{W^+-W^-}$  
and $F_{3}^{W^++W^-}$ for charged current reactions. Here we considered the deep-inelastic region, 
which at least requests scales $Q^2 \gsim m_c^2$ or larger and $W^2 > 4~\GeV^2$. For the charged 
current 
non-singlet combinations of structure functions also the Cabibbo suppressed Wilson coefficient
$H_{i,q}$ contributes, which we have considered in the asymptotic region starting at $O(a_s^2)$ as 
an approximation. Since the deep-inelastic structure functions are inclusive observables, the 
formerly considered tagged flavor case \cite{Buza:1995ie,Buza:1996xr} is not sufficient. We have 
accomplished the calculation for the inclusive case, to which at $O(a_s^2)$ also virtual 
corrections and real corrections with massless final states, containing massive virtual corrections, 
contribute. We present detailed numerical results for the different unpolarized and polarized 
structure functions for the charm and bottom contribution in the neutral current case and the 
charm contributions in the charged current case, which are the most important. We compared in all 
cases to the formerly calculated asymptotic corrections in the region $Q^2 \gg m^2$, showing that
except for the structure function $F_2^{\rm NS}(x,Q^2)$ this approximation holds only at higher 
scales, while for  $F_2^{nc,\rm NS}(x,Q^2)$ a very good agreement for $Q^2 \simeq 25 \GeV^2$ is 
obtained in the case of charm. In those cases in which the asymptotic 3-loops corrections are 
available, we have partly compared to these corrections as well. The $O(a_s^2)$ non-singlet heavy 
flavor effects are of the order of several per cent of the whole non-singlet structure function
and are of relevance in precision measurements reaching this accuracy. The corrections will become 
even more important in the case of planned high-luminosity measurements at facilities like the EIC 
\cite{EIC}, future neutrino factories \cite{Kaplan:2014xda} or the LHeC 
\cite{AbelleiraFernandez:2012cc}.

We also investigated the heavy flavor corrections for deep-inelastic scattering sum rules, such as 
the Adler, polarized Bjorken, unpolarized Bjorken and Gross-Llewellyn Smith sum rule. While the 
corrections  vanish in case of the Adler sum rule, finite corrections are obtained to the other
three sum rules. They turn out to be of the same size as the massless $O(a_s^4)$ corrections 
which have been calculated recently and complete the picture from the side of the heavy 
quarks. Here it is important to refer to the inclusive rather than to the tagged heavy flavor 
case, since in the latter, logarithmic terms in the region of larger $Q^2$ would remain, after
having already performed the renormalization completely (e.g. in the $\overline{\rm 
MS}$ scheme) \cite{Bierenbaum:2009mv}. In the inclusive case, on the other hand, the transition from 
$N_F \rightarrow N_F+1$ proceeds smoothly. We also quantified the effect of target mass corrections
to the deep inelastic sum rules. In general it turns out that for the sum rules the transition 
$N_F \rightarrow N_F+1$ proceeds slowly in $\xi =  Q^2/m^2$. Therefore  assuming scale matching 
at $Q^2 = m^2$ is, at least here, not appropriate.

\newpage
\appendix
\section{\boldmath Calculation of the Compton Contribution to $g_1^{{\rm NS},(2)}$}
\label{sec:A}

\vspace*{1mm}
\noindent
In this appendix we calculate the contribution of the subprocess
\begin{equation}\label{subprocess}
 q(p_1)+\gamma^*(q)\longrightarrow q(p_2)+Q(p_c)+\overline{Q}(p_{\overline{c}})
\end{equation}
to the non-singlet coefficient function $L_{g_1,q}^{{\rm NS},(2)}$, which is given by Compton 
scattering diagrams shown in Figure~\ref{QCDCOMP}. The structure function $g_1(z)$ is extracted 
from the antisymmetric part of the hadronic tensor $W^{\mu\nu}$, given by
\begin{equation}
 \widehat{W}_A^{\mu\nu}(p,q,s)=-\frac{m_0}{2p\cdot q}
\epsilon^{\mu\nu\alpha\beta}q_\alpha\bigg[\widehat{g}_{1,0}(z,Q^2)\,s_\beta+\widehat{g}_{2,0}(z,Q^2)\,
\bigg(s_\beta-\frac{q\cdot s}{q\cdot p}p_\beta\bigg)\bigg],
\end{equation}
where $z=\frac{Q^2}{2p.q}$ and $m_0$, $p$ and $s$ are the mass, momentum and spin of 
the incoming light quark, with
\begin{equation}
 p.s = 0,\qquad s. s = -1.
\end{equation}
Later on we will consider the limit $m_0 \rightarrow 0$.
The term $\widehat{g}_{1,0}(z,Q^2)$ can be obtained by \cite{Zijlstra:1993sh}
\begin{align}\label{g1pro}
\begin{split}
\widehat{g}_{1,0}(z,Q^2)&=\frac{2}{(d-2)(d-3)}\frac{1}{p\cdot q}\,\epsilon_{\mu\nu\rho\sigma}
p^\rho q^\sigma\,\widehat{W}^{\mu\nu}_A \left(z,q,s=\frac{p}{m_0}\right).
\end{split} 
\end{align}
The hadronic tensor is given by
\begin{eqnarray}\label{Wamunu}
\widehat{W}^{\mu\nu}_A 
&=& 
4 \pi \alpha_s^2 C_F T_F
\int\,\frac{d^dp_2}{(2\pi)^d}\frac{d^dp_c}{(2\pi)^d}\frac{d^dp_{\overline{c}}}
{(2\pi)^d}(2\pi)^d\delta^d(p+q-p_2-p_c-p_{\overline{c}})\nonumber\\
&&\bigg\{\text{Tr}\bigg[\frac{\gamma_5\slashed{s}}{2}(\slashed{p}+m_0)
\bigg(\frac{\gamma^{\nu}(\slashed{p}+\slashed{q}+m_0)\gamma_\sigma(\slashed{p_2}+m_0)
\gamma_{\rho}(\slashed{p}
+\slashed{q}+m_0)\gamma^{\mu}}{\big[(p+q)^2-m^2_0\big]^2}\nonumber\\
&&\hspace{2cm}+\frac{\gamma^{\nu}(\slashed{p}+\slashed{q}+m_0)\gamma_\sigma(\slashed{p_2}+m_0)
\gamma^{\mu}(\slashed{p}_2-\slashed{q}+m_0)\gamma_\rho}{\big[(p+q)^2-m^2_0\big]
\big[(p_2-q)^2-m^2_0\big]}\nonumber\\
&&\hspace{2cm}+\frac{\gamma_\sigma(\slashed{p}_2-\slashed{q}+m_0)\gamma^\nu(\slashed{p}_2+m_0)
\gamma_{\rho}(\slashed{p}+\slashed{q}+m_0)\gamma^{\mu}}{\big[(p+q)^2-m^2_0\big]
\big[(p_2-q)^2-m^2_0\big]}\nonumber\\
&&\hspace{2cm}+\frac{\gamma_\sigma(\slashed{p}_2-\slashed{q}+m_0)\gamma^\nu(\slashed{p}_2+m_0)
\gamma^{\mu}
(\slashed{p}_2-\slashed{q}+m_0)\gamma_\rho}{\big[(p_2-q)^2-m^2_0\big]^2}\bigg)\bigg]
\nonumber\\
&&\hspace{2cm}\times\frac{4}{\big[q_2^2\big]^2}\,\bigg[p_c^{\rho}p_{\overline{c}}^{\sigma}
+p_c^{\sigma}
p_{\overline{c}}^{\rho}-(p_c\cdot p_{\overline{c}}+m^2)\,g^{\rho\sigma}\bigg]\bigg\}\nonumber\\
&&\hspace{2cm}
\times(2\pi)^{3}\delta_+(p_2^2-m_0^2)\delta_+(p_c^2-m^2)\delta_+(p_{\overline{c}}^2-m^2)\\
&&\hspace{-1.5cm}
\equiv 
{4 \pi \alpha_s^2 \,C_F T_F}\; \int\,\frac{d^dp_2}{(2\pi)^d}
\frac{d^dp_c}{(2\pi)^d}\frac{d^dp_{\overline{c}}}{(2\pi)^d}(2\pi)^d\delta^d(p+q-p_2-p_c-p_{\overline{c}})
\nonumber\\
&&\hspace{-1cm}\times(2\pi)^{3}\delta_+(p_2^2-m_0^2)\delta_+(p_c^2-m^2)\delta_
+(p_{\overline{c}}^2-m^2)\,\mathcal{T}^{\mu\nu}_{\;\rho\sigma}(p_2,q_2)\,
J_c^{\rho\sigma}(p_c,p_{\overline{c}}),
\nonumber
\end{eqnarray}
where $m$, ($p_{\overline{c}}$) $p_{c}$ are the mass and four momenta of the heavy (anti-)quark, $q_2=p_c+p_{\overline{c}}$ and 
$p_2$ denotes the momentum of the light quark in the final state. 
The distribution $\delta_+$ is defined by
\begin{equation}
 \delta_+(p^2-m^2_1)=\delta(p^2-m^2_1)\theta(p^0).
\end{equation}
The last line of Eq.~(\ref{Wamunu}) emphasizes that the phase space integral is factorized. 
The tensors $\mathcal{T}^{\mu\nu}_{\;\rho\sigma}$ and 
$J_c^{\rho\sigma}$ read
\begin{align}
\begin{split}
 \mathcal{T}^{\mu\nu}_{\;\rho\sigma}&=\frac{1}{\big[q_2^2\big]^2}\,\bigg\{\text{Tr}
\bigg[\frac{\gamma_5\slashed{s}}{2}(\slashed{p}+m_0)\bigg(\frac{\gamma^{\nu}(\slashed{p}
+\slashed{q}+m_0)
\gamma_\sigma(\slashed{p_2}+m_0)\gamma_{\rho}(\slashed{p}+\slashed{q}+m_0)\gamma^{\mu}}
{\big[(p+q)^2-m^2_0\big]^2}\\
&\hspace{2cm}+\frac{\gamma^{\nu}(\slashed{p}+\slashed{q}+m_0)\gamma_\sigma(\slashed{p_2}+m_0)
\gamma^{\mu}(\slashed{p}_2
-\slashed{q}+m)\gamma_\rho}{\big[(p+q)^2-m^2_0\big]\big[(p_2-q)^2-m^2_0\big]}\\
&\hspace{2cm}+\frac{\gamma_\sigma(\slashed{p}_2-\slashed{q}+m_0)\gamma^\nu(\slashed{p}_2+m_0)
\gamma_{\rho}(\slashed{p}
+\slashed{q}+m)\gamma^{\mu}}{\big[(p+q)^2-m^2_0\big]\big[(p_2-q)^2-m^2_0\big]}\\
&\hspace{2cm}+\frac{\gamma_\sigma(\slashed{p}_2-\slashed{q}+m_0)\gamma^\nu(\slashed{p}_2+m_0)
\gamma^{\mu}(\slashed{p}_2
-\slashed{q}+m_0)\gamma_\rho}{\big[(p_2-q)^2-m^2_0\big]^2}\bigg)\bigg]\bigg\},\\
J_c^{\rho\sigma}&=4\;\left[p_c^{\rho}p_{\overline{c}}^{\sigma}+p_c^{\sigma}p_{\overline{c}
}^{\rho}-(p_c\cdot p_{\overline{c}}+m^2)g^{\rho\sigma})\right].
\end{split} 
\end{align}
We consider the incoming quark directed in the $\widehat{z}$ axis. The momentum 
and spin of a longitudinally polarized quark are then given by
\begin{align}
\begin{split}
p&= (\sqrt{p^2+m^2_0},0,0,p) 
\longrightarrow 
p(1,0,0,1)+\left(\frac{m^2_0}{2p},\overrightarrow{0}\right),\\
s&=\frac{1}{m_0} (p,0,0,\sqrt{p^2+m^2_0})
\longrightarrow \frac{p}{m_0}(1,0,0,1)+\left(0,0,0,\frac{m_0}{2p}\right),~~~~\text{for}~~~m_0 
\rightarrow 0.
\end{split} 
\end{align}
Therefore, it is important to retain the linear terms in the mass $m_0$, which is finally 
canceled by the normalization of the spin vector.

The integrals are finite in $d=4$ dimensions to which we turn from now on. The phase space 
integrals can be carried out analytically leading to
\begin{align}\label{g1d4}
\begin{split}
\widehat{g}_{1,0}(z,Q^2)&=-\frac{g_s^4 C_F T_F}{12\pi^3}\int_{4 
m^2}^{Q^2\left(\frac{1-z}{z}\right)}\,
\frac{d\mu^2}{(2\pi)}\sqrt{\frac{\mu^2-4m^2}{\mu^2}}\bigg[\frac{\mu^2+2m^2}{2}\bigg]
\frac{Q^2(1-z)-z\mu^2}{8Q^4(1-z)\mu^4}\\
&\hspace{-1cm}\times\bigg\{\frac{2}{(1-z)^2\,(z\mu^2-Q^2)}\bigg[\mu^4\,z^2(4z^2-6z+3)
+\mu^2Q^2\,z(-8z^3+14z^2-9z+2)\\
&+3Q^4(2z^3-4z^2+3z-1)\bigg]\\
&\hspace{-1cm}+4\frac{2\mu^4\,z^3-2\mu^2Q^2\,z(2z^2-z+1)+Q^4(1+z^2)}{(Q^2(1-z)-\mu^2z)}\,
\ln\left(\frac{\mu^2z^2}{(1-z)(Q^2-z\mu^2)}\right)\bigg\}.
\end{split} 
\end{align}
It is convenient to perform the $\mu^2$-integral over 
\begin{align}\label{mutobeta}
 \beta=\sqrt{1-\frac{4 m^2}{\mu^2}}.
\end{align}

After applying this transformation, the integral becomes
\begin{align}\label{betaint}
\begin{split}
\widehat{g}_{1,0}(z,Q^2)&=\frac{4 a_s^2 C_F T_F}{3}\int_0^{\sqrt{1-4\frac{z}{\xi(1-z)}}}d\beta\, 
\frac{\beta ^2 \left(\beta^2-3\right)}{\left(1-\beta ^2\right)^3 
\xi^3 (1-z)^3 \left(\beta^2-1 + \frac{4 z}{\xi}\right)} \\
&\bigg\{\bigg[16 \left(1-\beta^2\right) \xi  \left(4 (z-1) z^2+1\right) z^2
-64 \left(4 z^2-6 z+3\right) z^3\\
&-3 \left(1-\beta ^2\right)^3 \xi ^3 (z-1)^2 (2 (z-1) z+1)+4 \left(1-\beta ^2\right)^2 
\xi ^2 (z-1)
\nonumber
\end{split}
\end{align}
\begin{align}
\begin{split}
&\times(z (4 z (2 z-5)+15)-5) z\bigg]+2(1-z)^2(\beta^2-1+\frac{4z}{\xi})\xi\bigg[32z^3\\
&-8z(1-z+2z^2)\xi(1-\beta^2)+(1+z^2)\xi^2(1-\beta^2)^2\bigg]
\ln\left(
\frac{ \frac{4z}{\xi(1-z)} - \frac{4z}{\xi}}
{1-\frac{4z}{\xi}-\beta^2}
\right)\bigg\}.
\end{split}
\end{align}
This term can finally be integrated analytically to yield (\ref{eq:LNS1C}).
\section{The virtual corrections}

\vspace*{1mm}
\noindent
The interaction of an on-shell fermion and the electromagnetic current is parameterized in terms of 
the Dirac and Pauli form factors $F_1(q^2)$, $F_2(q^2)$, respectively. In the space-like 
case one has
\begin{align}\label{Dirac}
\begin{split}
\langle p_2|J^\mu(q^2)|p_1\rangle &= \overline{u}(p_2)\,\Gamma^\mu\,u(p_1)
= \overline{u}(p_2)\,\bigg[\gamma^\mu\,F_1(q^2)-\frac{F_2(q^2)}{4m_0}\sigma^{\mu\nu}
q_\nu\bigg]\,u(p_1),
\end{split}
\end{align}
where $\sigma^{\mu\nu}=\frac{i}{2}\big[\gamma^\mu,\gamma^\nu\big]$. The correction can be 
obtained by the subtracted dispersion relation for the Dirac form factor.

\begin{figure}[H]
  \centering
  \includegraphics[scale=0.3]{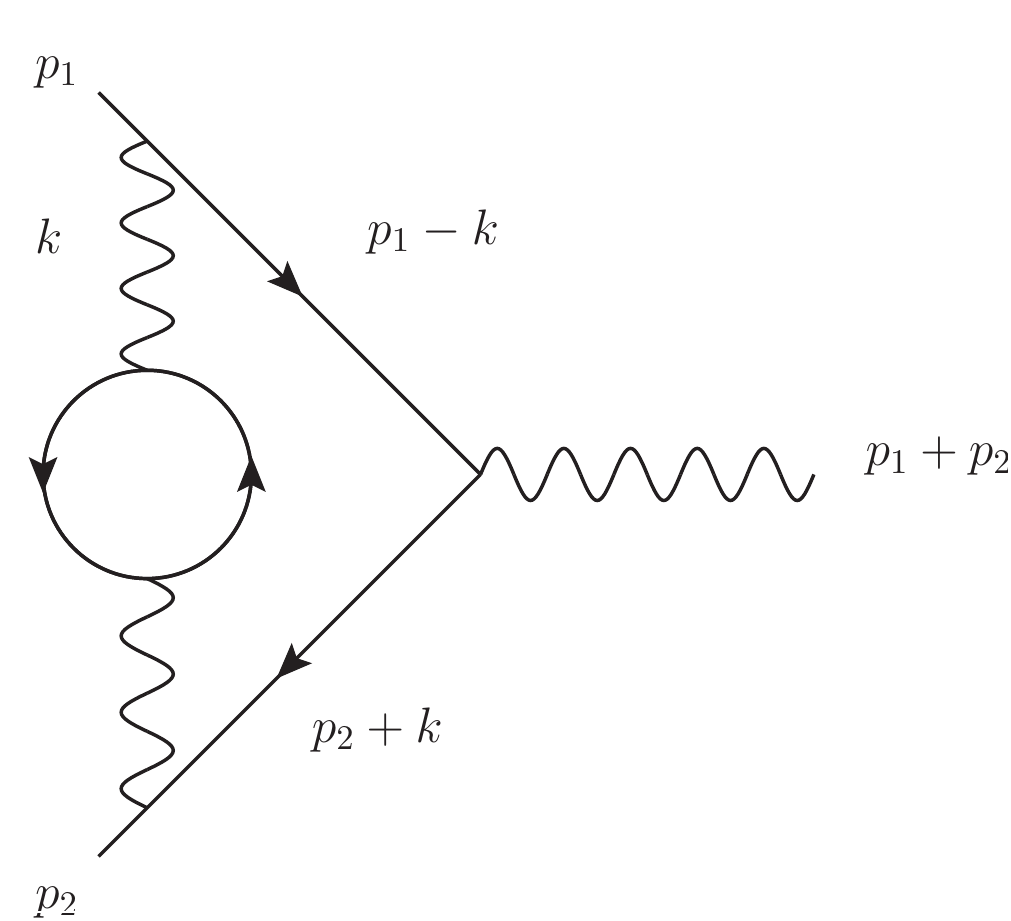}
  \caption{\sf \small Two-loop diagrams with vacuum polarization insertions.}
  \label{vpc}
\end{figure}

\begin{figure}[H]
\centering
\includegraphics[scale=0.3]{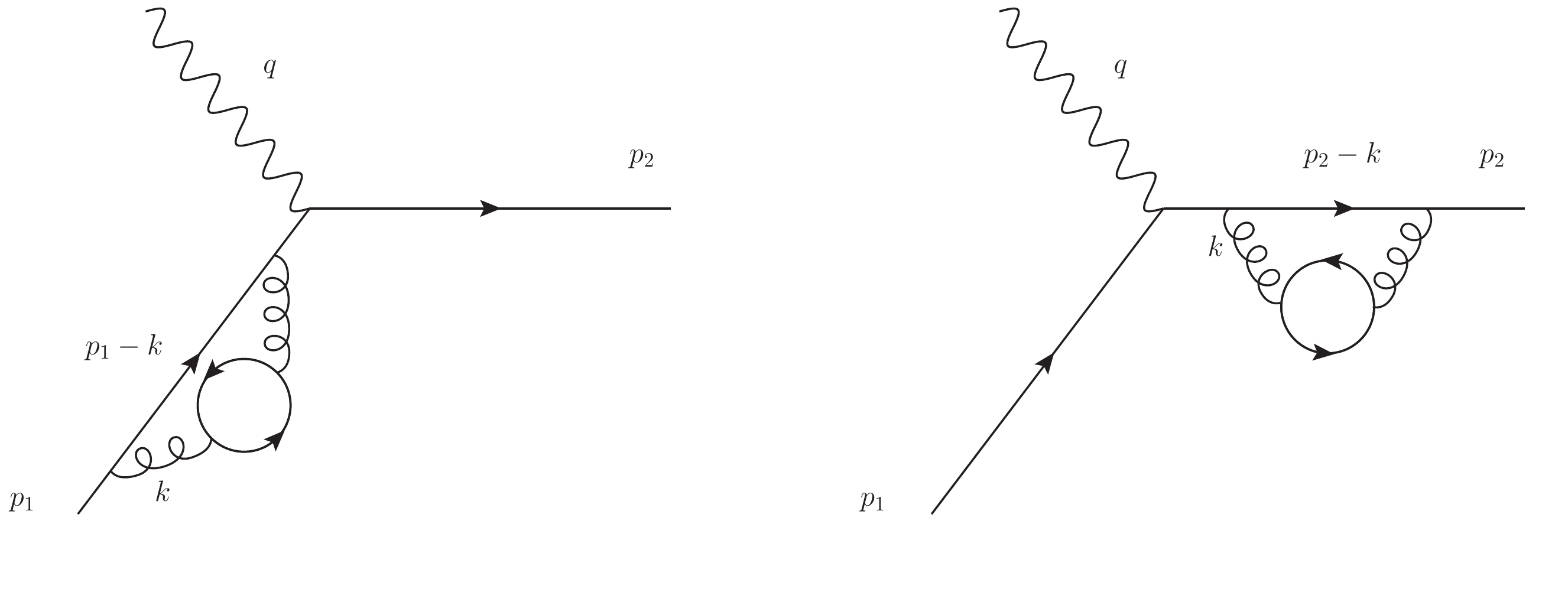}
\caption{\sf \small Reducible diagrams with fermion self energy insertion.}
\label{se}
\end{figure}

We will first perform the calculation in the time-like case and obtain then the space-like result by
analytic continuation. One has
\begin{equation}
F_1(s) - F_1(0) = \frac{s}{\pi} 
\int_{0}^\infty dz \frac{{\rm Im}(F_1(z))}{z(z-s)},
\label{eq:subdis}
\end{equation}
which can be calculated from the diagrams in Figures~\ref{vpc}, \ref{se} and applying the Ward identity of 
Subsection~\ref{sec:Ward}, with $s = (p_1+p_2)^2$. 
\noindent
Note that the integral (\ref{eq:subdis}) in the case of the unsubtracted dispersion relation 
diverges.  
The calculation proceeds in a similar way as in Refs.~\cite{Barbieri:1972as,Barbieri:1972hn}.
We will initially work in $d$ dimensions and use quarks of mass $m_0$ for the external particles
and $m$ for the heavy quark in the loop. Later on we take  the limit $d=4$ and $m_0=0$.

The Dirac form factor is projected to
\begin{equation}\label{genf1}
F_1(s)=-\frac{1}{2(d-2)(s-4m^2_0)}\,\text{Tr}\bigg[\left(\gamma^\mu
+\frac{2m_0(d-1)}{s-4m^2_0}(p_1-p_2)^\mu\right)(\slashed{p}_2-m_0)\Gamma^\mu\,(\slashed{p}_1+m_0)\bigg].
\end{equation}
The contribution of the diagram in Figure~\ref{vpc} to the vertex $\Gamma^\mu$ is given by
\begin{align}\label{twoloopgamma}
\begin{split}
 \overline{v}(p_2)\Gamma^{\mu}u(p_1)=&\! C_F T_F \int\bigg\{\frac{\overline{v}(p_2)
(ig_s\gamma^\rho)(-i)(\slashed{p}_2+\slashed{k}-m)\gamma^\mu\. 
i(\slashed{p}_1-\slashed{k}+m)(ig_s\gamma^\sigma)u(p_1)}{\big[(p_2+k)^2-m^2_0+i0\big]
\big[(p_1-k)^2-m^2_0+i0\big]}\\
 &\!\times \left(\frac{-i}{k^2+i0}\right)^2\, i\Pi(k^2)\left(k^2g_{\rho\sigma}-k_\rho 
k_\sigma\right)\bigg\}\;\frac{d^dk}{(2\pi)^d},
\end{split}
\end{align}
where $\Pi(k^2)$ denotes the vacuum polarization 
\begin{align}\label{kallen}
\begin{split} 
\Pi(k^2)-\Pi(0)&= 
- \frac{\alpha_s}{\pi}\frac{k^2}{3}\int_{4m^2}^{\infty}\frac{dx}{x^2}\frac{(x+2m^2)}{x-k^2}\sqrt{1-\frac{4m^2}{x}}
\end{split}
\end{align}
leading to
\begin{align}
\begin{split}
F_1(s)&=- i \frac{g_s^2 C_F T_F}{2(d-2)(s-4m^2_0)} \int \frac{d^dk}{(2\pi)^d}.
\bigg\{\text{Tr}\bigg[\left(\gamma^\mu+\frac{2m(d-1)}{s-4m^2}(p_1-p_2)^\mu\right)\\
& \times
(\slashed{p}_2-m_0)\gamma^\rho (\slashed{p}_2+\slashed{k}-m_0)\gamma_\mu
(\slashed{p}_1-\slashed{k}+m_0)\gamma^\sigma\,
(\slashed{p}_1+m_0)\bigg]\\
& \times
\frac{\left(k^2g_{\rho\sigma}-k_\rho 
k_\sigma\right)\Pi(k^2)}{\big[k^2+i0\big]^2\big[(p_1-k)^2-m^2_0+i0\big]
\big[(p_2+k)^2-m^2_0+i0\big]}
\bigg\}.
\end{split}
\end{align}
Its imaginary part is found putting the propagators
\begin{align}
\begin{split}
 \frac{1}{\big[(p_1-k)^2-m^2_0+i0\big]}&\rightarrow -(2\pi i)\delta_+((p_1-k)^2-m^2_0),\\
 \frac{1}{\big[(p_2+k)^2-m^2_0+i0\big]}&\rightarrow -(2\pi i)\delta_+((p_2+k)^2-m^2_0),
\end{split} 
\end{align}
on shell, cf.~\cite{Barbieri:1972as,Barbieri:1972hn}. Note that when taking the cut a minus 
sign has to be introduced \cite{Veltman:1963th,Kniehl:1996rh}
\begin{align}
\begin{split}
 2\text{Im}\big(F_1(s)\big)&= \frac{g_s^2 C_F T_F}{2(d-2)(s-4m_0^2)}
 \int\bigg\{\text{Tr}\bigg[\left(\gamma^\mu+\frac{2m(d-1)}{s-4m_0^2}(p_1-p_2)^\mu\right)\\
& \hspace{1.cm}
\times (\slashed{p}_2-m_0)\gamma^\rho (\slashed{p}_2+\slashed{k}-m_0)\gamma_\mu
(\slashed{p}_1-\slashed{k}+m_0)\gamma^\sigma\,(\slashed{p}_1+m_0)\bigg]\frac{\left(k^2g_{\rho\sigma}
-k_\rho k_\sigma\right)\Pi(k^2)}{\big[k^2+i0\big]^2}\bigg\}\\
 &\hspace{1.cm} 
\times (2\pi)^2\delta_+((p_1-k)^2-m^2_0)\delta_+((p_2+k)^2-m^2_0)\frac{d^dk}{(2\pi)^d}.
\end{split} 
\end{align}
The longitudinal parts of the photon polarization can be shown to vanish. One finally obtains
\begin{align}\label{Imd}
\begin{split}
\text{Im}\big(F_1(s)\big)
&=\frac{\alpha^2_s}{6} C_F T_F \frac{1}{\sqrt{s}}\left(\frac{s}{4}-m^2_0\right)^{\frac{d-3}{2}}
\frac{\Omega_{d-3}}{(2\pi)^{d-2}}\int_{4m^2}^\infty\frac{dx^2}{x^2}\sqrt{1-\frac{4m^2}{x}}
(x+2m^2)\\
&\int_0^\pi d\theta\big(\sin\theta\big)^{d-3} 
\frac{\cos^2\theta(s+4m_0^2(d-2))+\cos\theta(s(6-d)-4dm^2_0)+s(d-3)}{(s-4m^2_0)(1-\cos\theta)
+2x},
\end{split} 
\end{align}
with \cite{FICHTENHOLZ}
\begin{equation}
\Omega_{d} = \frac{2\pi^{\frac{d+1}{2}}}{\Gamma\left(\frac{d+1}{2}\right)}.
\end{equation}

We are now turning to $d=4$ and obtain
\begin{align}
\begin{split}
\text{Im}\big(F_1(s)\big)&=-\frac{\alpha^2_s}{24 \pi} C_F T_F \sqrt{1-\xi_1} \int_0^\pi 
d\theta\sin\theta\,\big[\cos^2\theta(1+2\xi_1)+2\cos\theta(1-2\xi_1)+1\big]\\
&\int_0^1d\beta\frac{\beta^2(\beta^2-3)}{(1-\cos\theta)(1-\xi_1)+2\xi_2-\beta^2 (1-\xi_1)(1-\cos\theta)},
\end{split} 
\end{align}
with 
\begin{equation}
\xi_2 = \frac{4 m^2}{s}. 
\end{equation}
Furthermore we consider the limit $\tfrac{4 m_0^2}{s} = \xi_1 
\rightarrow 0$. After integrating over $\beta$ we have
\begin{align}\label{ImF1m0}
\begin{split}
\text{Im}\big(F_1(s,m_0=0)\big)&=-\frac{\alpha^2_s}{24\pi}\int_{-1}^1dX 
\frac{(1+X)^2 }{(1-X)^2}\bigg[\frac{5-5X-6\xi_2}{3}
\\
&-\sqrt{1+\frac{2\xi_2}{1-X}}(1-X-\xi_2)\,
\ln\left(\frac{\sqrt{1+\frac{2\xi_2}{1-X}}+1}{\sqrt{1+\frac{2\xi_2}{1-X}}-1}\right)\bigg],
\end{split} 
\end{align}
where $X = \cos\theta$. Trading the root in (\ref{ImF1m0}) as a new integration variable one obtains
\begin{align}
 \begin{split}
\text{Im}\big(F_1(s,m_0=0)\big)&=-\frac{\alpha^2_s}{24\pi}\bigg[-\frac{5}{9}(53+33\xi_2)
+\frac{1}{3}\sqrt{1+\xi_2}(38+23\xi_2)\ln\left(\frac{\sqrt{1+\xi_2}+1}{\sqrt{1+\xi_2}-1}\right)\\
&\hspace{1cm}+\left(-2+\frac{3\xi^2_2}{4}\right)\ln^2\left(\frac{\sqrt{1+\xi_2}+1}{\sqrt{1+\xi_2}-1}
\right)\bigg].
\end{split}
\end{align}
In order to determine the complete expression of the form factor in the time-like region, 
we use again the subtracted relation (\ref{eq:subdis})
\begin{align}\label{F1m0}
\begin{split}
F_1(s,m_0=0)&= -\frac{\alpha^2_s}{24\pi^2}\bigg\{-\frac{1213}{54}-\frac{119}{3}s_\xi^2
+\left(\frac{200}{9}+\frac{110s_\xi^2}{3}\right)\ln(2)-\frac{5}{9}(20+33s_\xi^2)\ln(1-s_\xi^2)\\
&-s_\xi\frac{15+23s_\xi^2}{3}\left[\text{Li}_2\left(-\frac{1-s_\xi}{1+s_\xi}\right)-\text{Li}_2
\left(-\frac{1+s_\xi}{1-s_\xi}\right)\right]\\
&-\left(\frac{5}{2}+3s_\xi^2-\frac{3s_\xi^4}{2}\right)\left[\text{Li}_3\left(-\frac{1-s_\xi}{1+s_\xi}
\right)+\text{Li}_3\left(-\frac{1+s_\xi}{1-s_\xi}\right) - 2 \zeta_3\right]\bigg\},
\end{split} 
\end{align}
where 
\begin{equation}
s_\xi=\sqrt{1+\xi_2}~. 
\end{equation}
The analytic continuation of (\ref{F1m0}) to the space-like region is obtained by replacing
\begin{equation}
s_\xi \rightarrow \hat{s}_\xi = \sqrt{1 - \frac{4 m^2}{Q^2}} \equiv \tilde{\lambda}~,
\end{equation}
cf.~Eq.~(\ref{eq:LqNSg1V}). The real part of Eq.~(\ref{F1m0}), if considered in the time-like region,
may be compared with a result in \cite{Rijken:1995gi}, Eq.~(A1), for the Drell-Yan process and agrees.  
\subsection{A Ward identity}
\label{sec:Ward}

\vspace{1mm}
\noindent
In the following we derive the relation of the self-energy insertions to $F_1(0)$ 
through a Ward identity.
The graphs of Figure~\ref{se} obey
\begin{align}\label{secont}
\begin{split}
\frac{1}{2}\bigg[(a)+(b)\bigg] &= (Z_2-1)\, \overline{u}(p_2)\,(-ie\,\gamma_\mu)\,u(p_1), 
\end{split} 
\end{align}
where $e$ denotes the electric charge.

\begin{figure}[h!]
\centering
\includegraphics[scale = 0.3]{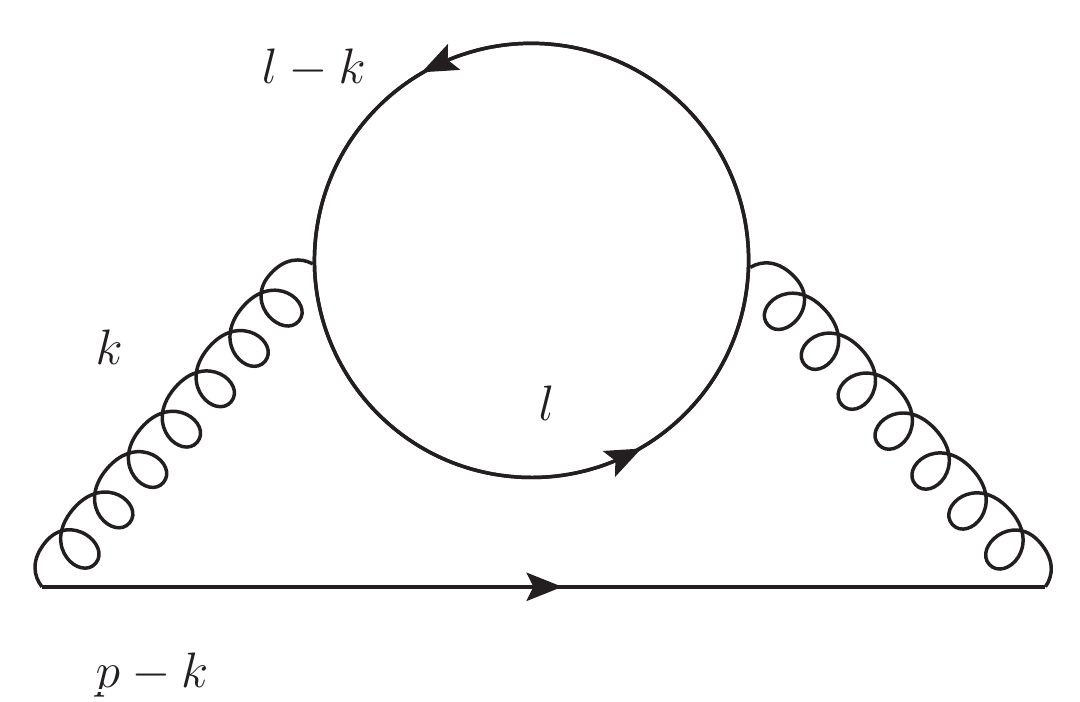}
\caption{\sf \small Proper self energy diagram.}
\label{properse}
\end{figure}

By following the notation of \cite{Yndurain:1999ui}, we define $-i\Sigma(p^2)$ as the proper 
self 
energy diagram in Figure~\ref{properse} and we find the renormalization constant by taking the 
residue of the complete quark propagator on the mass shell. In the case of massless external 
fermions we have
\begin{equation}
i S_F =\frac{i}{\slashed{p} -\Sigma(p)}\;{\underset{p^2\rightarrow 0}{\longrightarrow}}\; i\frac{Z_2}{\slashed{p}},
\end{equation}
so that $Z_2=1+\frac{\Sigma(p^2)\slashed{p}}{p^2}$ at leading order. The next step is to extract 
the spin structure of the self energy $\Sigma(p^2)=\slashed{p}\overline{\Sigma}(p^2)$, where the 
latter object is a scalar function, so that we can write the renormalization constant $Z_2$ as
\begin{equation}\label{Z2}
Z_2=1+\overline{\Sigma}(p). 
\end{equation}
By introducing (\ref{Z2}) into Eq.~(\ref{secont}), we obtain the contribution of self energy 
diagrams. 

We consider the derivative of the self energy 
\begin{align}
\begin{split}
 -i\Sigma(p^2)&= -ig_s^4C_F T_F\int 
\frac{d^dk}{(2\pi)^d}\,\frac{\gamma^\alpha(\slashed{p}-\slashed{k})\gamma^\beta}{(p-k)^2(k^2)^2}\bigg[k^2g_{\alpha\beta}-k_\alpha k_\beta\bigg]\,\Pi(k^2),
\end{split} 
\end{align}
with $\Pi(k^2)$ given in Eq.~(\ref{kallen}). It reads
\begin{align}\label{seward}
\begin{split}
\frac{\partial\Sigma}{\partial p_\mu}&=g_s^4C_FT_F\,\int\frac{d^dk}{(2\pi)^d} \Pi(k^2)\bigg[k^2g_{\alpha\beta}-k_\alpha k_\beta\bigg]\,\bigg\{\frac{\gamma^\alpha\,\gamma_\mu\,\gamma^\beta}{(p-k)^2\big[k^2\big]^2}-2(p-k)_\mu\frac{\gamma^\alpha(\slashed{p}-\slashed{k})\gamma^\beta}{\big[(p-k)^2\big]^2\big[k^2\big]^2}\bigg\}\\
&=g_s^4C_FT_F\,\int\frac{d^dk}{(2\pi)^d} \Pi(k^2)\bigg[k^2g_{\alpha\beta}-k_\alpha k_\beta\bigg]\,\bigg\{-\frac{\gamma^\alpha(\slashed{p}-\slashed{k})\gamma_\mu(\slashed{p}-\slashed{k})\gamma^\beta}{\big[(p-k)^2\big]\big[k^2\big]^2}\bigg\}.
\end{split} 
\end{align}
The vertex function is given by 
\begin{align}\label{vertward}
\begin{split}
-ie\,\Lambda_\mu(p_2,p_1)\equiv 
-ie\,F_1(q^2)\gamma_\mu=(-ie)\,g_s^4C_FT_F\,&\int\frac{d^dk}{(2\pi)^d} 
\Pi(k^2)\bigg[k^2g_{\alpha\beta}-k_\alpha k_\beta\bigg]\\
&\times\bigg\{\frac{\gamma^\alpha(\slashed{p}_2-\slashed{k})
\gamma_\mu(\slashed{p}_1-\slashed{k})\gamma^\beta}{\big[(p_2-k)^2\big]
\big[(p_1-k)^2\big]\big[k^2\big]^2}\bigg\},
\end{split} 
\end{align}
with $(p_2-p_1)^2 = q^2$.
In the limit of zero momentum transfer, $q^2 \rightarrow 0$, the vertex function becomes 
$\Lambda_\mu(0) = F_1(0)\gamma_\mu$. By comparing (\ref{vertward}) and (\ref{seward}) one 
gets
\begin{align}
\begin{split}
\Lambda_\mu(0)\rightarrow g_s^4C_FT_F\,&\int\frac{d^dk}{(2\pi)^d} \Pi(k^2)\bigg[k^2g_{\alpha\beta}
-k_\alpha k_\beta\bigg]\bigg\{\frac{\gamma^\alpha(\slashed{p}-\slashed{k})\gamma_\mu(\slashed{p}
-\slashed{k})\gamma^\beta}{\big[(p-k)^2\big]^2\big[k^2\big]^2}\bigg\}
=-\frac{\partial\Sigma(p^2)}{\partial p^\mu},
\end{split}
\end{align}
i.e.
\begin{align}\label{wardid2}
\begin{split}
F_1(Q^2=0)\,\gamma_\mu&=-\gamma_\mu\,\overline{\Sigma}(p)-\slashed{p}\,\frac{\partial\overline{\Sigma}}{\partial p^\mu}. 
\end{split} 
\end{align}
Eq.~(\ref{wardid2}) allows to write the function $\overline{\Sigma}(p)$ in terms of the form 
factor at zero momentum transfer. Therefore (\ref{secont}) can be written as
\begin{align}
\begin{split}
\frac{1}{2}\bigg[(a)+(b)\bigg]&=\bigg[\frac{1}{2}\overline{\Sigma}(p_1)+\frac{1}{2}\overline{\Sigma}(p_2)\bigg]\;\overline{u}(p_2)\,(-ie\gamma_\mu)\,u(p_1)\\
&=\frac{1}{2}\overline{u}(p_2)\,\bigg(-ie\,\overline{\Sigma}(p_1)\gamma_\mu\bigg)\,u(p_1)+\frac{1}{2}\overline{u}(p_2)\,\bigg(-ie\,\overline{\Sigma}(p_2)\gamma_\mu\bigg)\,u(p_1)\\
&=\frac{1}{2}\,\overline{u}(p_2)\,\bigg[-ie\,\bigg(-F_1(0)\gamma_\mu-\slashed{p}_1\frac{\partial\overline{\Sigma}(p_1)}{\partial{p}_1^\mu}\bigg)\bigg]\,u(p_1)\\
&\,+\frac{1}{2}\,\overline{u}(p_2)\,\bigg[-ie\,\bigg(-F_1(0)\gamma_\mu-\slashed{p}_2\frac{\partial\overline{\Sigma}(p_2)}{\partial{p}_2^\mu}\bigg)\bigg]\,u(p_1),
\end{split} 
\end{align}
where in the last identity we introduced $\overline{\Sigma}(p_i)$ in terms of $F_1(0)$, according to 
the relation (\ref{wardid2}). In the expression above, all the terms proportional to 
$\frac{\partial\overline{\Sigma}(p)}{\partial p^\mu}$ vanish because of the Dirac equation, finally
\begin{equation}\label{secontfin}
\frac{1}{2}\bigg[(a)+(b)\bigg]=\overline{u}(p_2)\,\bigg(ie\,F_1(0)\gamma_\mu\bigg)\,u(p_1).  
\end{equation}
In conclusion, the scattering amplitude of a massless quark and an off shell photon with momentum $q$ 
is given by the sum of the virtual correction to the proper $qq$-gauge boson  vertex, 
depicted in 
Figure~\ref{vpc}, which we computed via dispersion relations up to an offset $F_1(0)$, and the 
self energies contributions above,
which proves that the results of the subtracted dispersive approach Eqs. (\ref{F1m0}, 
\ref{eq:subdis}) give the complete renormalized form factor.
\section{The first moment}
\label{sec:A2}

\vspace*{1mm}
\noindent
In the following we derive the heavy flavor contributions to the Bjorken sum rule to $O(a_s^2)$.
The Compton contribution is obtained by the integral
\begin{align}\label{bjsr}
\begin{split}
 A_{g_1}(\xi)&=\int_0^\frac{\xi}{\xi+4}dz\,\widehat{g}_1(z,Q^2),
\end{split} 
\end{align}
where $\widehat{g}_1(z,Q^2)$ is given by the integral (\ref{betaint}), see also~(\ref{eq:LNS1C}). 
To 
obtain the analytic expression, it is easier to step back one integral and to use
\begin{align}
\begin{split}
A_{g_1}(\xi)=\int_0^{\frac{\xi}{\xi+4}}dz\,\int_{4m^2}^{Q^2\left(\frac{1-z}{z}\right)}I(z,\mu^2)\,
d\mu^2&=\int_{4m^2}^{\infty}d\mu^2\int_0^{\frac{Q^2}{Q^2+\mu^2}} I(z,\mu^2)\; dz,
\end{split} 
\end{align}
where the integrand $I(z,\mu^2)$ is the same (including normalization) as in (\ref{g1d4}). 
The $z$-integral yields
\begin{align}
\begin{split}
A_{g_1}(Q^2)&={\hat{a}_s^2 C_F 
T_F}\int_1^{\infty}dx\,\sqrt{\frac{x-1}{x}}\frac{1+2x}{1152\xi^2x^4}\bigg\{4x\xi^3-1088x^3\xi+248x^2\xi^2\\
&\!\!+192x^2(\xi-4x)^2\ln^2\left(\frac{4x}{\xi}\right)-\bigg(1280x^4+512x^3\xi-16x\xi^3-\xi^4\bigg)\ln(\xi)\\
&\!\!-32x^2\ln\left(\frac{4x}{\xi}\right)\bigg[40x^2+40x\xi-9\xi^2+6(\xi-4x)^2\ln\left(\frac{4x-\xi}{\xi}\right)\bigg]\\
&\!\!+\bigg[1280x^4+512x^3\xi-16x\xi^3-\xi^4\bigg]\ln\left(\xi+4x\right)\\
&\!\!+192x^2(\xi-4x)^2\text{Li}_2\left(\frac{\xi}{4x}\right)\bigg\},
\end{split} 
\end{align}
where $x =\mu^2/(4 m^2)$. The result of the last integration can be conveniently 
expressed in terms of the variables
\begin{eqnarray}
 \lambda =\sqrt{1+\frac{4}{\xi}},~~~~~~~~
 \tilde{\lambda} = \sqrt{1-\frac{4}{\xi}}.
\end{eqnarray}
The $x$-integral results into
\begin{align}\label{taggedsr}
\begin{split}
A_{g_1}(\xi)&=\hat{a}_s^2 C_F T_F \bigg\{\left(\frac{2}{\xi^2}-\frac{1}{3}\right)\bigg[\text{Li}_3
\left(\frac{\tilde{\lambda}+1}
{\tilde{\lambda}-1}\right)+\text{Li}_3\left(\frac{\tilde{\lambda}-1}{\tilde{\lambda}+1}\right)-2\zeta_3\bigg]\\
&+\tilde{\lambda}\left(\frac{19}{18}-\frac{23}{9 \xi}\right)\bigg[\text{Li}_2\left(\frac{\tilde{\lambda}+1}
{\tilde{\lambda}-1}\right)-\text{Li}_2\left(\frac{\tilde{\lambda}-1}
{\tilde{\lambda}+1}\right)\bigg]-\frac{5}{12 \xi^2}\ln^2\left(\frac{\lambda+1}{\lambda-1}\right)\\
&+\lambda\ln\left(\frac{\lambda+1}{\lambda-1}\right) \bigg[-\frac{527}{2520}-\frac{401}{420 \xi}
-\frac{53\xi}{2520}-\frac{\xi^2}{1680}\bigg]\\
&+\ln(\xi)\bigg[\frac{265}{108}-\frac{55}{9 \xi}+\frac{\xi}{45}+\frac{\xi^2}{1680}\bigg]
+\frac{19591}{1260 \xi}+\frac{\xi}{840}-\frac{42047}{9072}\bigg\}
\end{split} 
\end{align}
We have still to add the virtual contribution (\ref{eq:LqNSg1V}) and the contribution due to the term
with massless final states (\ref{eq:massless_g1}) which yields 
\begin{align}\label{Arv2}
\begin{split}
 A_{g_1}(\xi) &= \hat{a}_s^2 C_F T_F \bigg\{-\frac{5}{12 
\xi ^2} \ln ^2\left(\frac{\lambda +1}{\lambda -1}\right)
-\frac{\lambda  \left(3 \xi ^3+106 \xi ^2+1054 \xi +4812\right)}{5040 \xi} 
\ln \left(\frac{\lambda +1}{\lambda -1}\right) \\
 &+\frac{6 \xi^2+2735 \xi+11724}{5040\xi}
+\frac{\xi  (3 \xi +112)}{5040} \ln(\xi) + \frac{1}{4} \ln(\xi) \bigg\}~.
\end{split} 
\end{align}
In the asymptotic limit one obtains
\begin{align}\label{Arv1}
\begin{split}
 A_{g_1}(\xi) &\propto \hat{a}_s^2 C_F T_F \frac{1}{2} + O\left(\frac{\ln^2(\xi)}{\xi}\right)~. 
\end{split} 
\end{align}
Note that the pure Compton contribution diverges like $\sim \ln^3(\xi)$. Adding the virtual 
corrections, the term still diverges $\sim \ln(\xi)$. This behaviour is obtained considering
tagged heavy quark production instead of the inclusive heavy flavor corrections, 
cf.~\cite{Blumlein:1998sh}.

\vspace{5mm}
\noindent
{\bf Acknowledgment.}~
We would like to thank A.~Behring, K.~Chetyrkin, C.~Schneider, and A.~Vogt for discussions. The graphs 
have been drawn using {\tt  Axodraw}~\cite{Vermaseren:1994je}. This work was supported in part by the 
European Commission through contract PITN-GA-2012-316704 ({HIGGSTOOLS}).

\clearpage

\end{document}